\documentclass[12pt,preprint]{aastex}              
\usepackage{natbib}
\usepackage{graphicx}

\bibpunct{(}{)}{;}{a}{}{,}

\newcommand{\new}{}
\newcommand{\newrsk}{}

\begin{document}                          

\title{The Formation of Stellar Clusters in Turbulent Molecular Clouds: \\
Effects of the Equation of State}    

\author{Yuexing Li\altaffilmark{1,2}, Ralf S. Klessen\altaffilmark{3} and
  Mordecai-Mark Mac Low\altaffilmark{1,2}} 

\affil{$^{1}$Department of Astronomy, Columbia University, New York,
NY 10027, USA}
\affil{$^{2}$Department of Astrophysics, American Museum of Natural
History, 79th Street at Central Park West, New York, NY 10024-5192, USA}
\affil{$^{3}$Astrophysikalisches Institut Potsdam, An der Sternwarte
16, D-14482 Potsdam, Germany} 
\email{yxli@astro.columbia.edu, rklessen@aip.de, mordecai@amnh.org} 

\begin{abstract}
  
  We study the effect of varying the equation of state on the formation of
  stellar clusters in turbulent molecular clouds, using three-dimensional,
  smoothed particle hydrodynamics simulations. Our results show that the
  equation of state helps determine how strongly self-gravitating gas
  fragments.  The degree of fragmentation decreases with increasing
  \new{polytropic exponent} $\gamma$ in the range $0.2 < \gamma < 1.4$,
  although the total amount of \new{mass accreted onto collapsed fragments}
  appears to remain roughly 
  constant through that range.  Low values of $\gamma$ are expected to lead to
  the formation of dense clusters of low-mass stars, while $\gamma>1$ probably
  results in the formation of isolated and massive stars. Fragmentation and
  collapse ceases entirely for $\gamma > 1.4$ as expected from analytic arguments.  
  The mass spectrum of overdense gas clumps is roughly log-normal for {\em
    non}-self-gravitating turbulent gas, but changes to a power-law under the
  action of gravity. The spectrum of collapsed cores, on the other hand,
  remains log-normal for $\gamma\le 1$, but flattens markedly for $\gamma >1$.
  The density PDFs approach log-normal, with widths that decrease with
  increasing $\gamma$.  Primordial gas may have effective $\gamma >
  1$, in which case these results could help explain why models of the
  formation of the first stars tend to produce isolated, massive
  objects.
\end{abstract}


\section{INTRODUCTION}

Stars form alone and in groups, associations, and clusters (see
\citealt{pud02} and \citealt{ward02} for recent reviews). Yet the
origin of this diversity in the process of star formation remains
poorly understood.  Recent work suggests that stars form by
gravitational collapse of shock-compressed density fluctuations
generated by the supersonic turbulence generally observed in molecular
clouds (e.g. \citealt{elm93, pn99, khm00, osg01}; see \citealt{mk03} for a
recent review).  If so, then star formation is controlled by turbulent
fragmentation and dynamical collapse of individual, Jeans-unstable,
protostellar clumps. 

Fragmentation of gas clouds has been studied for more than a century,
yet the process remains poorly understood. There have been analytical models
\citep{jeans02, low76}, and numerical investigations of the effects of various 
physical processes on the collapse, such as geometry and rotation of the
clouds (see \citealt{bob93} for a review), and magnetic fields
\citep{galli01}. Recently, the effects of turbulence have been studied
extensively in a series of three-dimensional simulations using 
both grid-based, Eulerian and particle-based, Lagrangian hydrodynamics
(\citealt{kbb98}; \citealt[hereafter Paper I]{kb00}; \citealt[Paper
II]{khm00}; \citealt[Paper III]{hmk01}; and \citealt[Paper IV]{rsk01}).    
These papers show that turbulence strongly influences the fragmentation of
molecular clouds. Fast, clustered collapse and star formation occurs
in regions with turbulence insufficient to support against
gravitational collapse, while slow, scattered star formation results
from strong turbulent support. However, these results are based on
isothermal models with \new{polytropic exponent} $\gamma=1$ only.  The role of
the equation of state (EOS), which is essential in understanding the
physical structure and stability of the turbulent clouds, remained
unexplored.

The balance of heating and cooling in a molecular cloud can be
approximately described by a polytropic EOS
\begin{equation}
P = K\rho^{\gamma},
\label{eq_eos}
\end{equation}
where $K$ is a constant, and $P, \rho$ and $\gamma$ are thermal pressure, gas
density and \newrsk{polytropic exponent}, respectively.  The density 
structure generated by supersonic motions in highly compressible
turbulence depends on the EOS.  The stiffness of the EOS can largely
control the density probability distribution function (PDF) in strongly
compressible turbulence \citep{scalo98, pv98, vg01}. In particular,
one-dimensional simulations by \citet{pv98} show the shape of  
the density PDF varying with $\gamma$. They found that the PDF is log-normal
when $\gamma = 1$ (isothermal), and develops a power-law tail for $\gamma \neq
1$. \new{\citet{ss00} showed that radiatively cooling gas can be described by a
piecewise polytropic EOS, in which the polytropic exponent $\gamma$ changes with gas
density \newrsk{$\rho$}. They conducted a detailed analysis including
chemistry, thermal balance and radiative transfer, and found that in the
interstellar medium the effective polytropic exponent $\gamma$ has a range of
$0.2 < \gamma < 1.4$. \newrsk{They further predicted that}
the initial mass function (IMF) of protostellar cores would vary with
$\gamma$, yielding a peaked IMF for $\gamma > 1$ and a power-law function for
$\gamma < 1$}.  

In this paper, we perform three-dimensional simulations of
self-gravitating, supersonic turbulence over the range of $\gamma$
identified by \citet{ss00} to investigate the \newrsk{effect of varying the 
EOS on the dynamical evolution  and fragmentation behavior} of turbulent
clouds and the subsequent formation of protostellar clusters. 
\newrsk{To quantify the dependency on $\gamma$, we  consider a true polytropic
EOS, with $\gamma$ being strictly constant in each simulation regardless of
the density, rather than the piecewise polytrope used by \citet{ss00} and
earlier workers.} 

In \S\ref{sec_com} we describe our
computational methods; in \S\ref{sec_result} we give results on
fragmentation, the resulting mass distribution of protostellar cores,
and the gas density PDF; and in \S\ref{sec_discussion} we conclude and
speculate about the implications or our work for the formation of star
clusters and primordial star formation.

\section{METHODS}
\label{sec_com}
In analytical work on the stability of a self-gravitating, isothermal medium,
\citet{jeans02} derived a relation between the oscillation frequency $\omega$
and the wavenumber $k$ of small perturbations,
\begin{equation}
\label{eq_jeans1}
\omega^2 - c^2_sk^2 + 4\pi G\rho =0
\end{equation}
where $c_s$ is the isothermal sound speed, $G$ the gravitational
constant, and $\rho$ is the gas density.  The medium is
unstable to fragmentation at all wavelengths greater than a critical
length $\lambda_J = 2\pi / k_J$, or equivalently all masses exceeding
the Jeans mass,
\begin{equation}
\label{eq_jeans2}
M_J = \rho\lambda^3_J = \left(\frac{\pi}{G}\right)^{3/2} \rho^{-1/2} c^3_s
\end{equation}
will collapse under their own weight. Note that we use a cubic
definition of $M_J$.
 
In a polytropic cloud with EOS given by equation~\ref{eq_eos},
the sound speed is
\begin{equation}
c_s = \left(\frac{dP}{d\rho}\right)^{1/2} =
(K\gamma)^{1/2}\rho^{(\gamma-1)/2}, 
\end{equation} 
so the Jeans mass will be given as a function of $\gamma$ and $\rho$,
\begin{equation}
\label{eq_jeans3}
M_J = \left(\frac{K\pi}{G}\right)^{3/2} \gamma^{3/2}
\rho^{\frac32(\gamma-\frac43)} \,.
\end{equation}
\new{The specific internal energy is $u = K\rho^{\gamma-1}/(\gamma_{ad} -1)$, where
$\gamma_{ad}$ is the adiabatic index, with $\gamma_{ad} = 5/3$ for ideal gas}. For
isothermal gas, $\gamma =1$, and $K = c^2_s$.

We carry out simulations to determine when and how fragmentation
occurs.  We use a smoothed particle hydrodynamics (SPH) code \citep{benz90} in
order to resolve several orders of magnitude in density during the collapse of
shock-compressed regions. Our version of the code includes periodic boundary
conditions \citep{rsk97} and can replace high-density cores with sink
particles \citep{bbp95}. Sink particles accrete surrounding gas particles
while conserving mass and momentum, but they only interact gravitationally. They
prevent time step from becoming prohibitively short in very dense
regions. This allows us to follow the dynamical evolution of the system over
many free-fall times.  

We set up a periodic cube with side length $L=2$ and total mass of
unity such that initial density $\rho_0 = 0.125$, and sound speed $c_s
= 0.1$.  The treatment of the polytropic EOS follows \citet{bob91}, where we
set $K =(R_g/\mu)T_0\rho^{1-\gamma}_0$.  We choose values consistent with the
sound speed of $R_g = 2/3$, $\mu = 1$, and $T_0 = 0.015$ for the gas constant,
mean molecular weight, and initial temperature of the cloud.  Uniform
turbulence is driven with the method described by \citet{mm98} and
\citet{mm99}, adding energy over a small range of wavenumbers $k$.  Driving in 
two wavenumber ranges is considered here, $1 \le k \le 2$ and $7 \le k \le
8$, corresponding to models $B1h$ and $B3$ in Paper II. In each case,
the driving strength is chosen to ensure the same rms velocity despite
the different driving wavenumbers, so that the average turbulent Jeans
mass $\langle M_{J}\rangle_{turb} = 3.2$ and Jeans Number $N_J = 64$,
are the same.  In a self-gravitating medium, the maximum resolvable
density is determined by the requirement that the local thermal Jeans
mass be resolved by the SPH kernel \citep{bab97}; our models all have the same 
resolution limit. 

The models presented here are computed in normalized units. If scaled
to mean densities of $n({\rm H}_2) \approx 10^2 $cm$^{-3}$, and a
temperature of $11.4 $K (i.e.\ an isothermal sound speed $c_s = 0.2
\,$km$\,$s$^{-1}$), values appropriate for a dark cloud like Taurus
\citep{hart98}, then our simulation cube holds a mass of $M \approx 4
\times 10^3\,$M$_{\odot}$ and has a size of $L \approx 9\,$pc. For details on 
the scaling relations, see Paper II.

We computed models with $0.2 \leq \gamma \leq 1.4$, covering the range
suggested for interstellar gas by \citet{ss00}.  We vary
$\gamma$ in steps of 0.1 in otherwise identical simulations for both
driving wavenumbers. Each simulation starts with a uniform density
$\rho_0 = 0.125$.  Driving begins immediately, while self-gravity is
turned on at t=2.0, after the turbulence is fully established (see
Paper II). Our models use $200\,000$ particles. Twenty-six simulations
were performed simultaneously using a serial version of the code on
single 1 GHz Pentium III processors of the Parallel Computing Facility
of the American Museum of Natural History (AMNH). Each computation
took about 6 months.

\section{RESULTS}
\label{sec_result}
\subsection{Turbulent Fragmentation}
\label{subsec_frag}
\begin{figure}
\begin{center}
\includegraphics[height=1.5in]{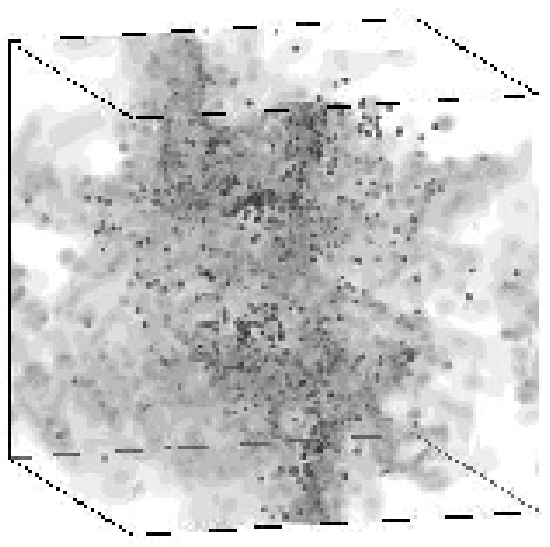}
\hspace{0.1cm}
\includegraphics[height=1.5in]{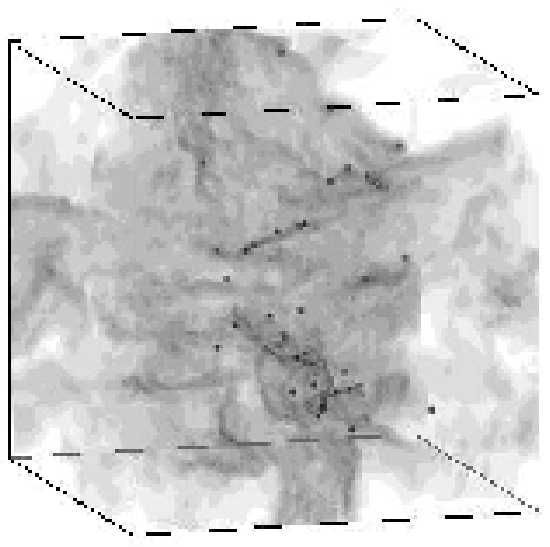}
\hspace{0.1cm}
\includegraphics[height=1.5in]{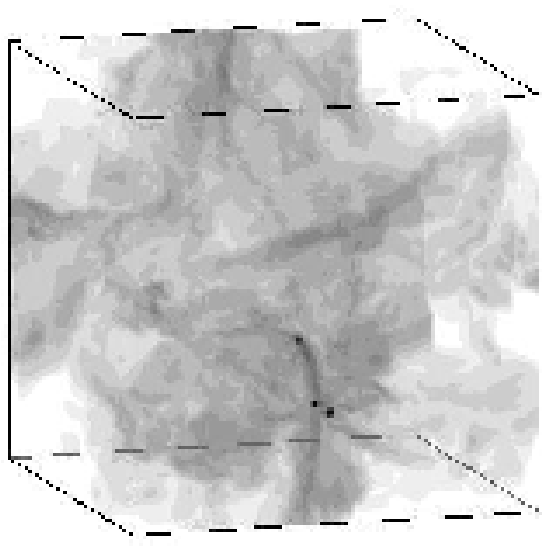}\\
\vspace{-0.5cm}
\includegraphics[height=1.5in]{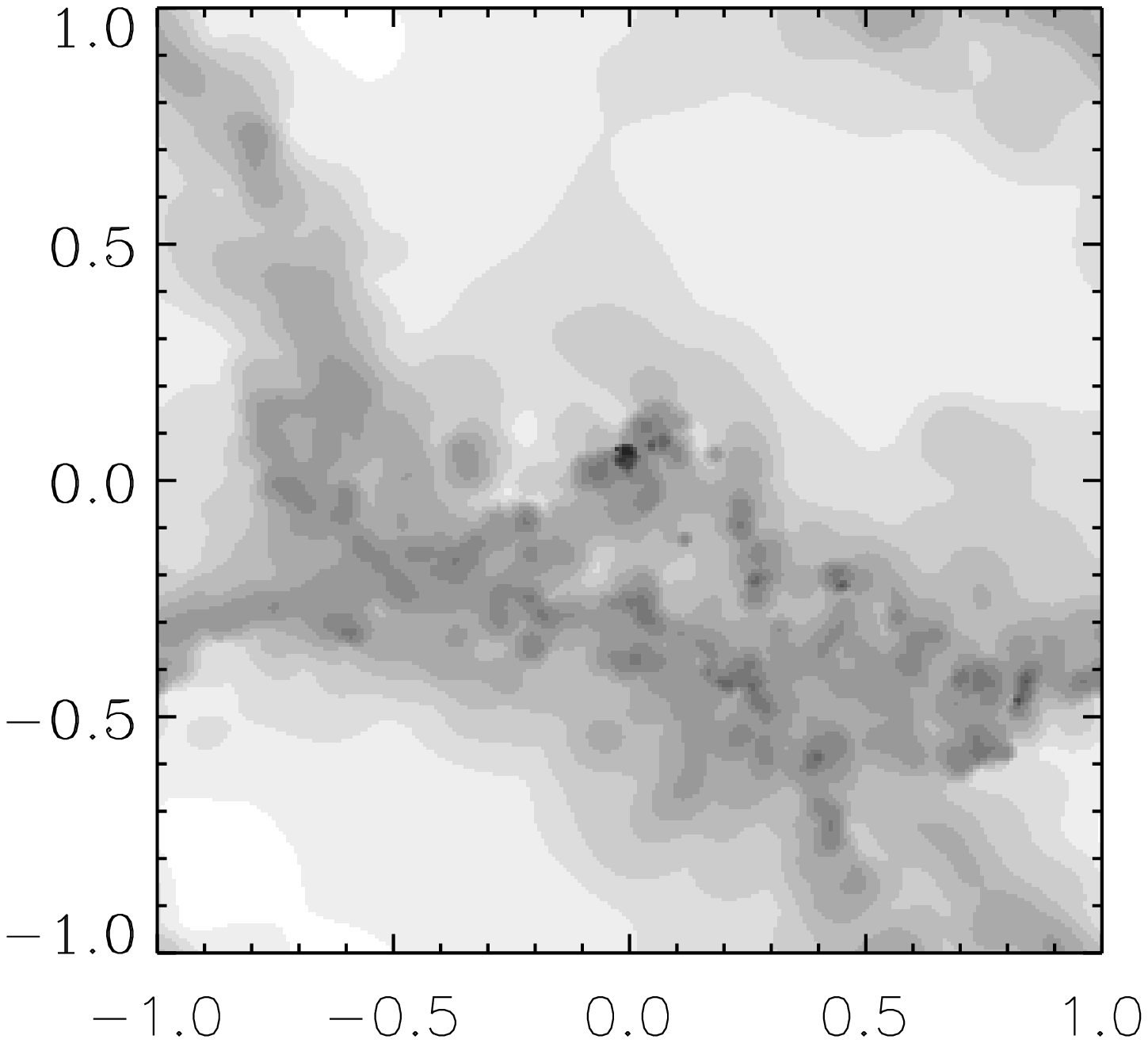}
\hspace{0.1cm}
\includegraphics[height=1.5in]{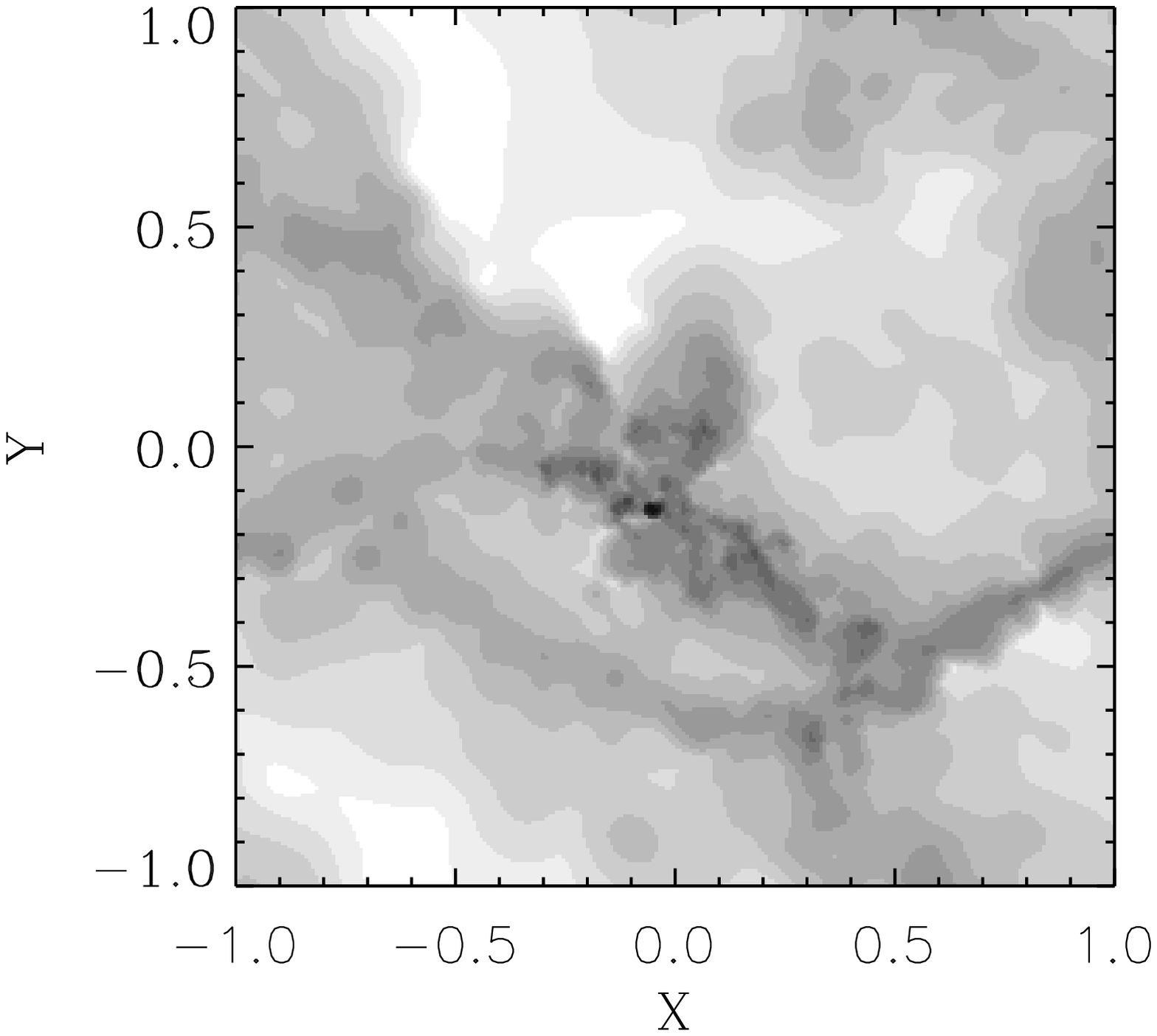}
\hspace{0.1cm}
\includegraphics[height=1.5in]{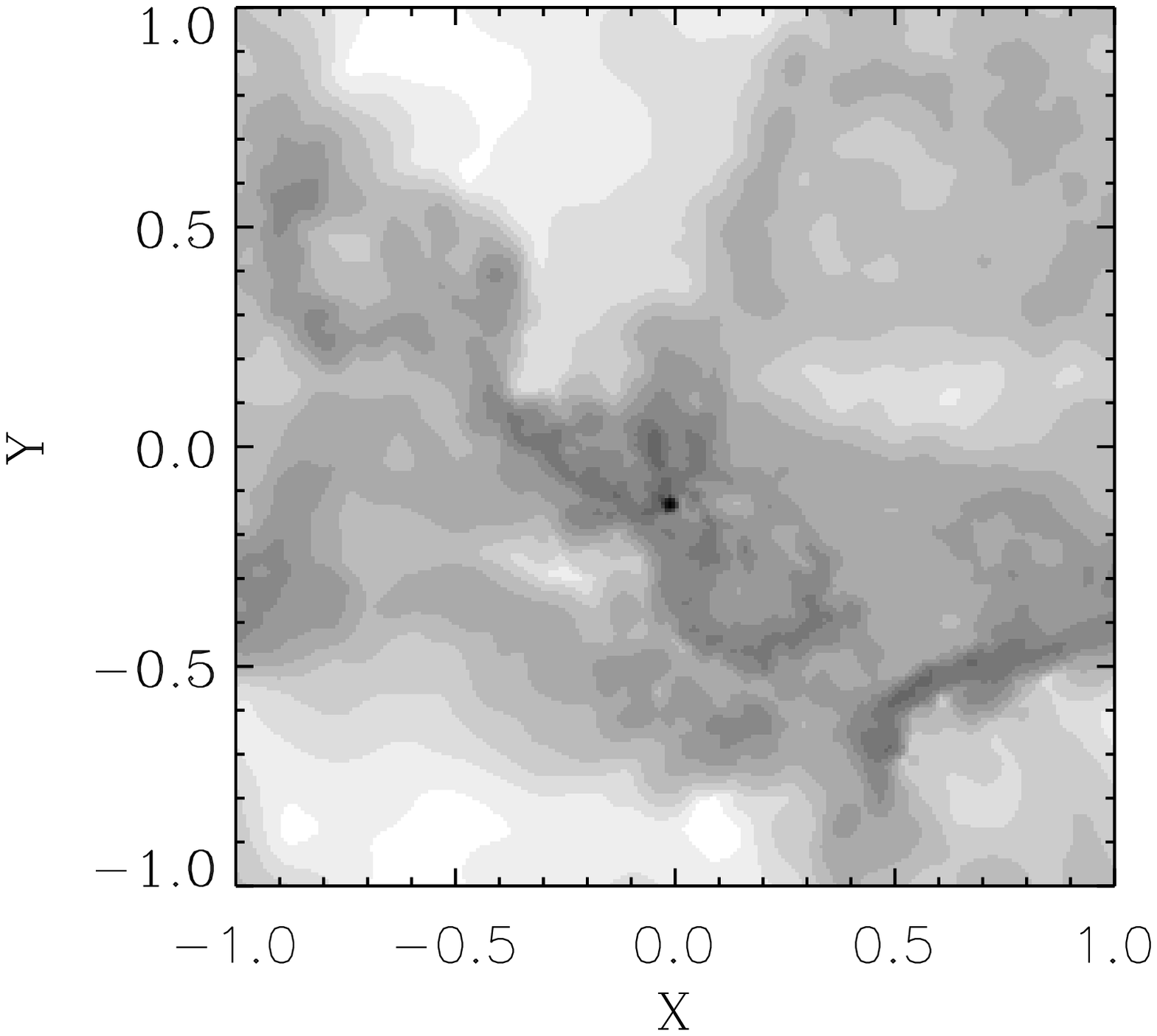}\\
\includegraphics[height=1.5in]{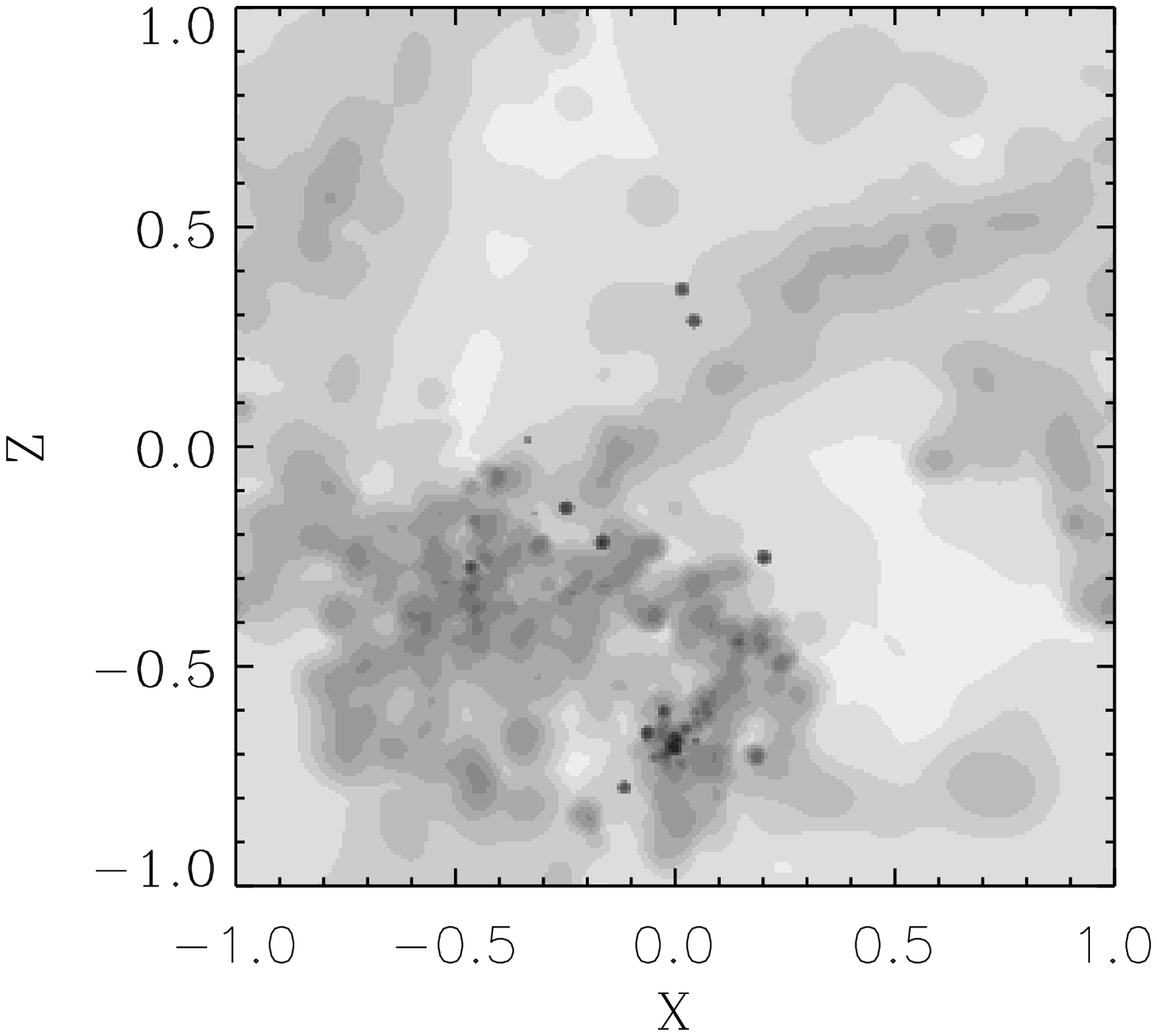}
\hspace{0.1cm}
\includegraphics[height=1.5in]{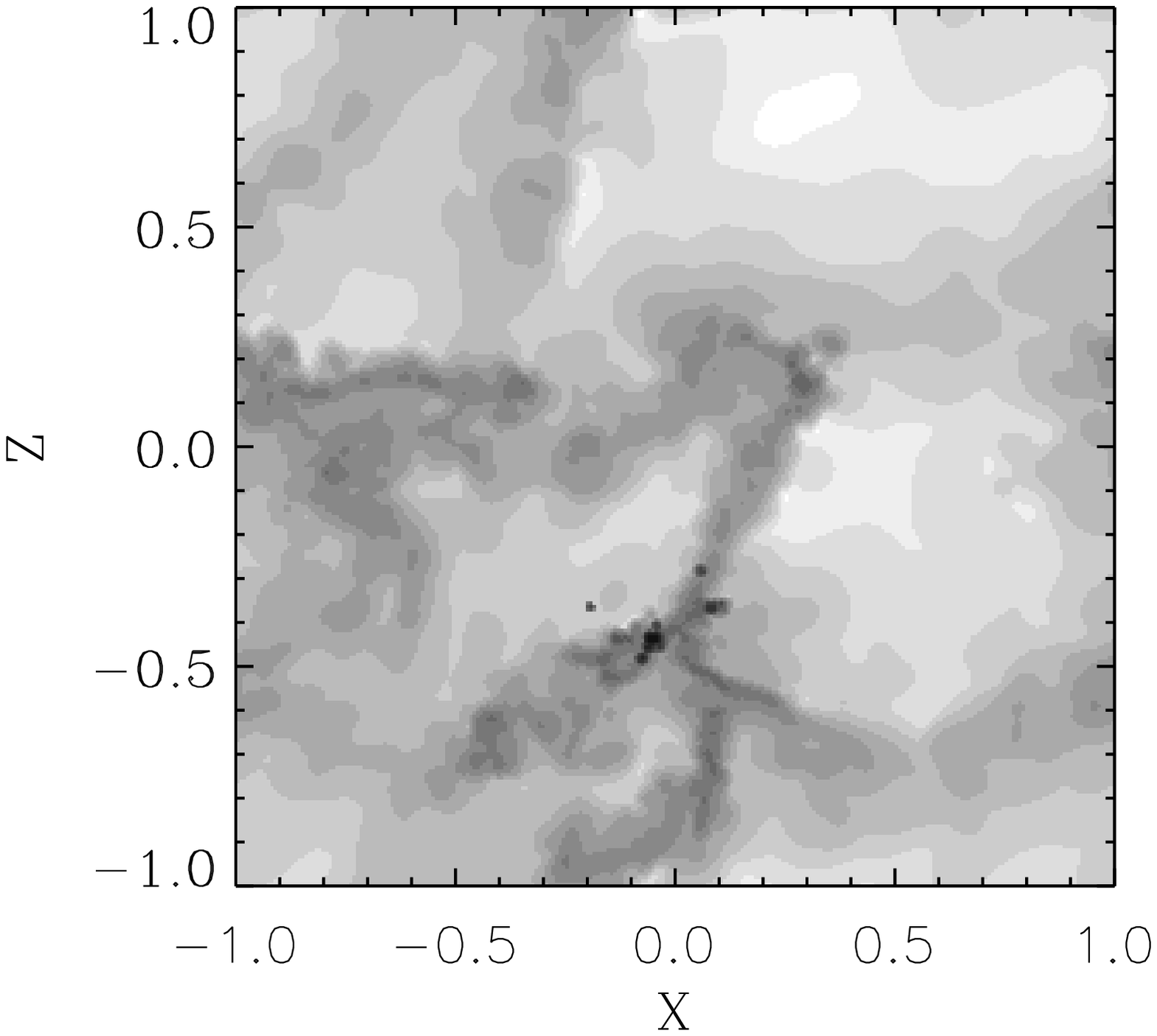}
\hspace{0.1cm}
\includegraphics[height=1.5in]{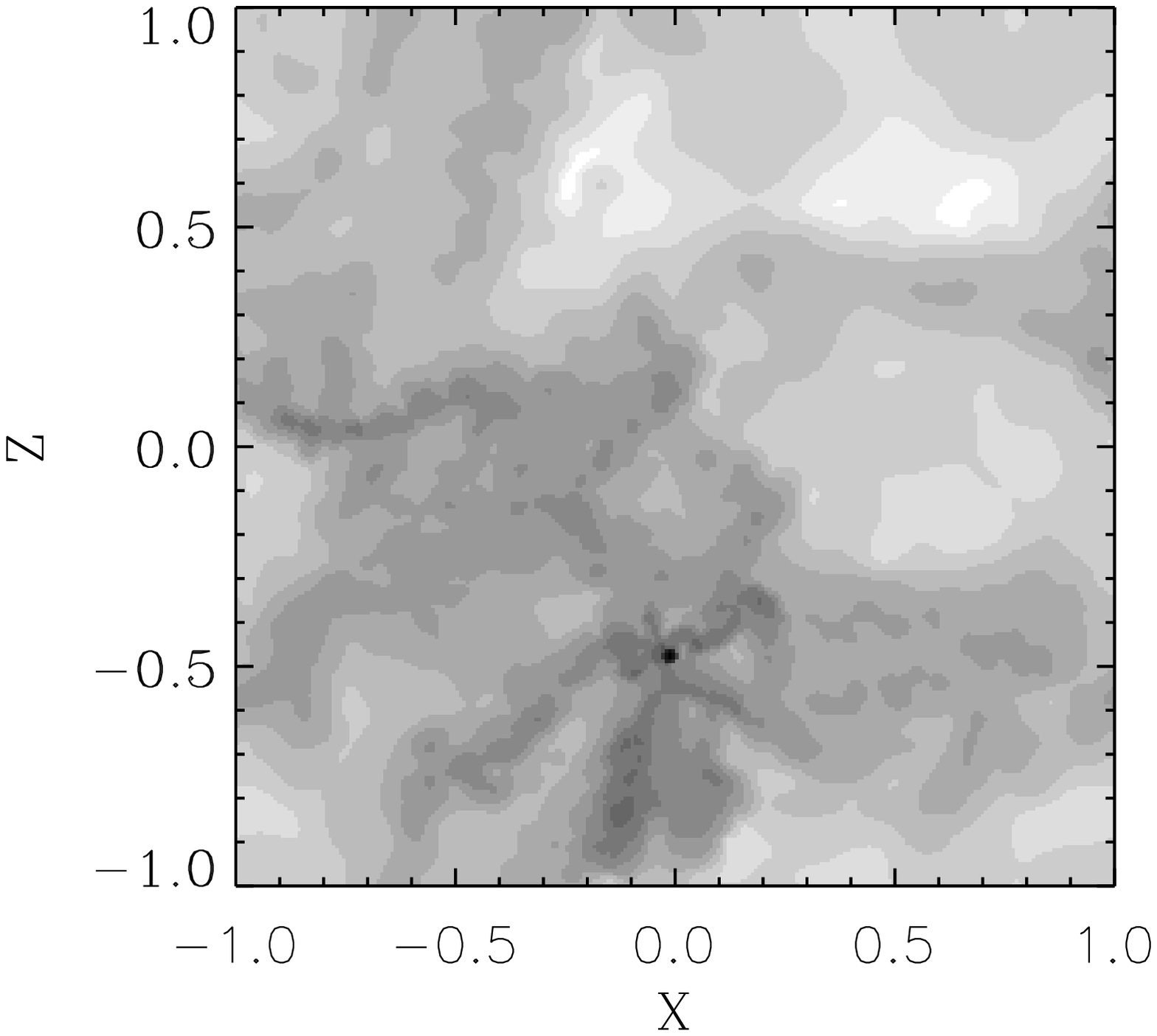}\\
\includegraphics[height=1.5in]{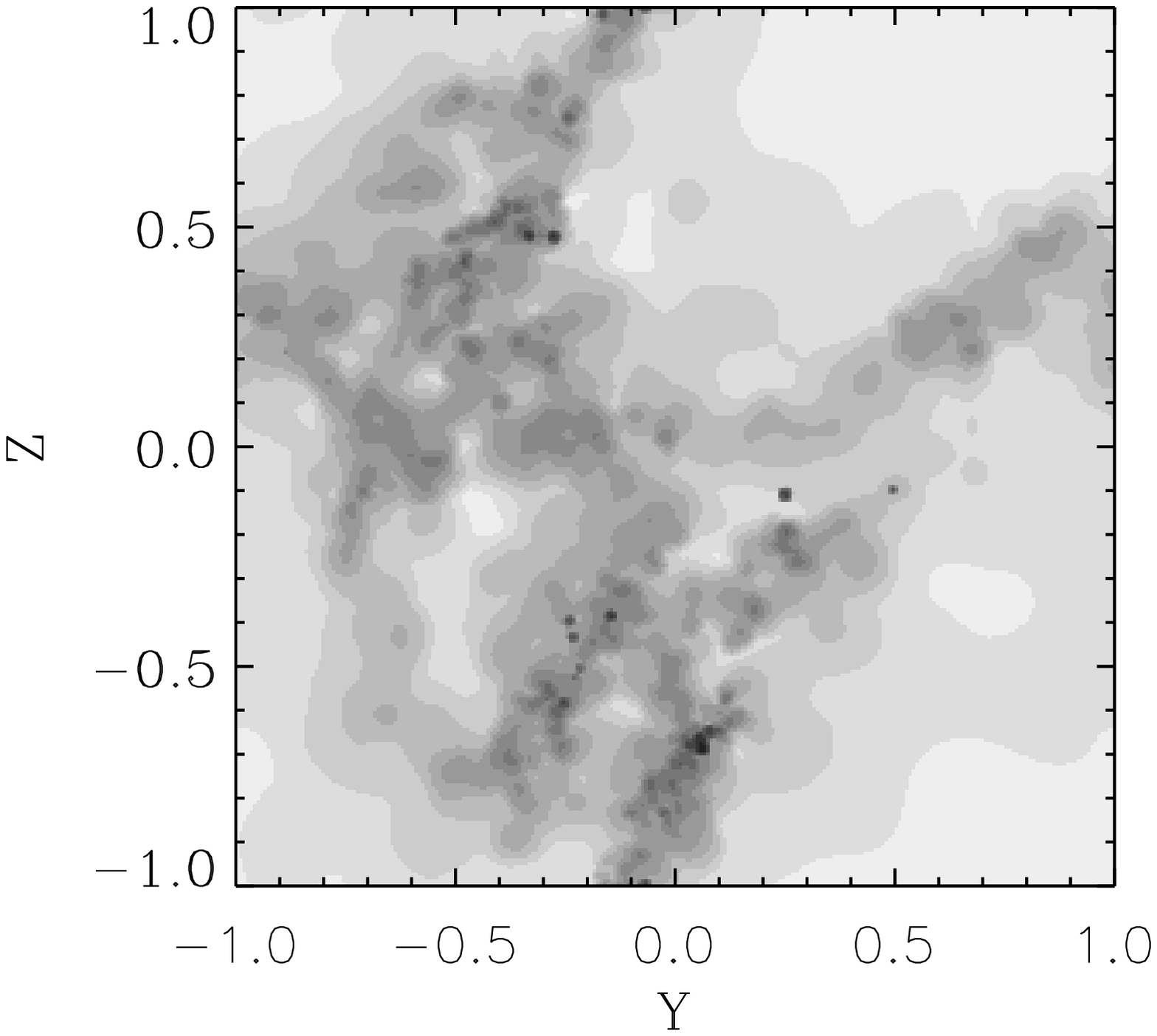}
\hspace{0.1cm}
\includegraphics[height=1.5in]{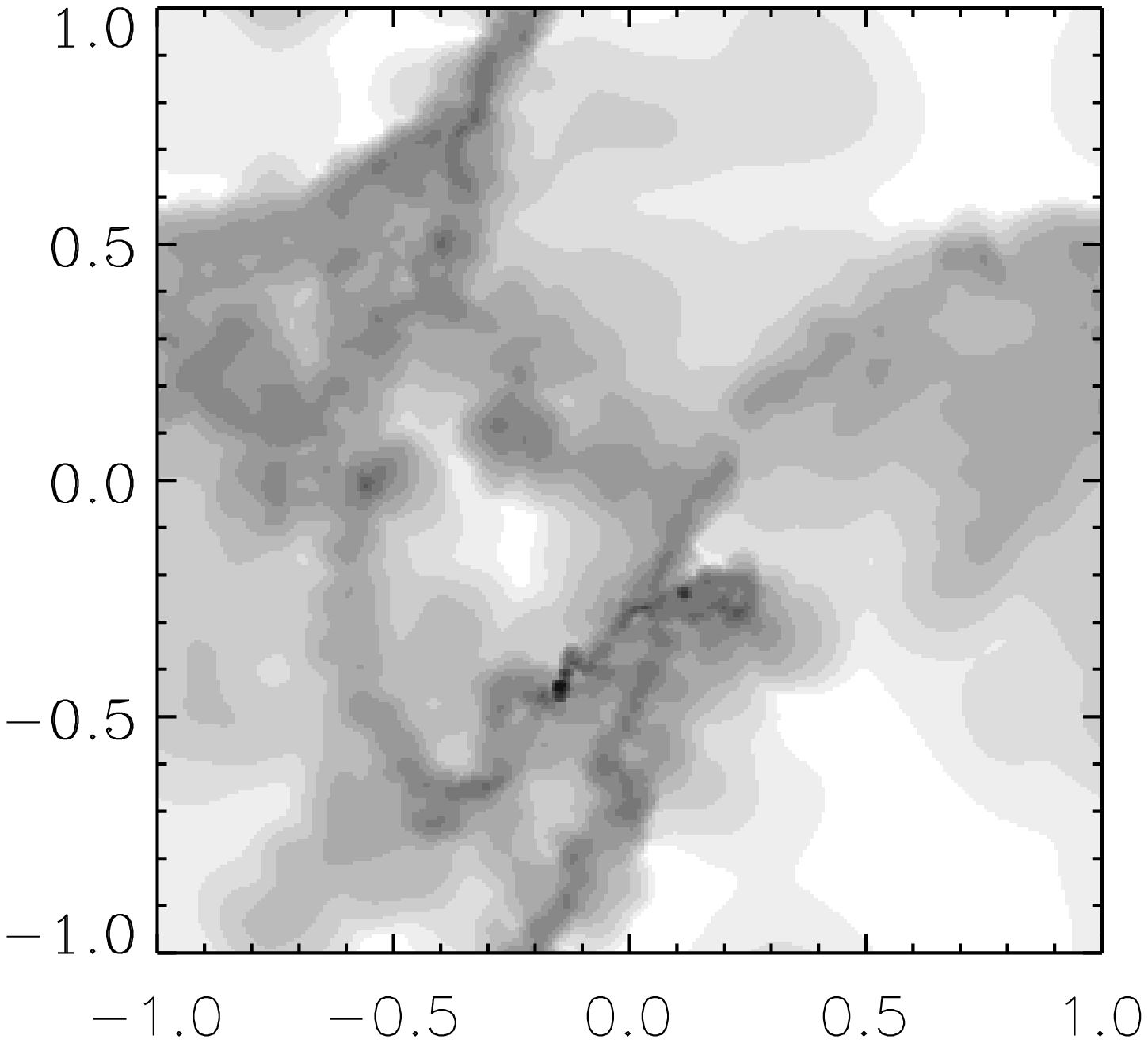}
\hspace{0.1cm}
\includegraphics[height=1.5in]{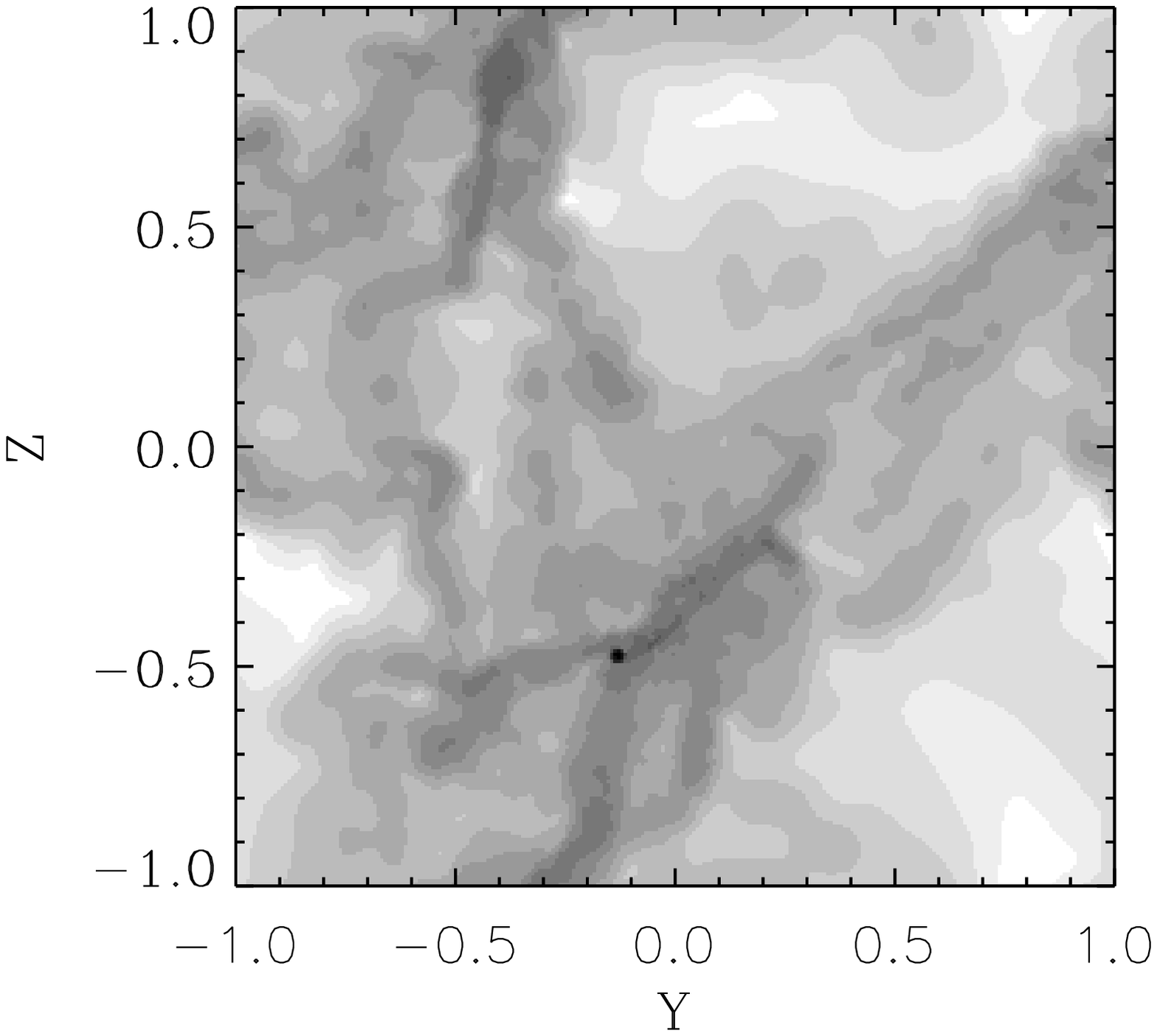}\\
\vspace{0.4cm}
\includegraphics[width=3.9in]{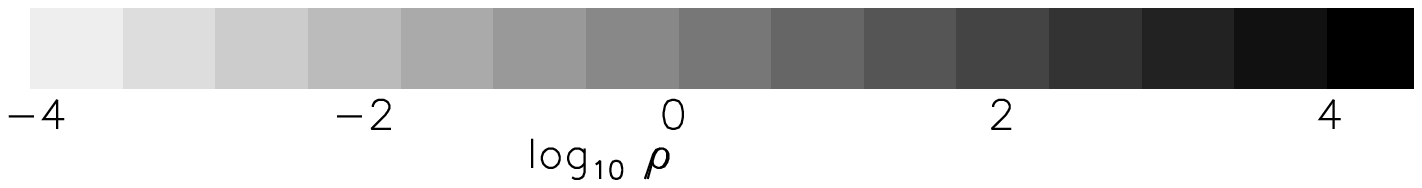}
\caption{\label{fig_sinkdis1}Density distribution of the gas shown on
a grid in three-dimensional projection (\textit{top row}) and in a
slice through maximum density, for driving with $k=1$--2 at 1 $\tau_{ff}$ 
after self-gravity is turned on. Values of $\gamma = 0.2$ (\textit{left
column}), $\gamma = 1.0$ (\textit{middle column}), and $\gamma = 1.3$
(\textit{right column}) are shown. \new{The color bar for the density scale \new{of the
slices} is at the bottom. Note the very high densities in these images apply
to the collapsed cores only. The distribution of gas density is shown in Fig.
\ref{fig_pdf_int}.}} 
\end{center}
\end{figure}

\begin{figure}
\begin{center}
\includegraphics[height=1.5in]{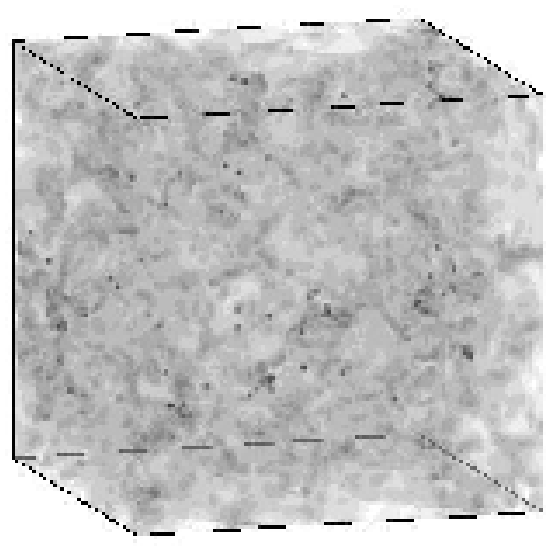}
\hspace{0.2cm}
\includegraphics[height=1.5in]{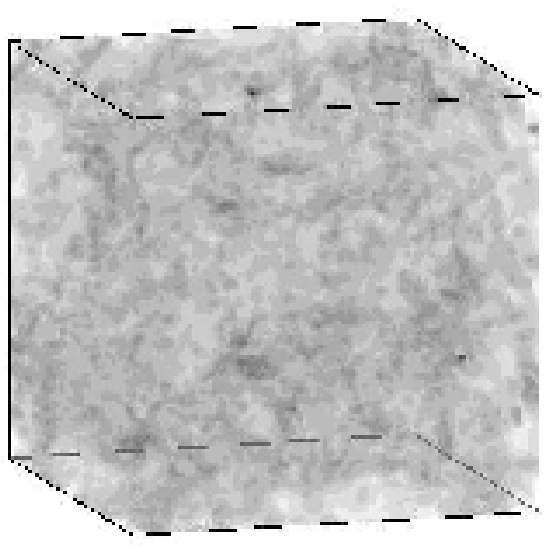}\\
\vspace{-0.5cm}
\includegraphics[height=1.5in]{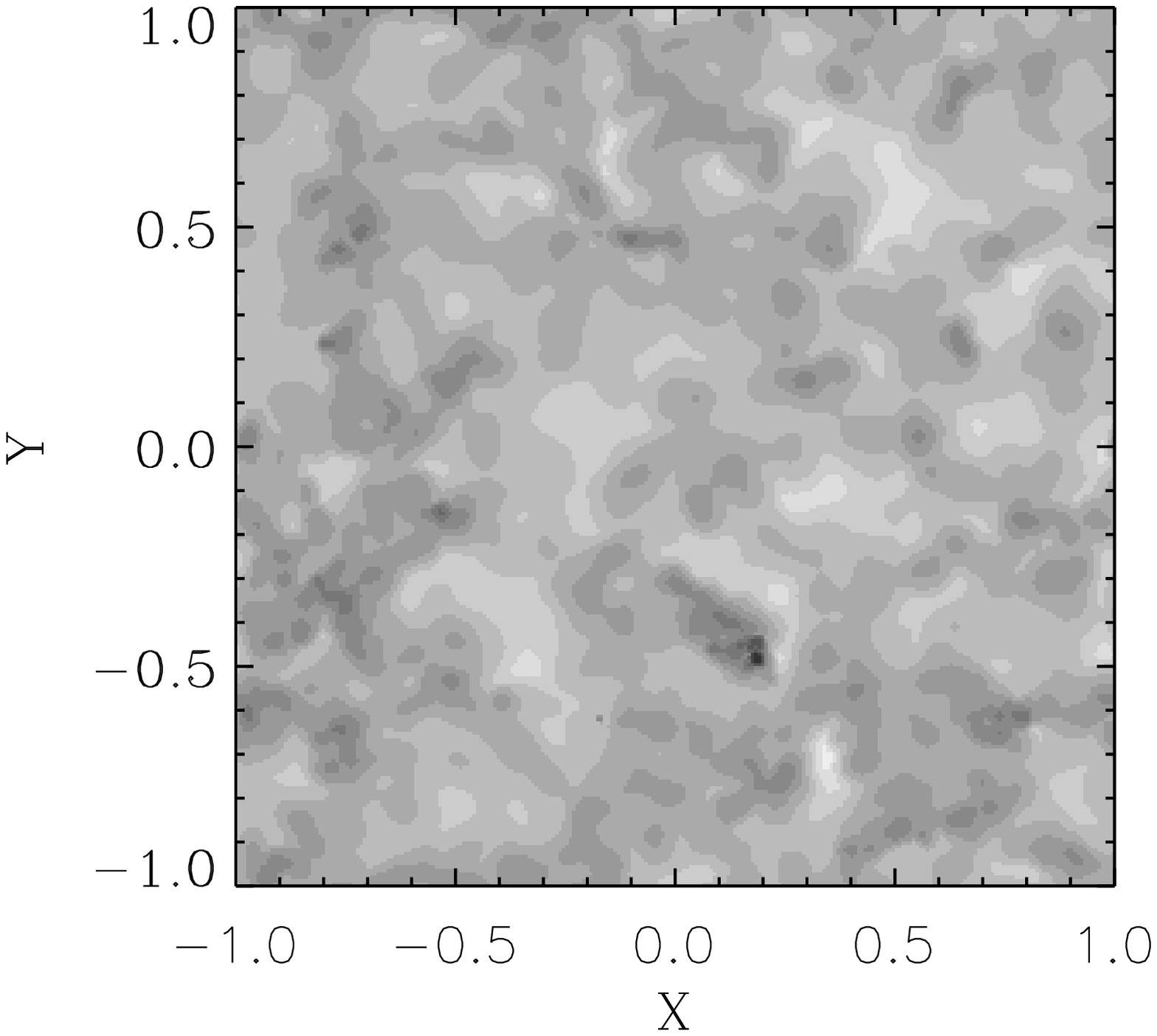}
\hspace{0.2cm}
\includegraphics[height=1.5in]{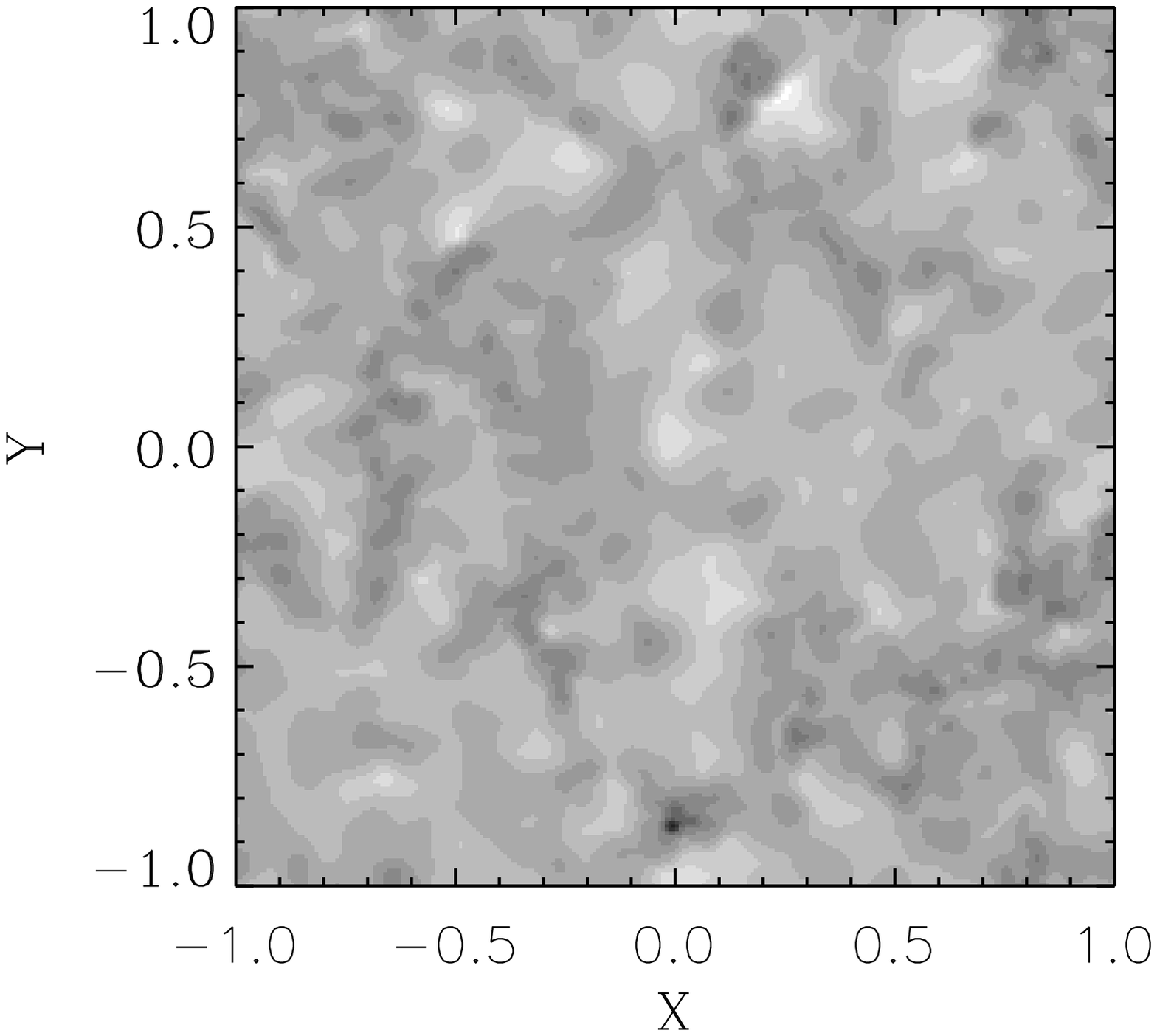}\\
\includegraphics[height=1.5in]{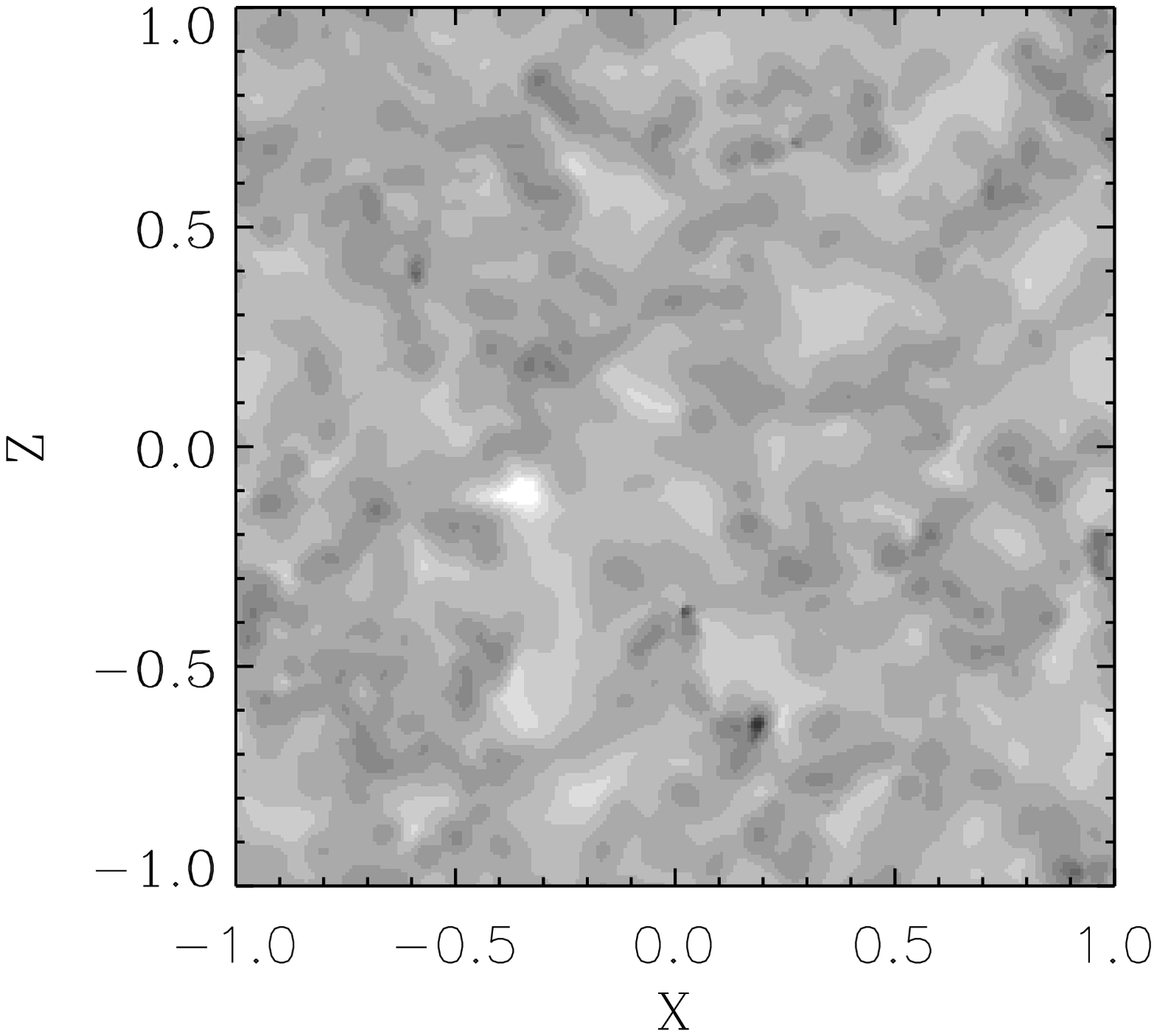}
\hspace{0.2cm}
\includegraphics[height=1.5in]{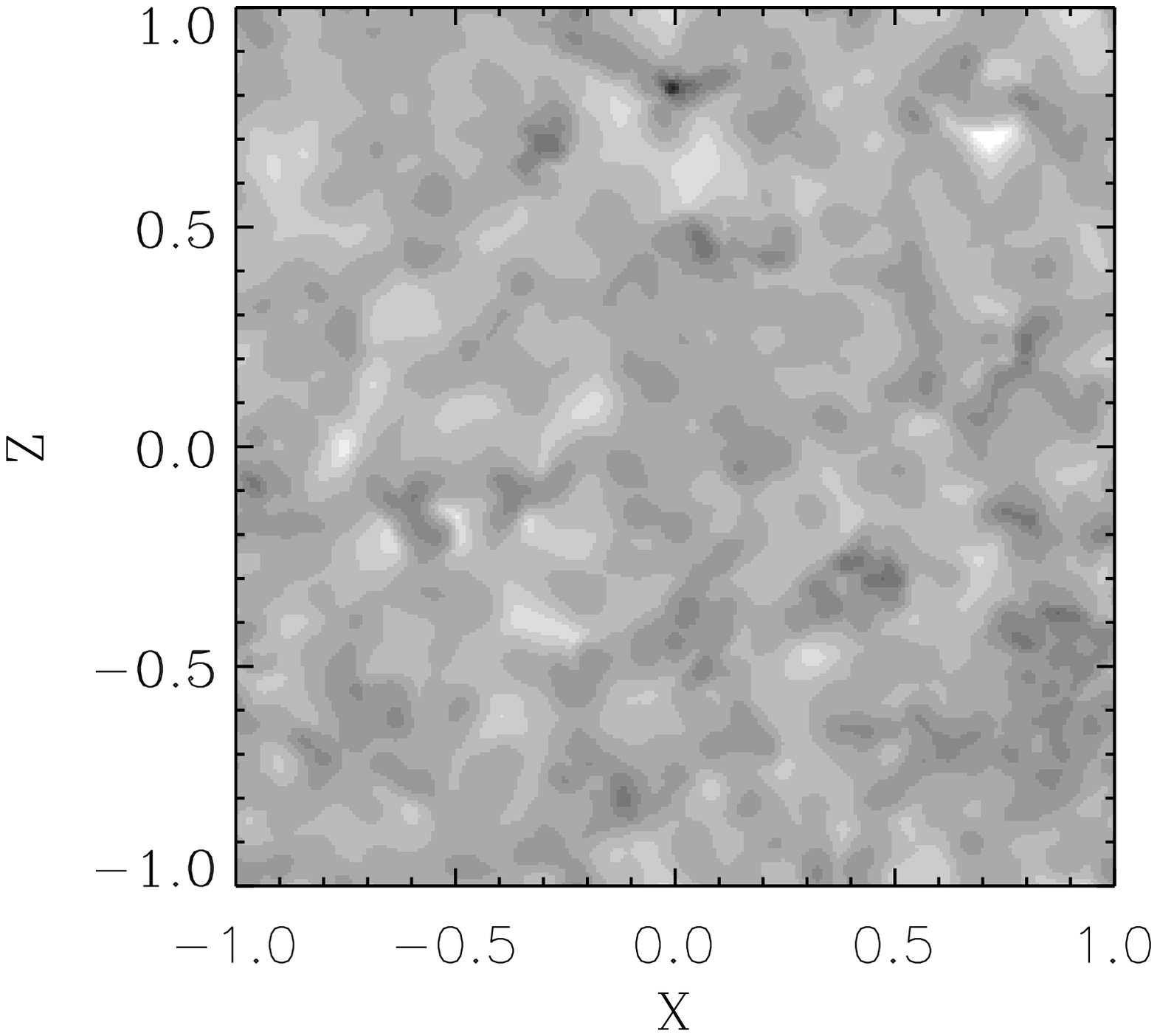}\\
\includegraphics[height=1.5in]{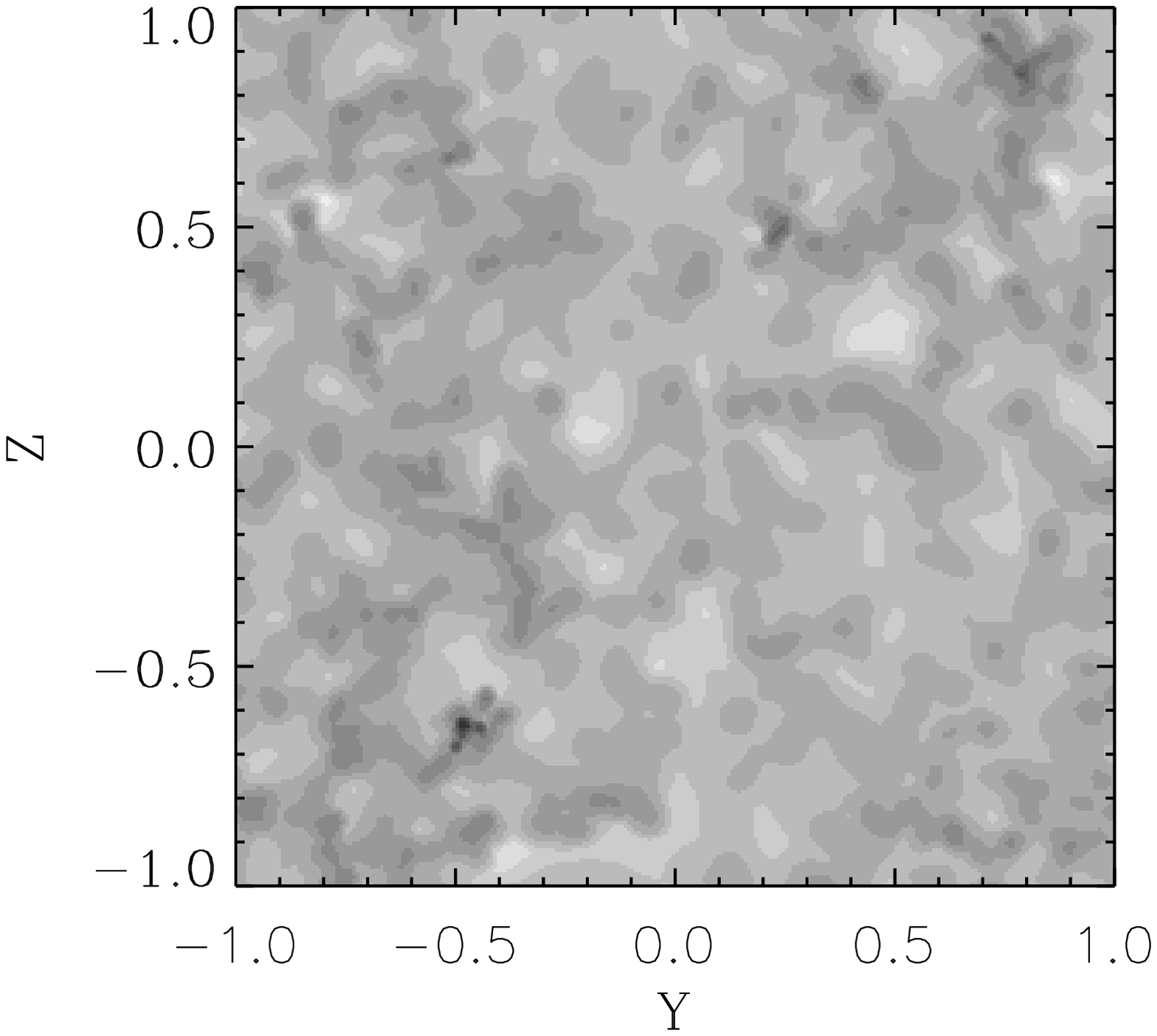}
\hspace{0.2cm}
\includegraphics[height=1.5in]{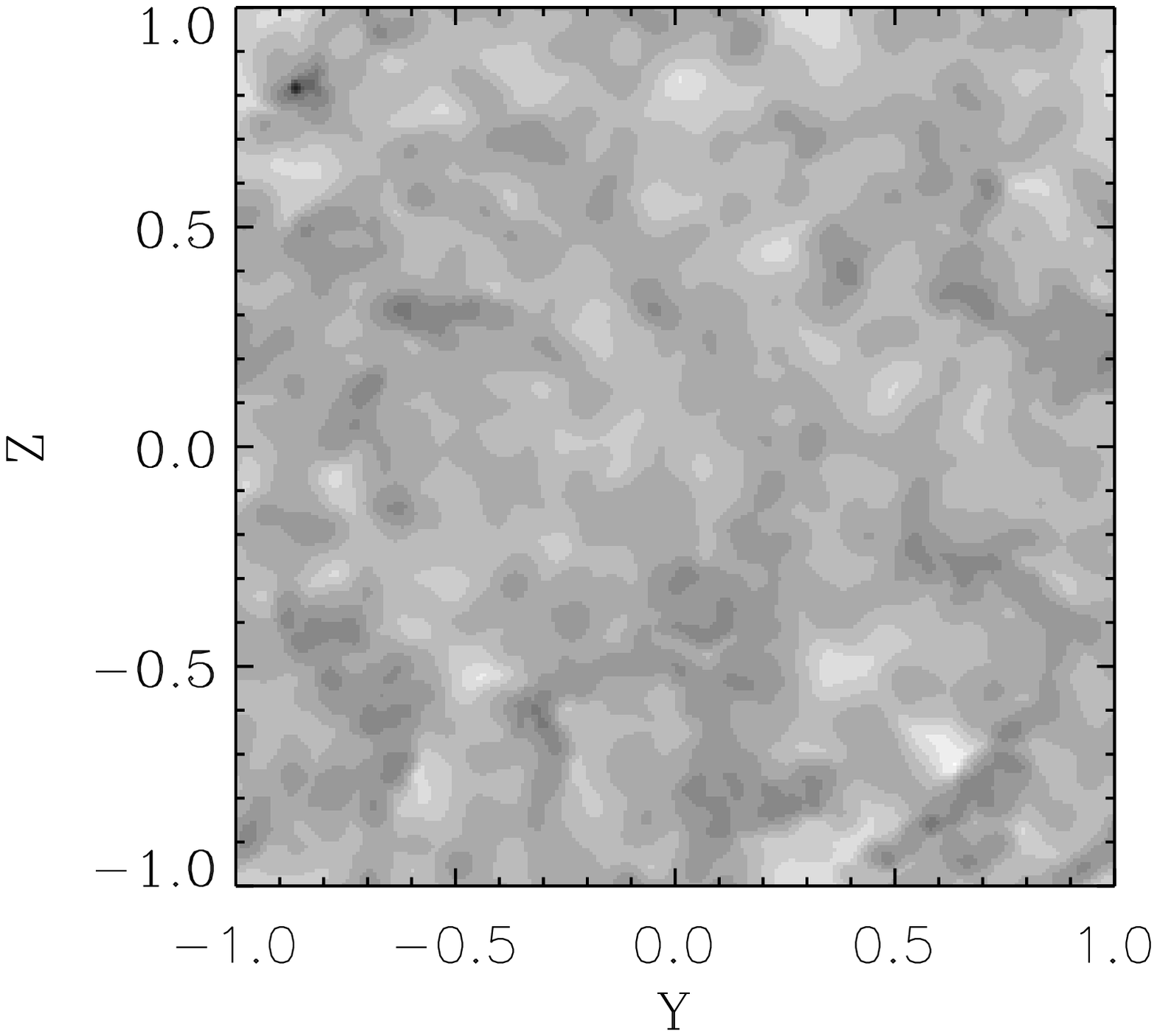}\\
\vspace{0.4cm}
\includegraphics[width=2.8in]{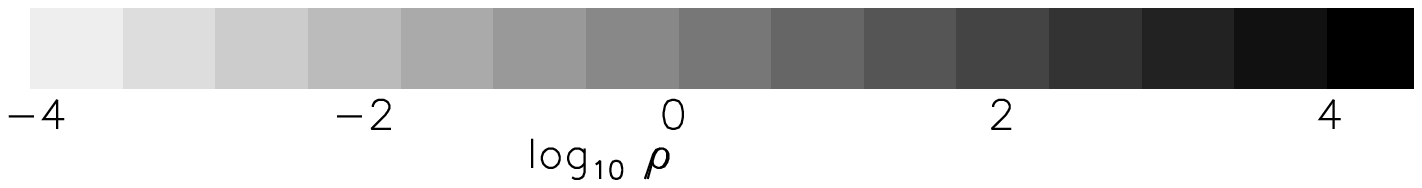}
\caption{\label{fig_sinkdis2}Density distribution of the gas shown on
a grid in three-dimensional projection (\textit{top row}) and in a
slice through maximum density, for selected $\gamma$ and for driving
with $k=7$--8 at 1 $\tau_{ff}$ after self-gravity is turned
on. Values of $\gamma = 0.2$ (\textit{left column}), and $\gamma = 1.0$
(\textit{right column}) are shown. The color bar for the density scale of the
slices is at the bottom. \new{Note the very high densities in these images apply
to the collapsed cores only}.} 
\end{center}
\end{figure}

Figure~\ref{fig_sinkdis1} and Figure~\ref{fig_sinkdis2} show the
density distribution of the gas.  To make the figures, a 256$^3$ grid
was filled with densities properly computed from the SPH kernel, and
the result is displayed in three-dimensional projection (\textit{top
row}) and in a slice through maximum density one free-fall time after
self-gravity is turned on for selected $\gamma$ and both driving
wavenumbers.  As found in previous studies (e.g.\ \citealt{mm99}), the 
driving wavenumber strongly influences the density distribution.
Driving with $k=1$--2 produces strong filamentary structure, while the
density distribution for $k=7$--8 remains more uniform at large scale.

We find that the value of $\gamma$ is as important as the driving in
determining collapse behavior. As $\gamma$ increases, the number of
collapsed cores replaced by sink particles decreases, and the cores
cluster less. For the same $\gamma$, driving with $k=1$--2 produces
both far more, and more clustered dense cores than driving with
$k=7$--8.  The dense cores tend to collapse predominantly in filaments
or at intersections of filaments. At $\gamma \geq 1.1$, no
fragmentation occurs by the time shown for driving with $k=7$--8. 

The isothermal case, $\gamma=1.0$, agrees well with previous results
in Papers~II and~III. Our current investigation agrees with the
results in Paper I--III, that collapse tends to form clusters with
high efficiency in regions with weak turbulence, while in regions with
strong turbulence, sparse, slow collapse occurs.  We now add the
additional criterion that collapse and fragmentation depends strongly on
$\gamma$. 

In regions with $\gamma < 1$, fragmentation
occurs earlier and more frequently, while in regions where $\gamma > 1$,
fragmentation is retarded and less frequent.

\placefigure{fig_sinkdis1}

\placefigure{fig_sinkdis2}

\begin{figure} 
\epsscale{1.0}\plottwo{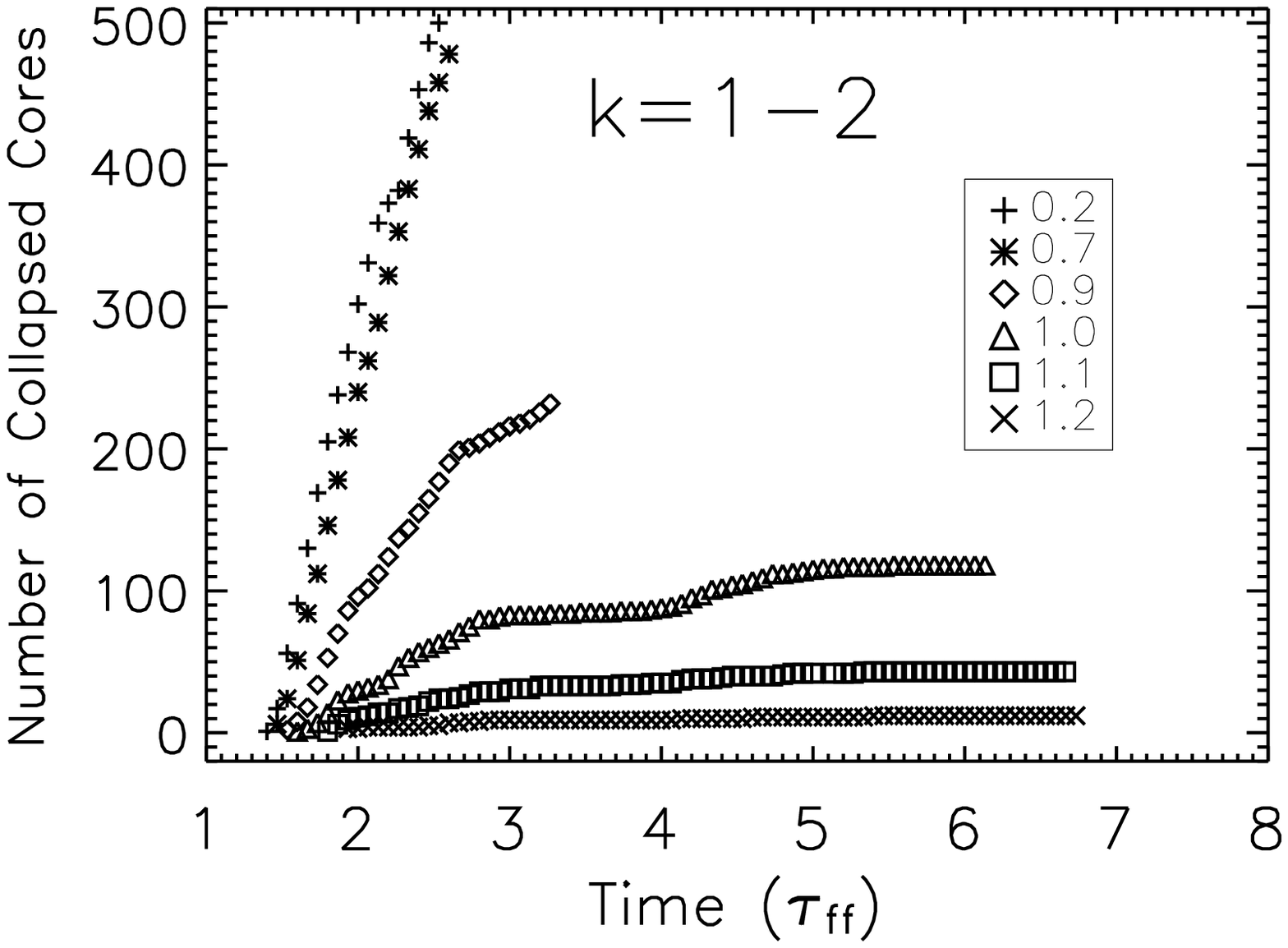}{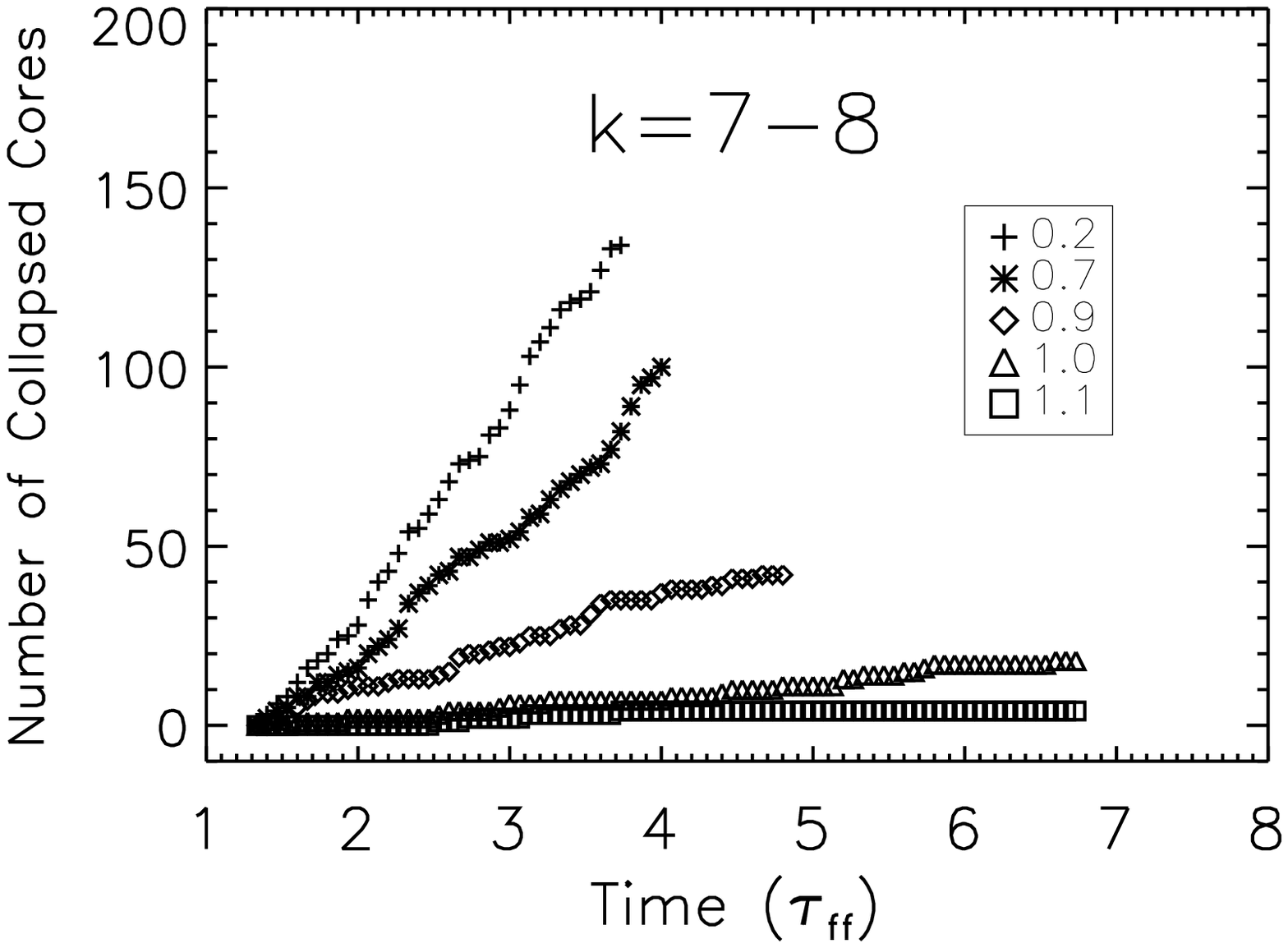}
\caption{\label{fig_sinknum}Comparison of number of collapsed cores
for models with different \new{polytropic exponent} $\gamma$, for driving with
wavenumber $k=1$--2 (\textit{left panel}) and $k=7$--8 (\textit{right
panel}). Note gravity was ``turned on'' at t = 2.0 $\simeq 1.33\tau_{ff}$. The
simulations of low-$\gamma$ cases (0.2 and 0.7 in this plot) of model $k=1$--2
were terminated after a few $\tau_{ff}$ due to prohibitively small time step.} 
\end{figure}

Figure \ref{fig_sinknum} compares the number of collapsed cores at
different $\gamma$ for models with different driving wavenumbers.
The rate at which new protostellar cores form differs for different
$\gamma$. Models with low $\gamma$ form cores quickly, while models
with large $\gamma$ form new cores more rarely.
Again we see more cores for driving with
$k=1$--2 than for $k=7$--8. 

\placefigure{fig_sinknum}

\begin{figure} 
\plotone{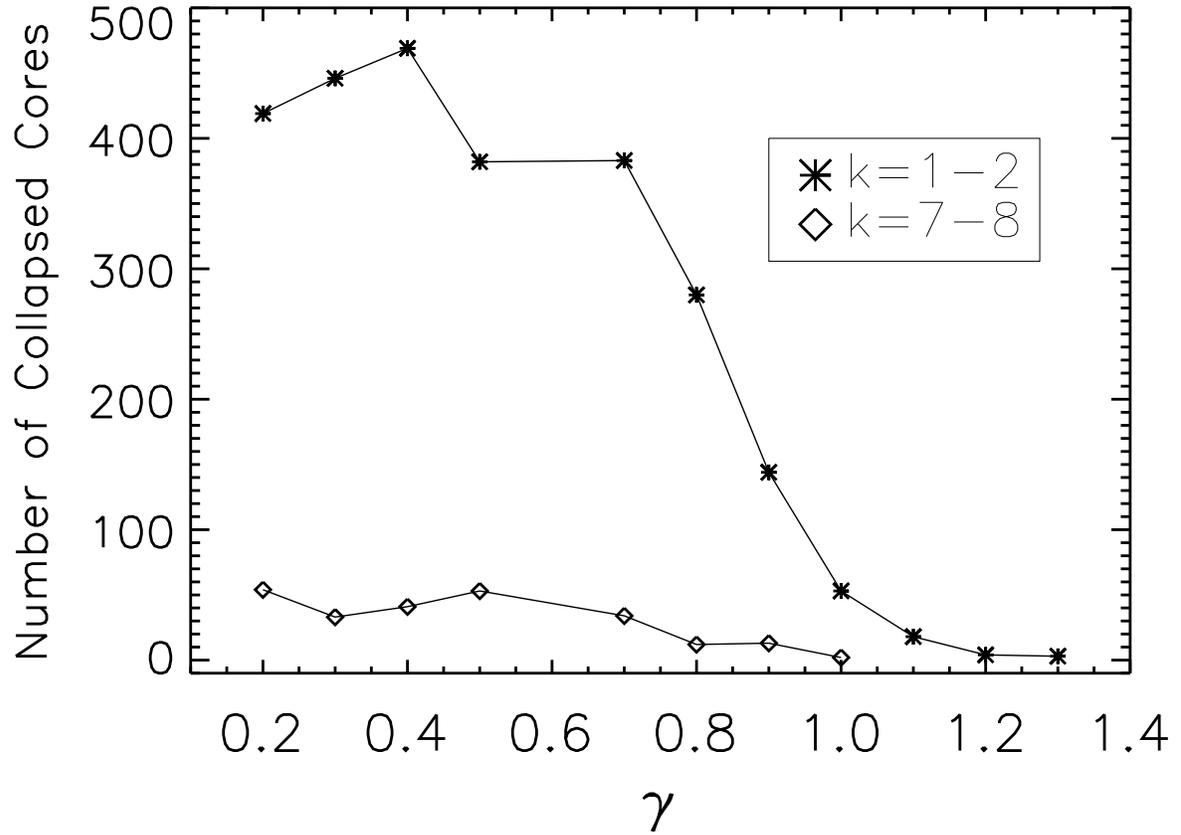}
\caption{\label{fig_sinkgamma}Relation of the value of $\gamma$ to the
number of collapsed cores one free-fall time after self-gravity
is turned on.}
\end{figure}

\citet{rb90} found an empirical scaling relation $J_c
\propto \gamma^2$ of the critical Jeans number $J_c = M/M_J$ for
collapse for elongated clouds.  The Jeans number should be related to
the number of fragments.  However, with higher resolution and more
statistics for a wide range of $\gamma$ in our simulations, we find
something more like the inverse, but strongly dependent on the details
of the driving. Fig \ref{fig_sinkgamma} shows the 
relation between the value of $\gamma$ and the number of fragments at
one free-fall time. The number of fragments drops quickly as $\gamma$
increases for driving with $k=1$--2, and more slowly for driving with
$k=7$--8, which produces fewer fragments to begin with.

\placefigure{fig_sinkgamma}

\begin{figure} 
\plottwo{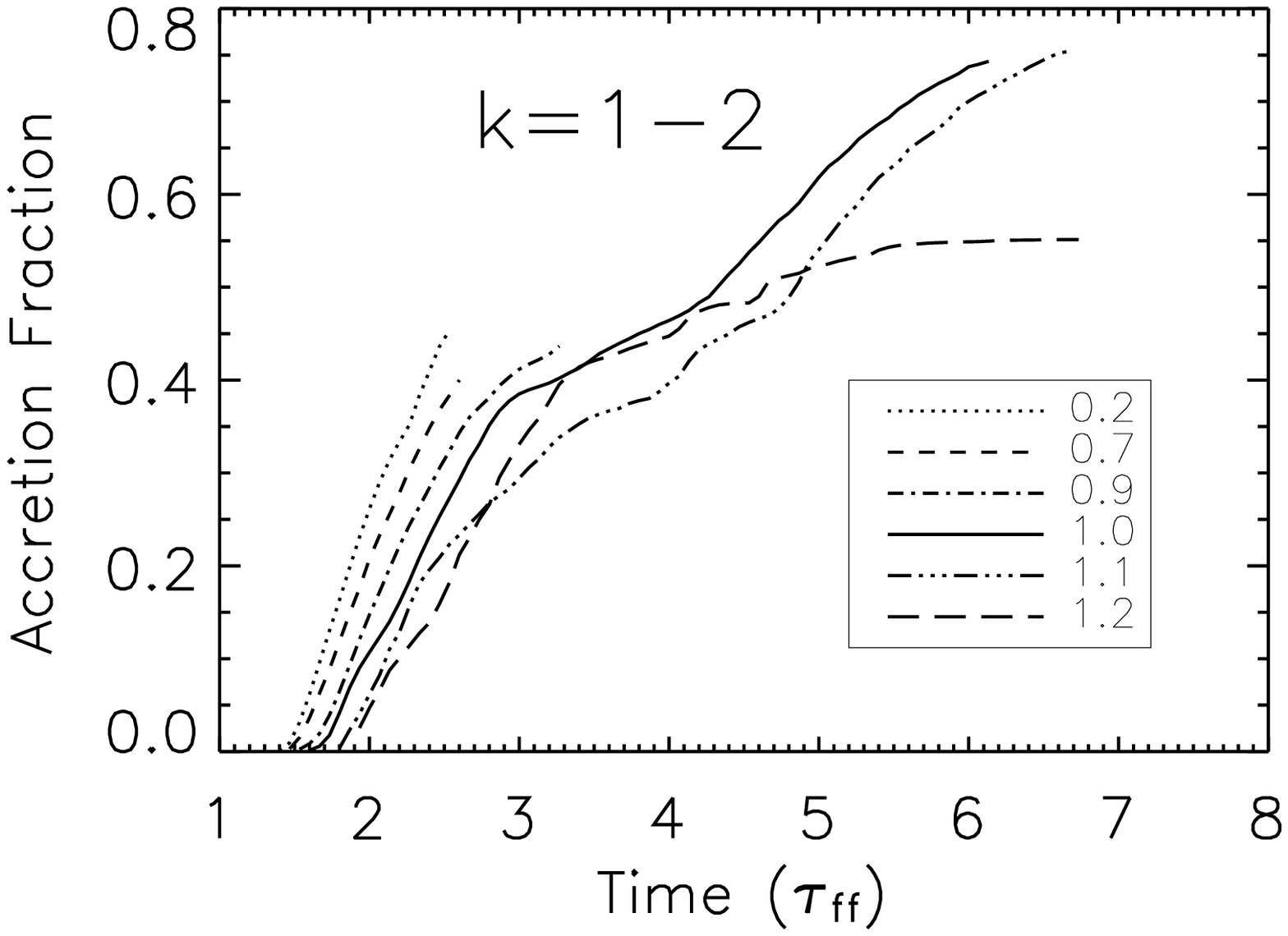}{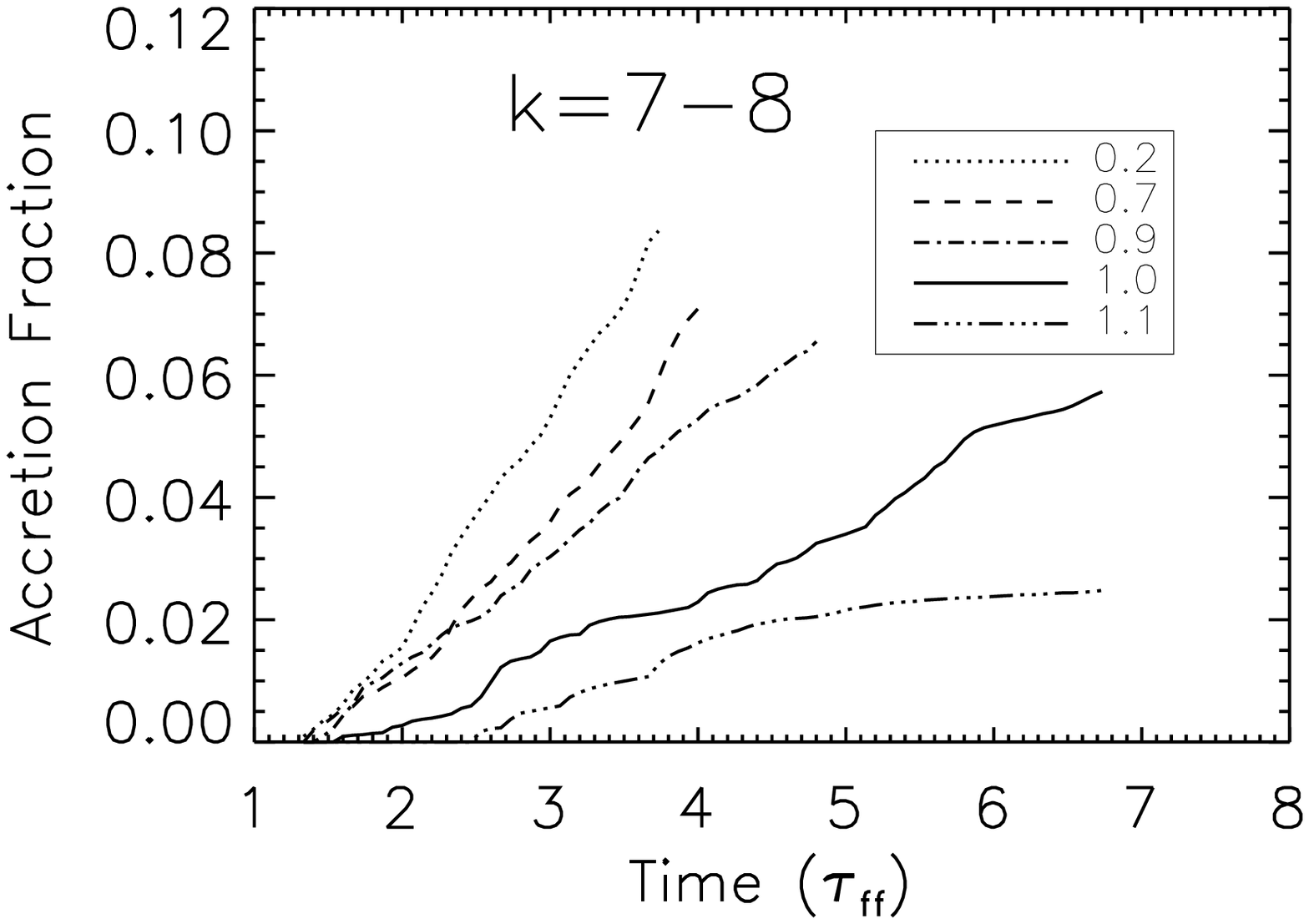}
\caption{\label{fig_sinkacc}Comparison of accretion rate of collapsing
cores for turbulence driven with wavenumber $k=1$--2 (\textit{left
panel}) and $k=7$--8 (\textit{right panel}). Note the different scales
in the two plots. Gravity is turned on at t = 2.0 $\simeq 1.33\tau_{ff}$. The
simulations of low-$\gamma$ cases (0.2 and 0.7 in this plot) of the model
driven with $k=1$--2 were terminated after a few $\tau_{ff}$ due to the
prohibitively small time step.}
\end{figure}

Figure \ref{fig_sinkacc} shows the accretion histories (the time
evolution of the combined mass fraction of all protostellar cores) for
all the individual $\gamma$ cases and for both driving
wavelengths. Note that much less mass is accreted when driving with
$k=7$--8.  We focus on the $k=1$--2 model.  The lower $\gamma$ is, the
earlier fragmentation occurs and the larger the number of fragments
that form. It also takes longer for the clouds with high $\gamma$ to
fragment. The slope of the accretion curves are roughly the same, but
the number of sinks is very different for different $\gamma$ at the
same time, which suggests that the average core mass is different for
different $\gamma$. A low-$\gamma$ environment produces a large number
of collapsed cores of relatively low mass, while $\gamma > 1$ results
in the formation of considerably fewer, but more massive, cores.
Since the overall mass growth rate of collapsing cores is closely
related to the expected star formation rate, our result suggests that
the overall star formation rate is higher in low-$\gamma$ clouds than
in high-$\gamma$ clouds.  It remains unclear what the final amount of
mass in stars will be, as higher mass stars may drive surrounding gas
away more quickly (see also \citealt{vbk03}).

\placefigure{fig_sinkacc}

We do not find any signs of collapse in the model with $\gamma=1.4$
even after several free-fall times.  This can be explained with a
Jeans mass analysis giving the critical condition for gravitational
collapse. From Equation~[\ref{eq_jeans3}], we have
\begin{equation}
\label{eq_jeans4}
\frac{\partial M_J}{\partial \rho} \propto \frac{3\gamma -
  4}{2}\rho^{(3\gamma - 6)/2}\,.
\end{equation} 
This implies that for $\gamma < 4/3$ the Jeans mass $M_J$ decreases as
density increases during collapse; for $\gamma = 4/3$, $M_J$ remains
constant; while if $\gamma > 4/3$, then $M_J$ \textit{increases} with
density.  Thus, any collapse will choke itself off when $\gamma >
4/3$, as found in our numerical results.

Fragmentation is a complex process depending on the local conditions
in the cloud, so the above analysis for an isolated spherical
perturbation is an approximation. Indeed, fragmentation already ceases
for $\gamma > 1.1$ in the model driven with $k = 7-8$, suggesting the
importance of the details of the turbulence in determining collapse
behavior. 

\subsection{Mass Spectrum}
\label{subsec_mas}
Observations suggest that the mass distribution of gas clumps in
molecular clouds follows a power law, $dN/dM = M^{\nu}$, with typical
values of the exponent being $\nu \approx -1.5$
(e.g. \citealt{wbm00}). \citet{salp55} derived a power law for the high mass   
stellar IMF in the same notation with $\nu = -2.35$ (see also \citealt{scalo98,
  kroupa02}).  

We measure clump-mass spectra in our models using a clump-finding
algorithm similar to the one described by \citet{wdb94} but working on all
three spatial coordinates and adapted to make use of the SPH kernel smoothing
procedure (for details see Appendix A in Paper I).  Figure \ref{fig_mas} shows
the spectra of gas clumps and collapsed cores for models with different values
of $\gamma$, driven with $k=1$--2.  Three different evolutionary phases
are shown: after turbulence has been fully established, but before
self-gravity has been turned on, at $t=2$; and when the fraction of
mass accumulated in collapsed cores (sink particles) has reached
$M_{\ast} \approx 20\%$ and $40\%$.

\placefigure{fig_mas}

The mass distribution of both clumps and collapsed cores changes with
$\gamma$, with the effect being most pronounced for the cores. In
low-$\gamma$ models, the core mass spectrum at the high-mass end is
roughly log-normal.  As $\gamma$ increases, fewer but more massive
cores emerge. When $\gamma > 1.0$, the distribution is dominated by a few
high mass cores, and the spectrum tends to flatten out. It is no
longer fit by either a log-normal or a power-law. The clump mass
spectra, on the other hand, do show power-law behavior on the high
mass side, even for $\gamma > 1.0$.

Our results suggest that massive stars can form in small groups or
even in isolation in gas with $\gamma > 1.0$. \citet{ss00}
suggested that a stiff EOS with $\gamma > 1.0$ should lead to a peaked IMF,
biased toward massive stars, while an EOS with $\gamma < 1.0$ results
in a power-law IMF, in general agreement with our simulations.

\subsection{Density Probability Distribution Function}
\label{subsec_pdf}

\placefigure{fig_pdf_int}

Figure \ref{fig_pdf_int} shows mass and volume weighted PDFs of gas
density $p_m$ and $p_v$ before self-gravity has been turned on for
selected values of $\gamma$ and both driving wavenumbers.  We also
show Gaussian fits to the PDFs together with their error $\epsilon =
\sum{|{p - p_{fit}}|/{p}}$, summed over all points above $10\%$ of
the peak values.

The mass-weighted density PDF $p_m(\rho)$ is calculated directly from
the local density associated with each SPH particle, as described by
\citet{rsk00}.  The volume-weighted density PDF $p_v(\rho)$ is
calculated on a cubic grid derived from the SPH density field by
taking a cube of $256^3$ cells and using the SPH kernel smoothing
procedure to find the density at the center of each cell.  Both
$p_m(\rho)$ and $p_v(\rho)$ are normalized so that
$\int_{-\infty}^{\infty} p_m(\rho)d\rho = \int_{-\infty}^{\infty}
p_v(\rho)d\rho = 1$.

The density behind shocks in supersonic turbulence depends on the
compressibility of the gas, 
\begin{equation}
d\rho \propto \frac{1}{\gamma}P^{\frac{1}{\gamma} - 1}dP\,,
\end{equation} 
So turbulent density fluctuations should increase as $\gamma$
decreases.  Indeed, we see that in Figure \ref{fig_pdf_int}, the width
of $p_m$ and $p_v$ increases as $\gamma$ decreases.

In models driven with wavenumber $k=7-8$, there are super-Gaussian
tails in the low-density end in the volume-weighted PDFs for all
$\gamma$ and in the mass-weighted PDF for $\gamma=0.2$.
One-dimensional simulations by \citet{pv98}
showed that the volume-weighted density PDF of supersonic turbulent
gas displays a power-law tail at high densities for $0 < \gamma < 1$,
becomes log-normal for $\gamma = 1$ (isothermal), and develops a
power-law tail at low density for $\gamma > 1$, which is only in
partial agreement with our results. The discrepancy may be due to the
low resolution in low-density regions in our simulations, the
difference between one- and three-dimensional simulations, or
different driving.  

A full investigation of the density PDFs under various conditions,
with different codes, different dimensions (three-dimensional vs.\
one-dimensional), different resolution, different turbulence driving,
different initial Mach number, and with and without magnetic field is
beyond the scope of this paper; these issues will be addressed
elsewhere.

\subsubsection{Comparison Between Mass and Volume Weighting}

In order to compare the mass and volume weighted PDFs, we will find it
useful to define $s=\ln (\rho/\rho_0)$; the mass and volume weighted
PDFs of $s$ are then $p_m(s)$ and $p_v(s)$, respectively.  We can
relate $p(s)$ to $p(\rho)$ if we remember that for any montonic
function $y(x)$, it can be shown that the PDF $|p(y)dy| =
|p(x)dx|$. Therefore,
\begin{equation}
p_m(\rho) = p_m(s)\frac{ds}{d\rho} = \frac{p_m(s)}{\rho}\,,
\end{equation}
and
\begin{equation}
p_v(\rho) = p_v(s)\frac{ds}{d\rho} = \frac{p_v(s)}{\rho} \,.
\end{equation}

Since $p_m(\rho) \propto dM/d\rho$, and $p_v(\rho) \propto dV/d\rho$
\citep{osg01}, it follows that 
\begin{equation}
p_m(\rho) \propto \frac{dM}{dV} \frac{dV}{d\rho} \propto \rho
 p_v(\rho).
\end{equation}
Then we can relate $p_m(s)$ to $p_v(s)$
\begin{equation}
\label{eq_pm_pv}
p_m(s) = \rho p_m(\rho) = C \rho^2 p_v(\rho) = C e^s p_v(s)\,.
\end{equation} 
where C is a normalization constant.

If $p_v(s)$ follows a normal distribution, 
\begin{equation}
\label{eq_pv}
p_v(s) = \frac{1}{\sqrt{2\pi} \sigma}\exp\left(-\frac{(s -
s_v)^2}{2\sigma^2}\right)\,, 
\end{equation} 
where $s_v$ is the average (in volume) value of $s$, and $\sigma$ is the
dispersion, then
\begin{equation}
\label{eq_pm_0}
p_m(s) = \frac{C}{\sqrt{2\pi} \sigma} \exp\left(-\frac{s^2_v - (s_v +
\sigma^2)^2}{2\sigma^2}\right) \exp\left(-\frac{\left[s - (s_v
+\sigma^2)\right]^2}{2\sigma^2}\right)\,. 
\end{equation} 
Since $\int_{-\infty}^{\infty} p_m(s)ds =1$, the normalization 
\begin{equation}
C = \exp\left(\frac{s^2_v - (s_v +
\sigma^2)^2}{2\sigma^2}\right)\,.
\end{equation} 
Equation~\ref{eq_pm_0} can thus be rewritten as:
\begin{equation}
\label{eq_pm}
p_m(s) = \frac{1}{\sqrt{2\pi} \sigma} \exp\left(-\frac{(s -
s_m)^2}{2\sigma^2}\right)\,. 
\end{equation} 
where $s_m = s_v + \sigma^2$. This is also a normal distribution, with the
same dispersion $\sigma$ as that of $p_v$ but shifted average value
$s_m$. Using the volume PDF normalization
$\int \rho \cdot p_v(s)ds = \int e^s p_v(s)ds = 1$,
it can be seen
\begin{equation}
\label{eq_sm_sv}
s_m = -s_v = \frac{\sigma^2}{2}\,.
\end{equation}
as noted, for example, by \citet{osg01}.  This
holds only if the density PDF is log-normal. (Note that in this case
$C= 1$.)

This derivation gives several properties of the density PDFs $p_m$
and $p_v$. If one of them is log-normal, so is the other; and if
they are log-normal, the width $\sigma$ of both profiles should be equal
(Eqs.~[\ref{eq_pv}] and~[\ref{eq_pm}]), and the peaks of the
PDFs $p_m(s)$ and $p_v(s)$ should lie symmetrically around zero
(Eq.~[\ref{eq_sm_sv}]).

In Figure \ref{fig_pdf_int} the volume-weighted PDFs lie to the left
of the mass-weighted PDFs, with $s_m \approx -s_v$, as predicted.  The
decent Gaussian fits to the peaks of the PDFs show that the assumption
of log-normal behavior is not bad.  Three-dimensional grid-based
simulations of decaying turbulence by \citet{osg01} show similar Gaussian
PDFs. 

A detailed examination of the moments of the density PDFs, shown in
Fig \ref{fig_mom_k1278}, shows that these PDFs are, in fact, not
perfect log-normals. The first moment (mean) shows that $p_m$ and
$p_v$ of models with both driving wavenumbers $k=1-2$ and $k=7-8$ are
not exactly symmetrically distributed.  This could be due to turbulent
intermittency in our small boxes producing deviations from a purely
Gaussian distribution.  The Gaussian behavior expected from a well
sampled distribution is more visible in models driven with $k=7-8$
compared to $k=1-2$. In fact, $p_m$ and $p_v$ of $k=7-8$ are more
equal in width, as seen from the second moment (variance). This is a
sign of better sampling -- more modes contribute to the overall
velocity field, thus the central limit theorem is more appropriately
applied to the $k=7-8$ than to the $k=1-2$ case \citep{rsk00}.

\placefigure{fig_mom_k1278}

\subsubsection{Effect of Self-Gravity}
\begin{figure} 
\begin{center}
\includegraphics[height=2.0in]{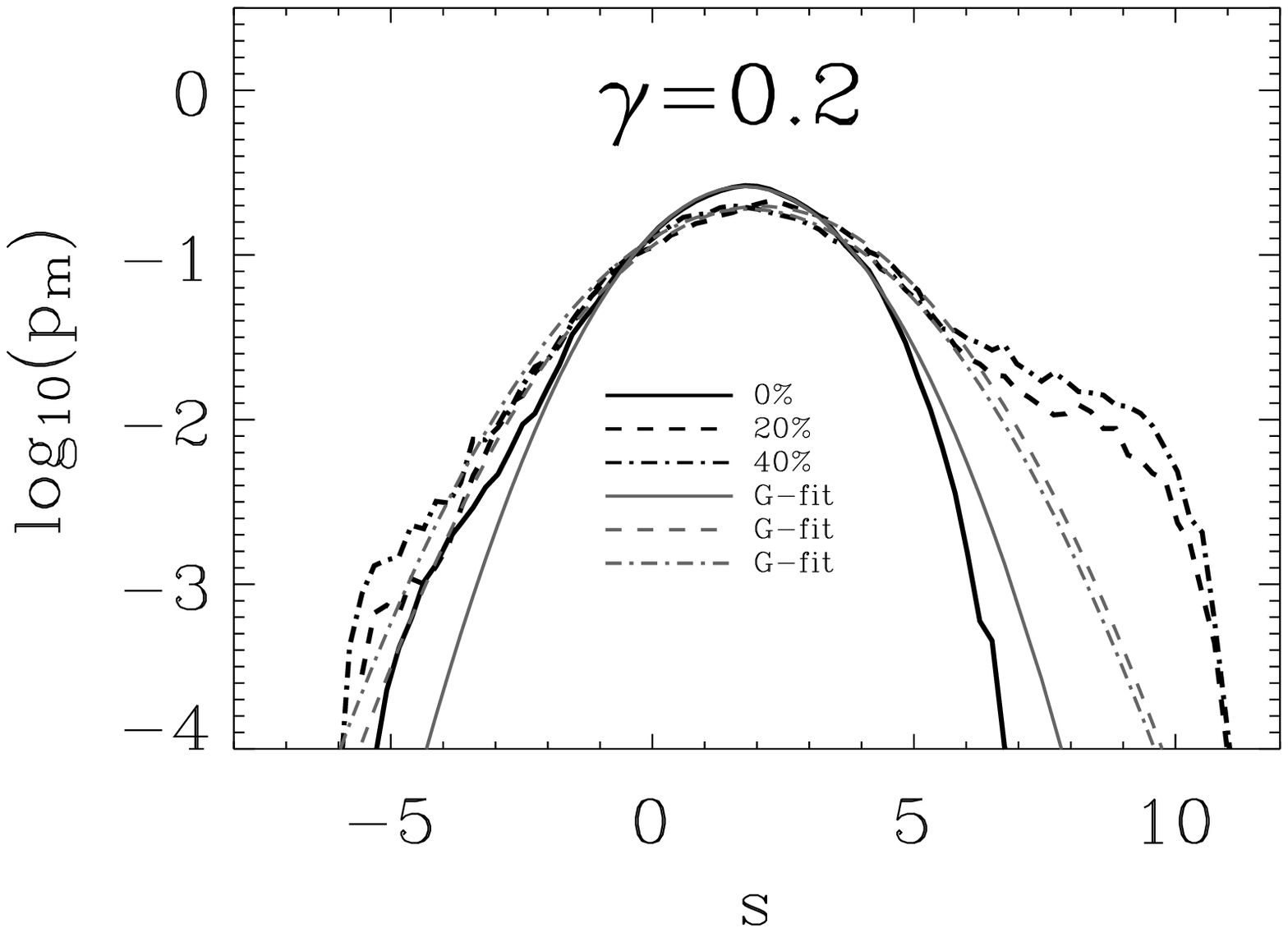}
\includegraphics[height=2.0in]{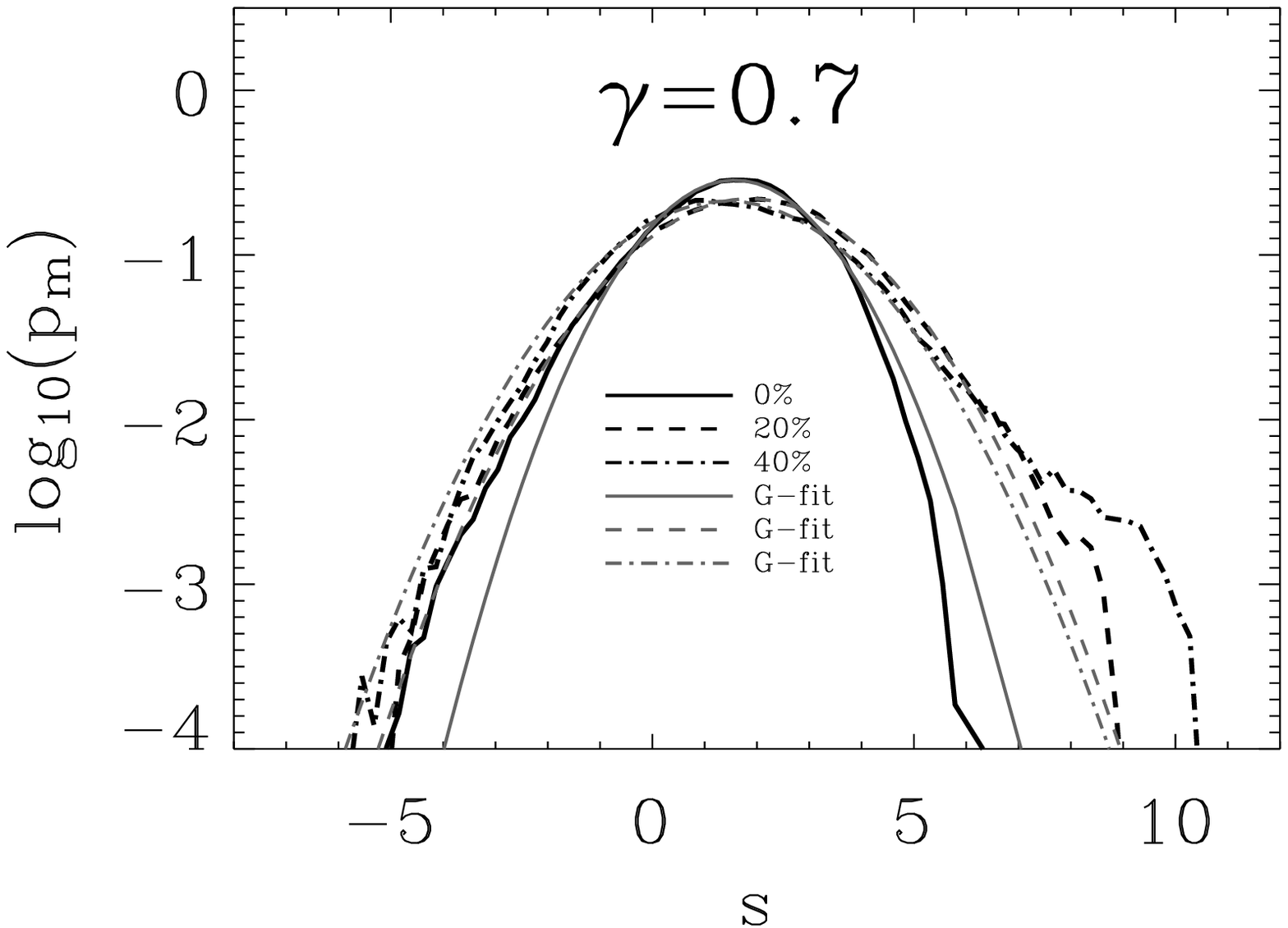}\\
\includegraphics[height=2.0in]{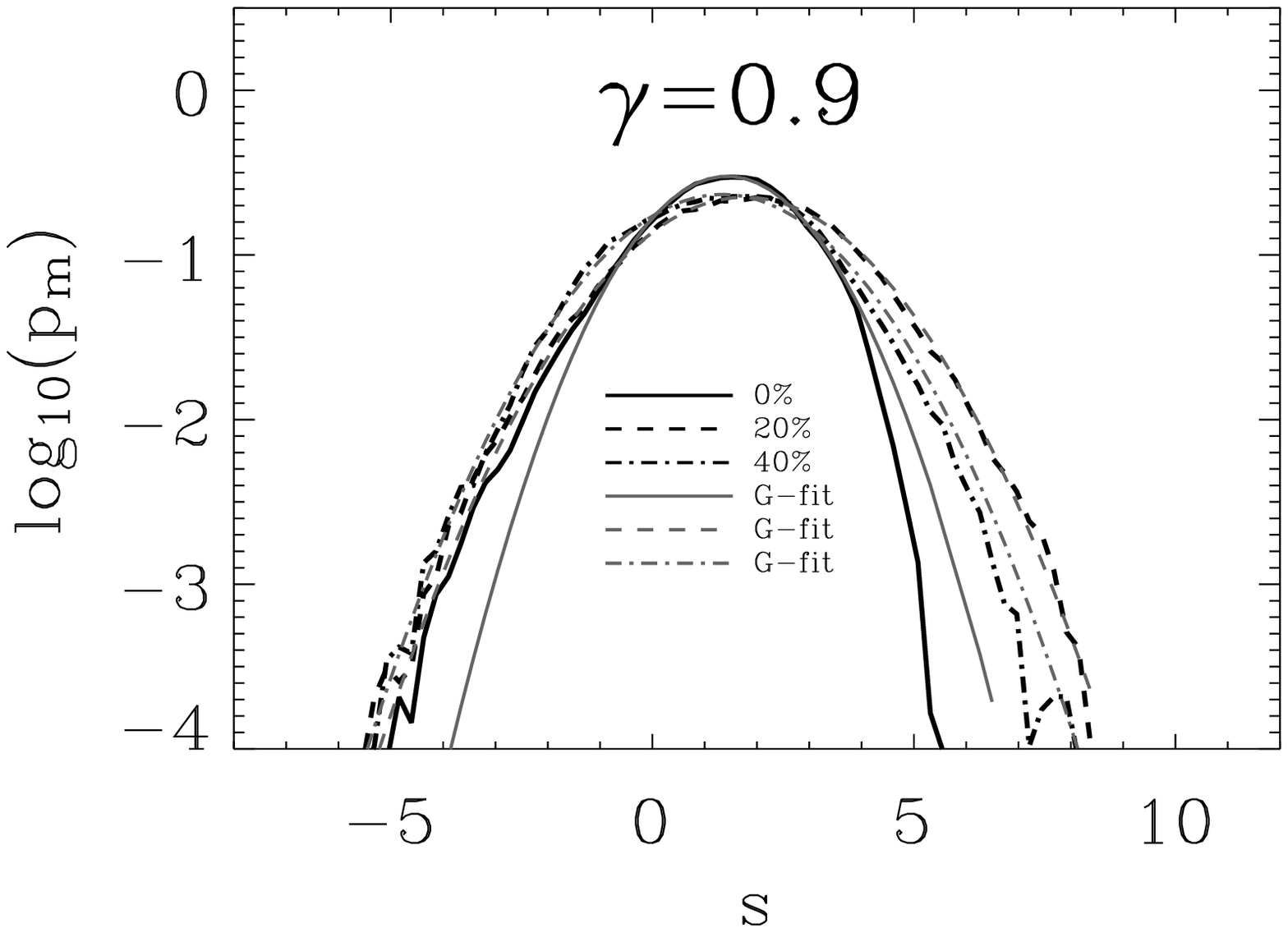}
\includegraphics[height=2.0in]{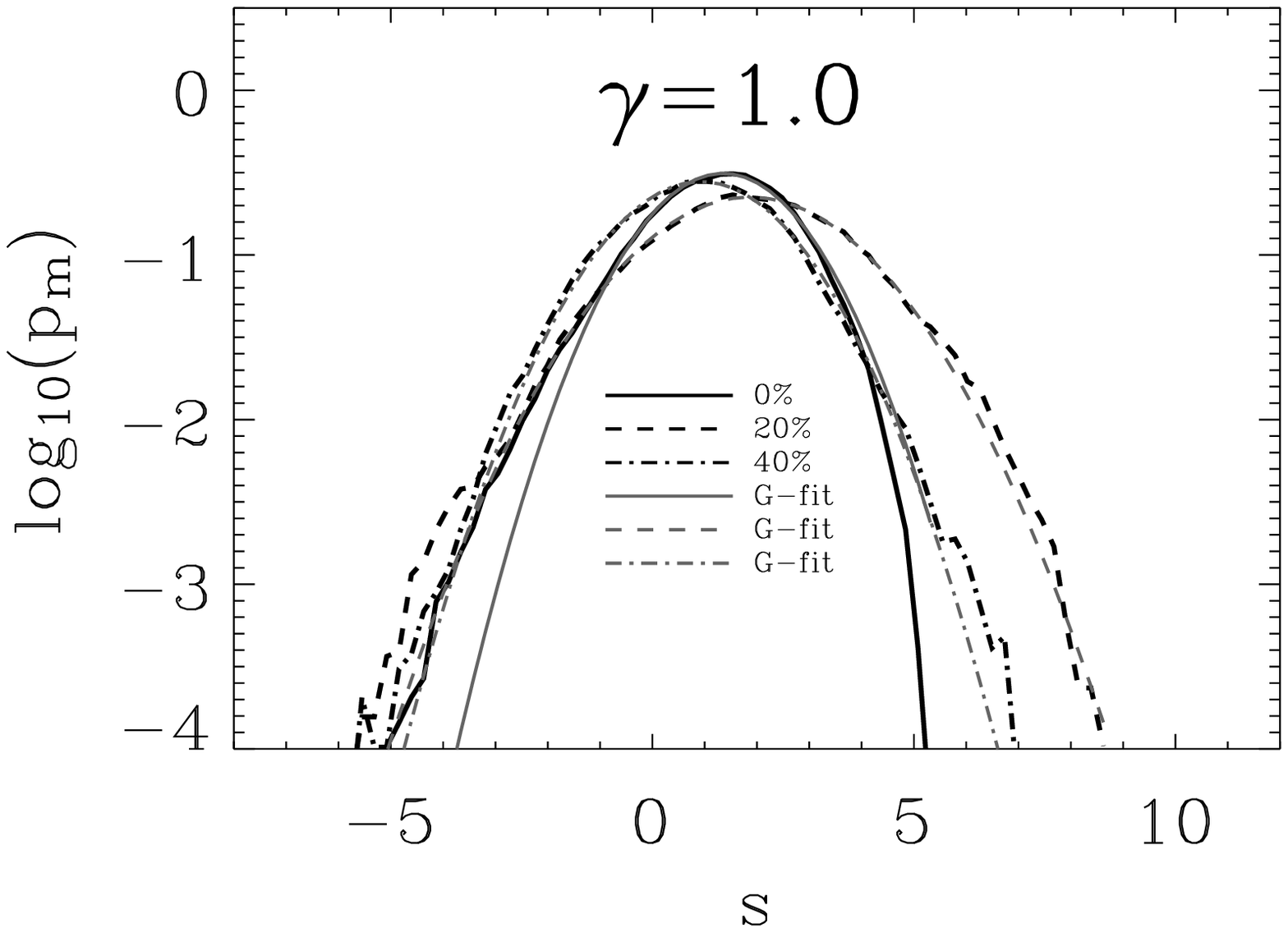}\\
\includegraphics[height=2.0in]{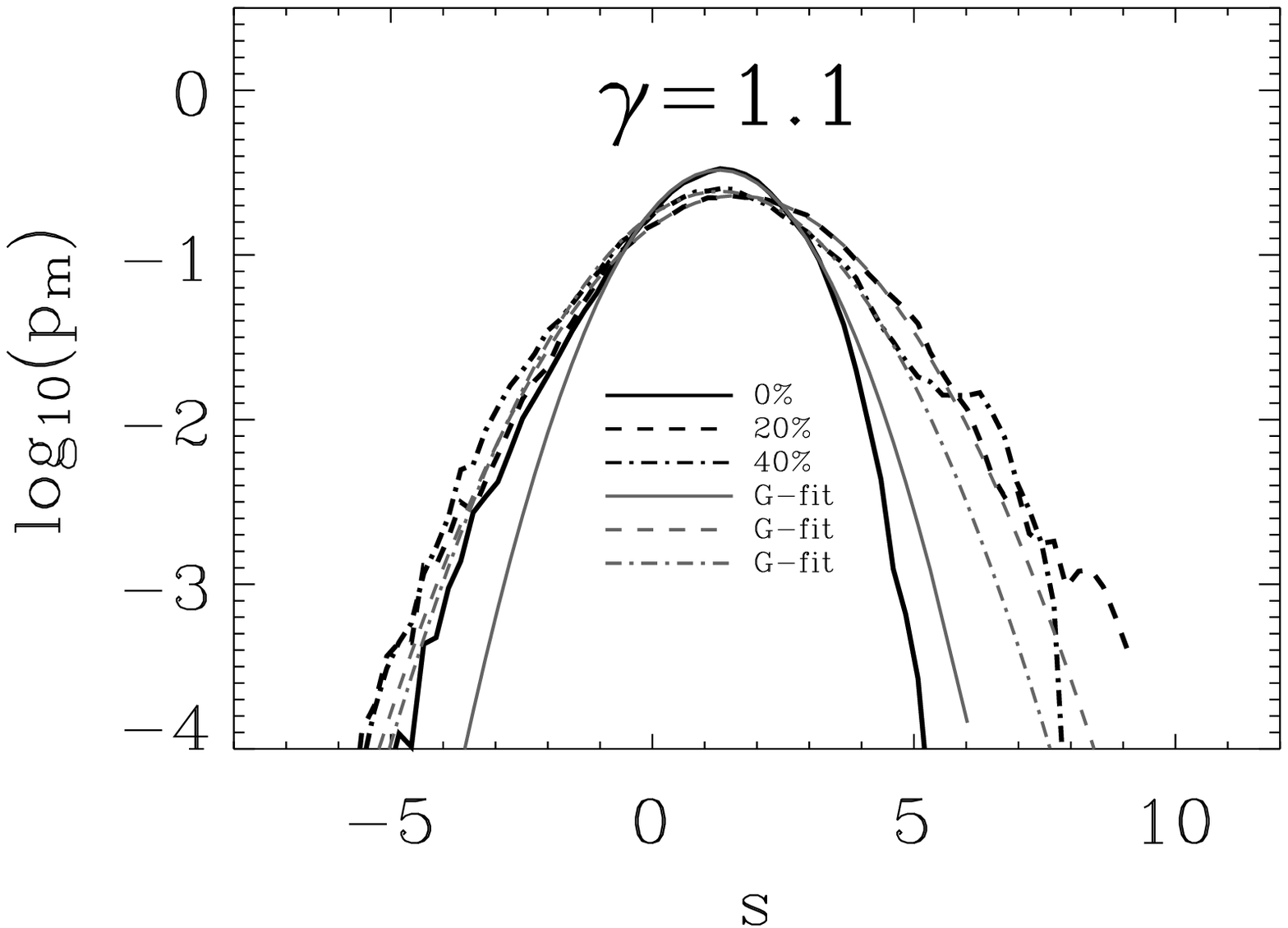}
\includegraphics[height=2.0in]{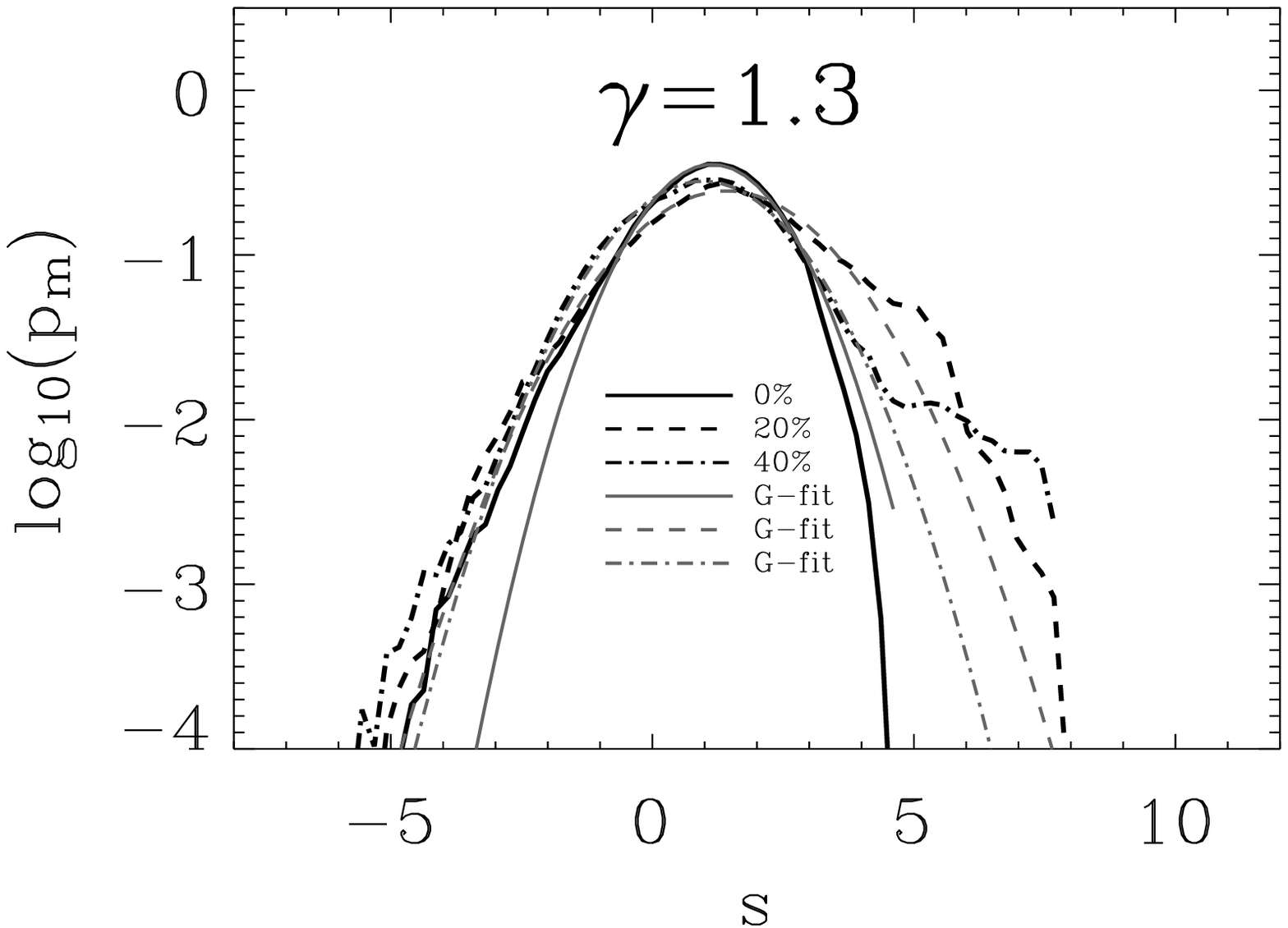}
\caption{\label{fig_pdf_3phases}Mass-weighted gas density PDFs $p_m$
of models driven with $k=1-2$ for selected $\gamma$. Only gas not
accreted into sink particles is included. Three evolutionary phases are
shown: the initial state (\textit{thick solid-line}), $M_{\ast}
\approx 20\%$ (\textit{thick dashed-line}), and $M_{\ast} \approx
40\%$ (\textit{thick dot-dashed line}). The thin lines (solid, dashed,
dot-dashed) are Gaussian fits to the corresponding PDFs above 10\% of
peak value. Again $s=\ln (\rho/\rho_0)$. Note the change of the
high-density tails with $\gamma$.}
\end{center}
\end{figure}

Figure \ref{fig_pdf_3phases} shows the evolution of the mass-weighted
gas density PDF $p_m$ of the model driven with $k=1-2$ for different
values of $\gamma$. The same three evolutionary phases as defined in
\S\ref{subsec_mas} are shown: the initial distribution, and times at
which dense cores have accreted $M_{\ast} \approx 20\%$ and $M_{\ast}
\approx 40\%$ of the mass.  These PDFs do not include gas accreted
into the cores.  We again show the best-fit Gaussian for the part of
each PDF above 10\% of peak. 

During the dynamical evolution, the PDFs in the non-isothermal cases develop
pronounced positive deviations from the the Gaussian fit, while the PDF in the 
isothermal case remains well fit by a Gaussian (of increasing width),
consistent with the results in \citet{rsk00}.  The high-density tails
gradually diminish as $\gamma$ increases from 0.2 to 1.0, vanishing almost
entirely at $\gamma = 1.0$.  At $\gamma > 1.0$, though, the tails develop
again and get stronger as $\gamma$ continues to increase.  We do not fully
understand why the tails only appear in non-isothermal cases.  

\begin{figure} 
\begin{center}
\includegraphics[height=2in]{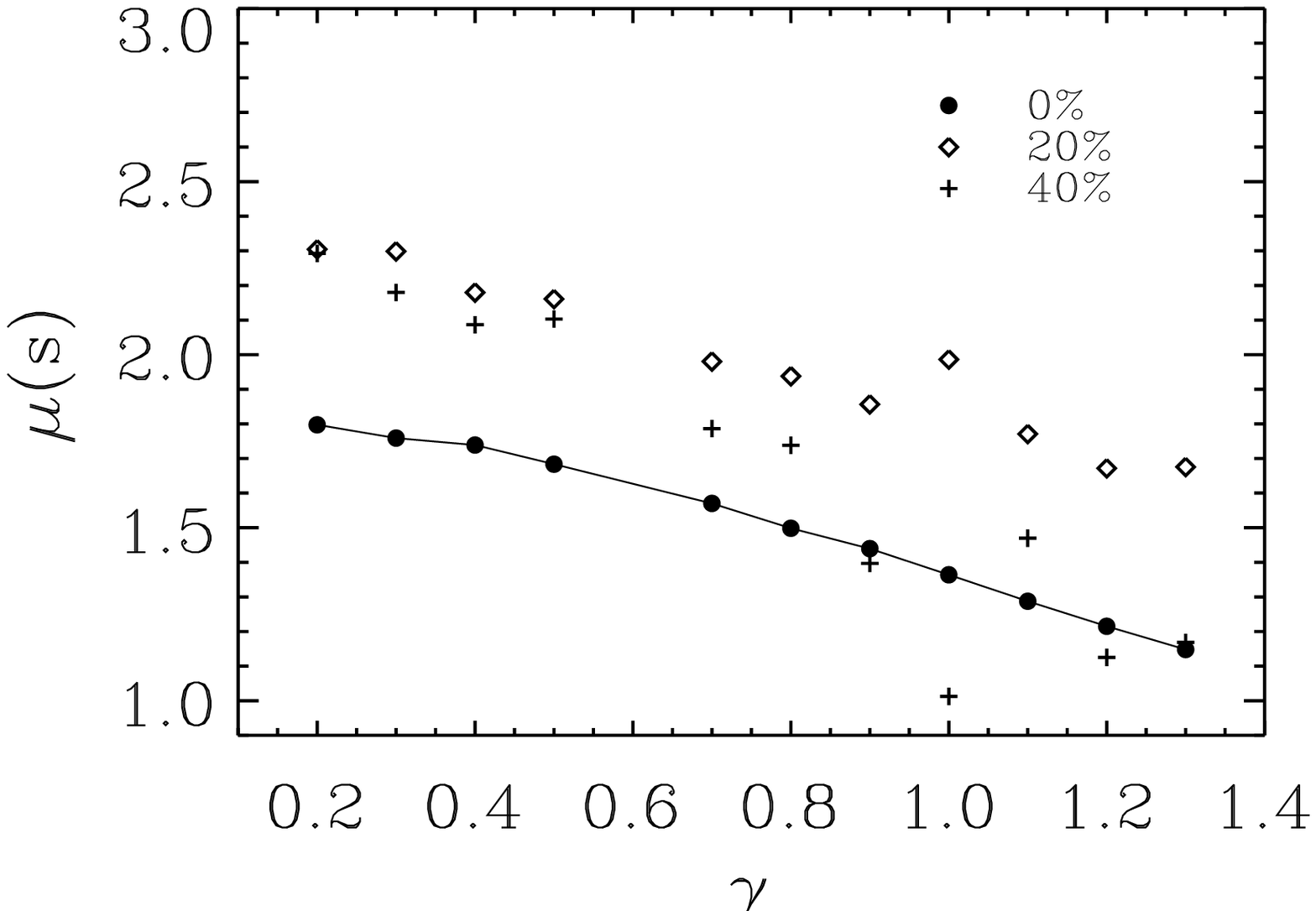}
\includegraphics[height=2in]{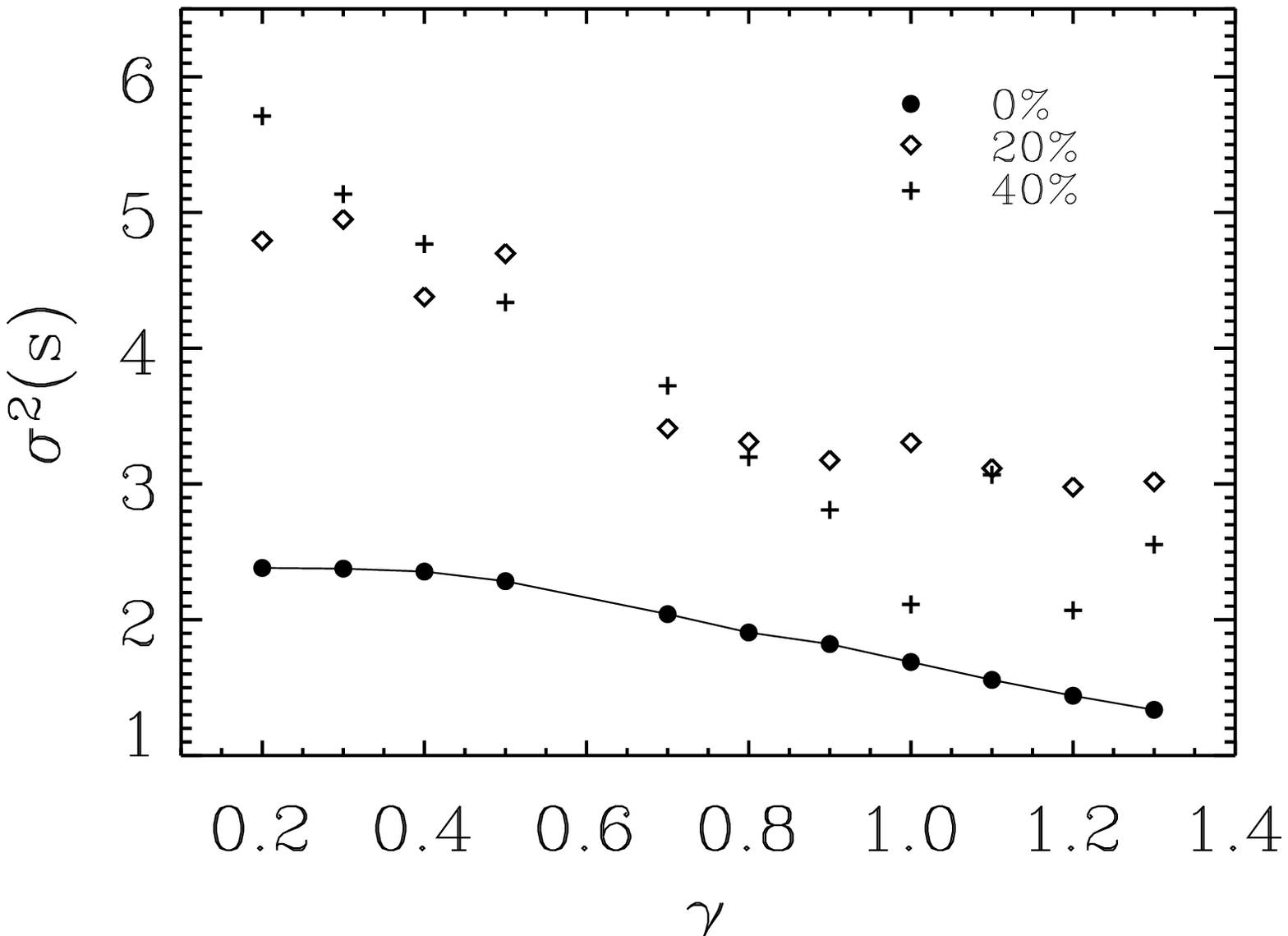}\\
\includegraphics[height=2in]{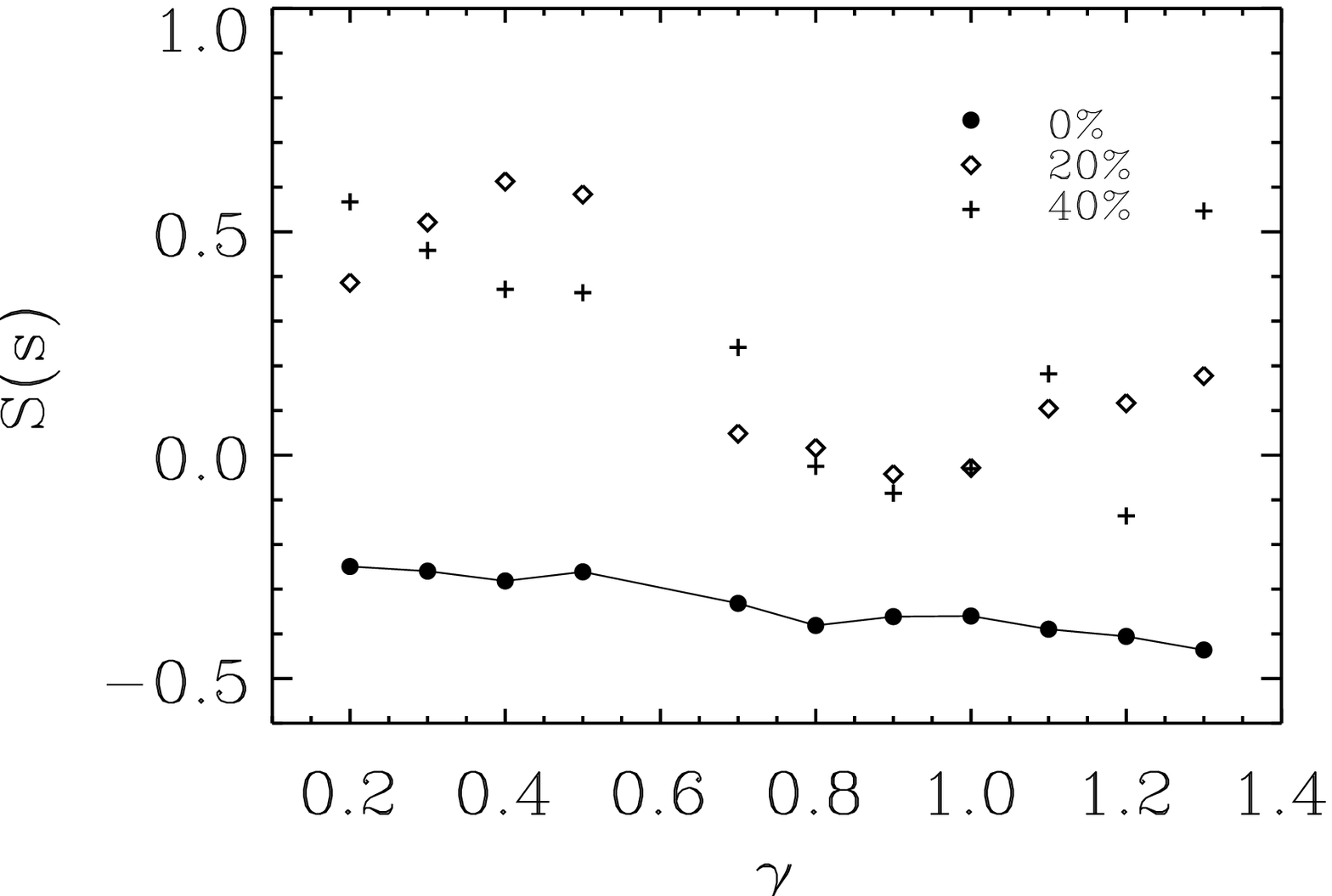}
\includegraphics[height=2in]{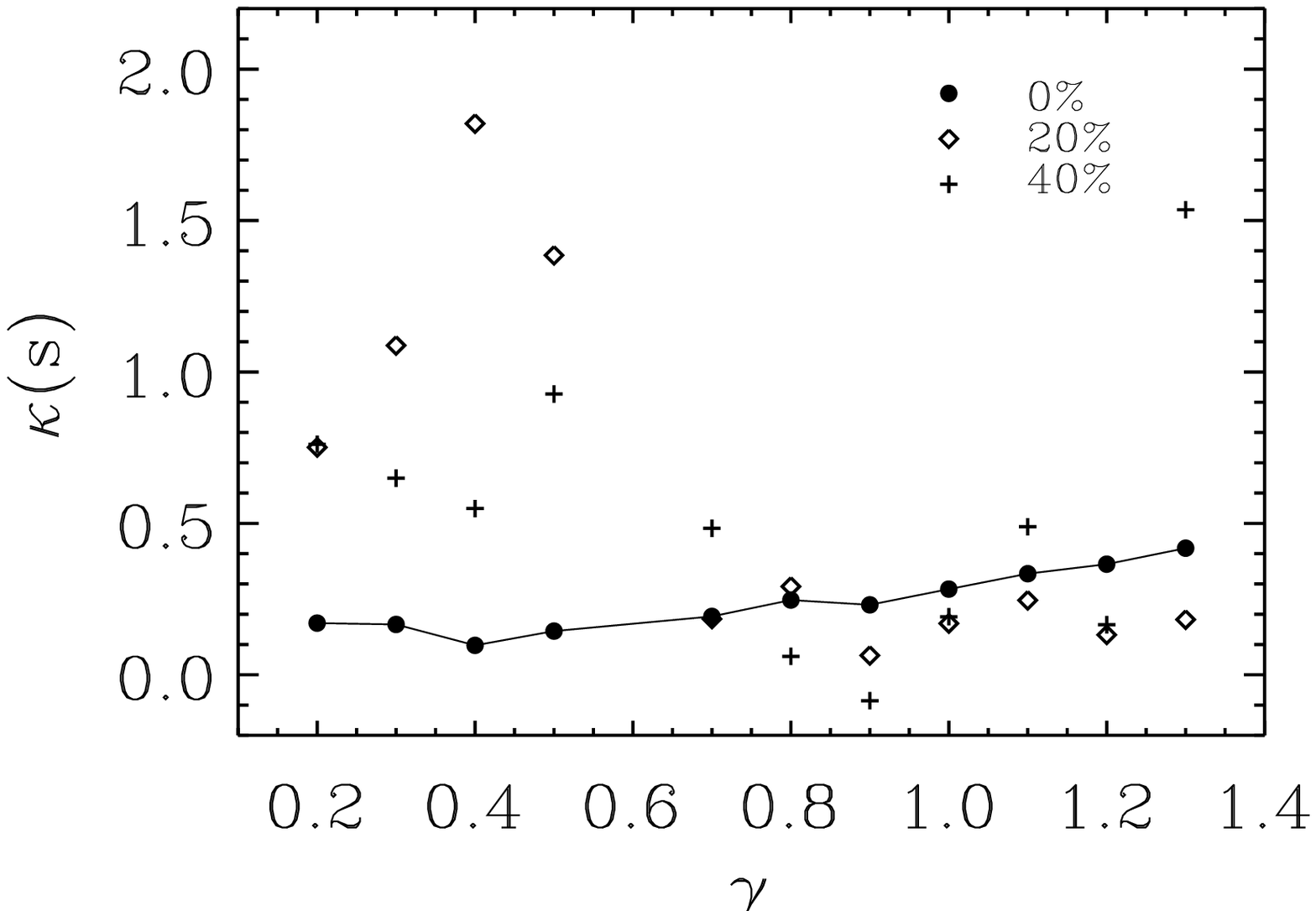}
\caption{\label{fig_mom}
The first four moments of three evolutionary phases of the
mass-weighted density PDF (corresponding to Figure
\ref{fig_pdf_3phases})of the model driven with $k=1-2$, as functions
of $\gamma$. , Shown are the mean $\mu$ (top left), variance
$\sigma^2$ (top right), the skewness $S$ (bottom left), and the
kurtosis $\kappa$, of $s=\ln (\rho/\rho_0)$. We define the
fourth moment $\kappa$ with a value of 3 subtracted, so that for a
Gaussian $\kappa =0$, while an exponential has $\kappa =3$.}
\end{center}
\end{figure}

The first four moments of the density PDFs for each of the three
evolutionary phases discussed previously are shown in
Figure~\ref{fig_mom}. We can see that the moments vary as collapse
proceeds, demonstrating that self-gravity plays an important role in
shaping the density PDF.

The plot of the skewness shows that the PDFs of turbulence
without self-gravity are always skewed to low densities, regardless of
$\gamma$. Shock compression comes at a cost: the gas swept up in shock
fronts has to come from somewhere, and those regions now contain only
low-density gas. As the filling factor of shocks is low \citep{smh00} most of
the volume is occupied by these low-density regions resulting in PDFs with
negative skewness. The situation changes, however, when self-gravity comes
into play. Gravitational contraction produces high densities in addition to
the turbulent shock compression, the PDF's in the later stages of evolution
are thus strongly biased towards high-density gas and the skewness becomes
positive.

\placefigure{fig_pdf_3phases} 

\placefigure{fig_mom}

\section{SUMMARY AND DISCUSSION}
\label{sec_discussion}

We have investigated the effect of varying the equation of
state (EOS) on the formation of stellar clusters in turbulent
molecular clouds using a set of three-dimensional numerical
simulations.  Our results show that the EOS plays an important role in
the fragmentation of the clouds, the determination of the initial
mass function (IMF) of the protostellar cores, and the shape of the
gas density PDFs.

The ability of interstellar gas clouds to fragment under the action of
self-gravity {\em decreases} with {\em increasing} \new{polytropic exponent}
$\gamma$ in the range $0.2 < \gamma < 1.4$ relevant for Galactic molecular
clouds. The total amount of mass in collapsed cores, however, appears to
remain roughly constant through that range. At least half of this material is
expected to accrete onto the protostellar system at the center of each
collapsing core (e.g. \citealt{rsk01}).  Small values of $\gamma$ thus lead to
the formation of dense clusters of predominantly low-mass stars, while
$\gamma>1$ results in the formation of isolated and massive stars.
Fragmentation and collapse ceases entirely for $\gamma > 1.4$ as
expected from analytic arguments.

The mass spectra of both clumps and collapsed cores change with $\gamma$, with 
that effect being most pronounced for the cores. For $\gamma \le 1.0$, their
mass spectrum appears roughly log-normal.  As $\gamma$ increases, fewer but
more massive objects emerge. When $\gamma > 1.0$, the distribution is
dominated by high mass objects only, and the spectrum tends to flatten out. It
is fit by neither a log-normal nor a power-law. The clump mass spectra, on the
other hand, do show power-law behavior at the high mass end for all $\gamma$.

The density PDF changes with $\gamma$ as well. The width of the
profile decreases as $\gamma$ increases, as expected from analytic
arguments about shock compression. Both mass- and volume-weighted
PDFs show imperfect log-normal distributions. Self-gravity helps 
shape the PDF. During dynamical collapse, the PDF develops a
high-density tail, which is the imprint of local collapse of high
density regions.

\begin{figure} 
\begin{center}
\includegraphics[height=2.0in]{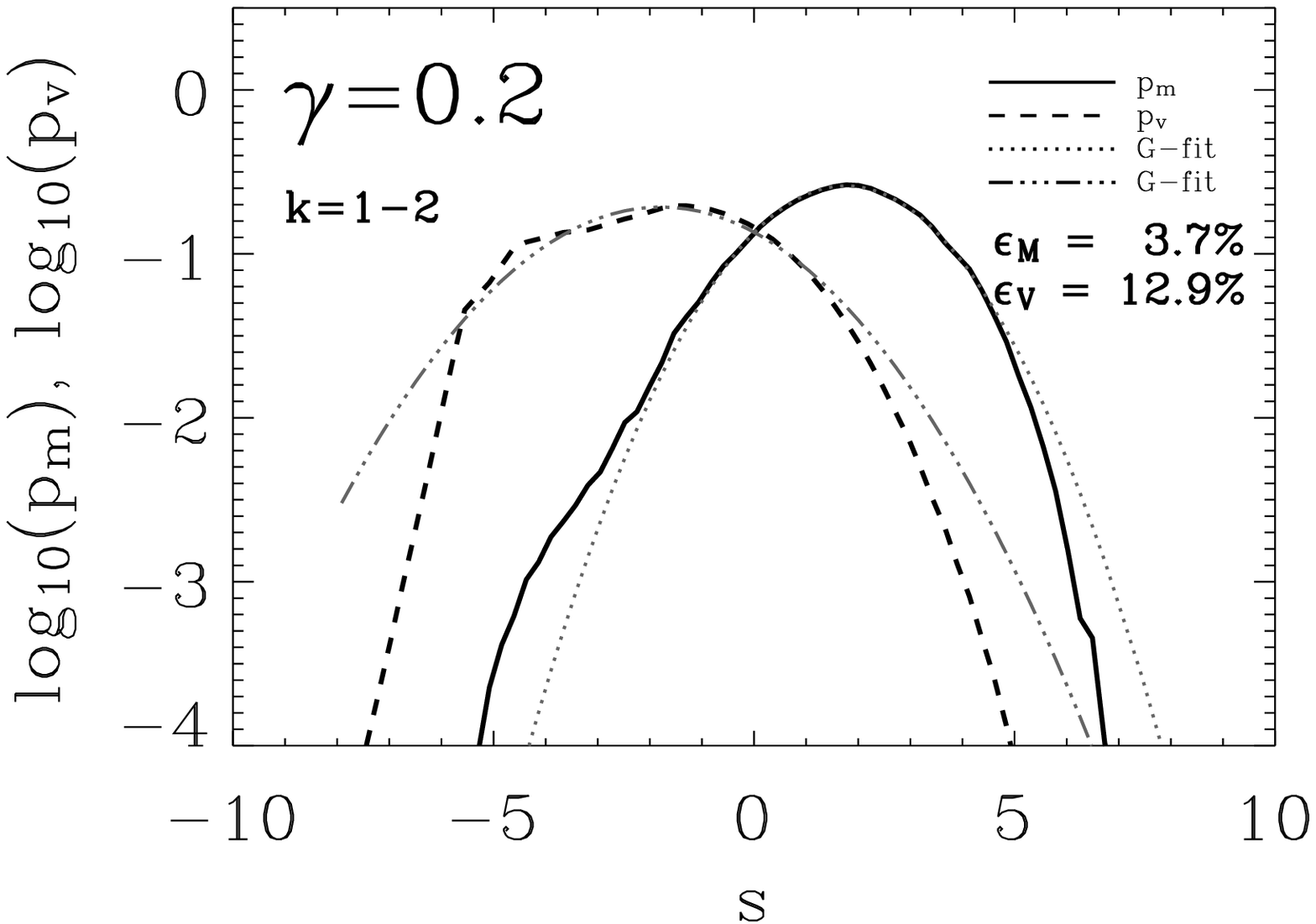}
\includegraphics[height=2.0in]{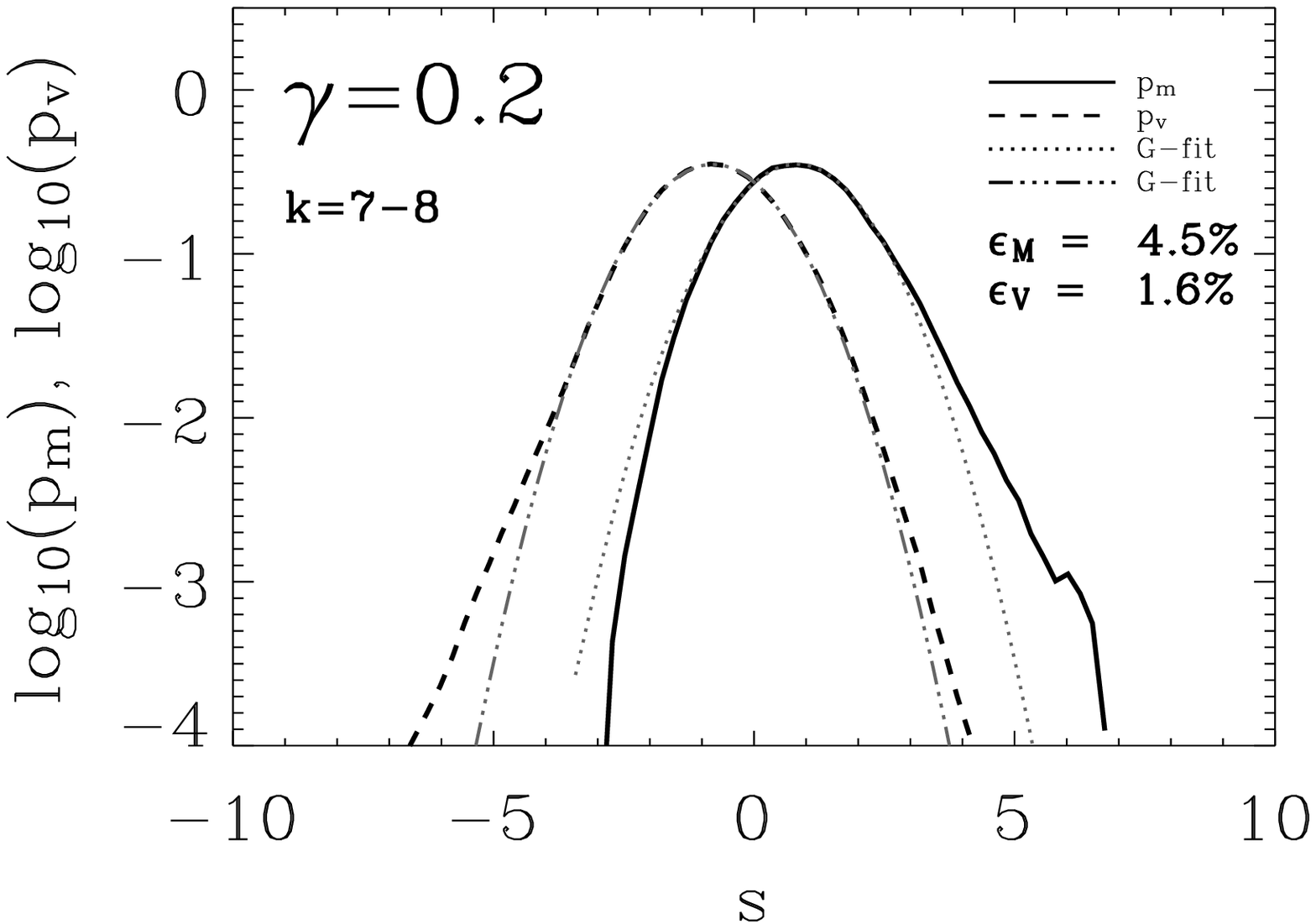}\\
\includegraphics[height=2.0in]{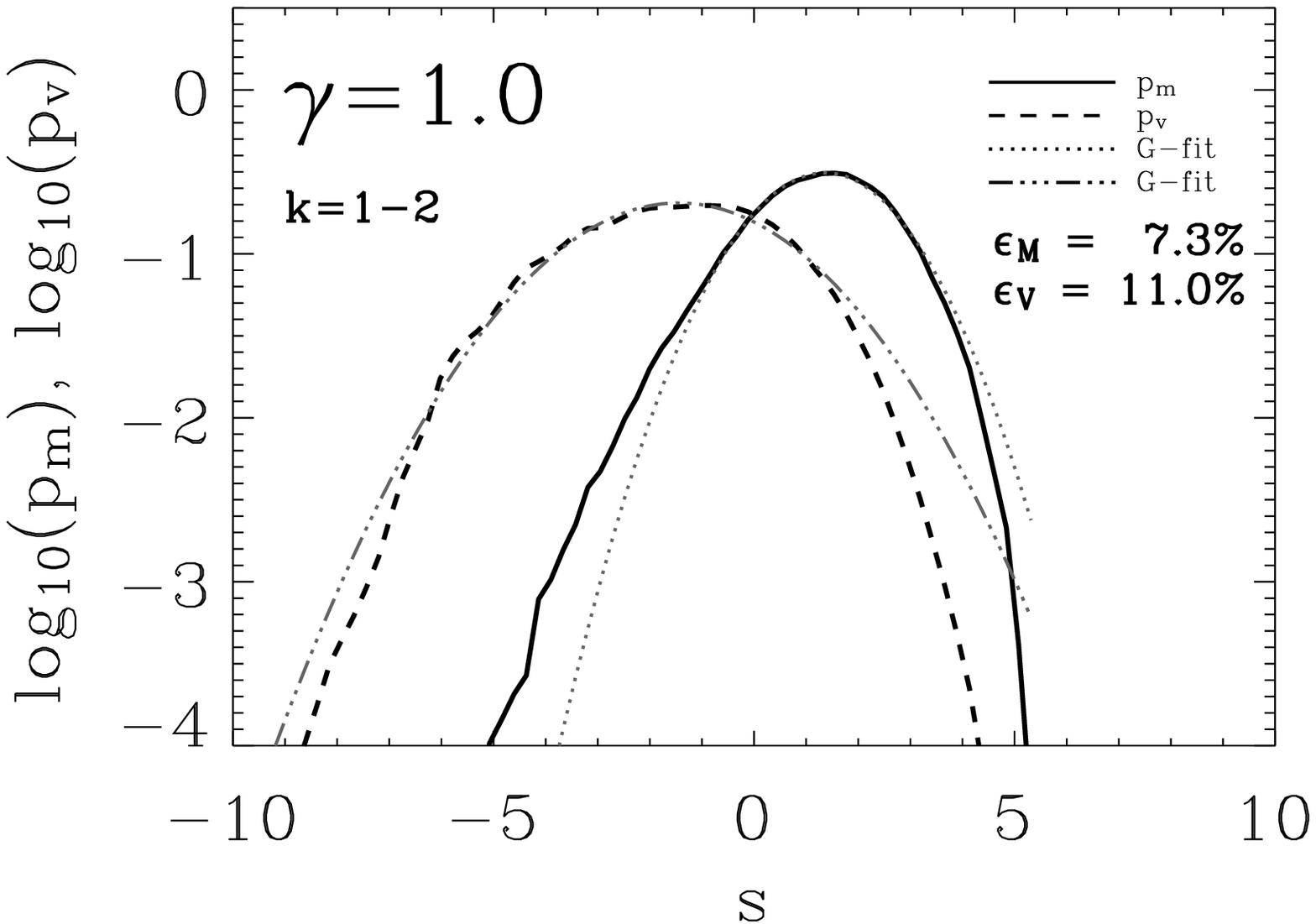}
\includegraphics[height=2.0in]{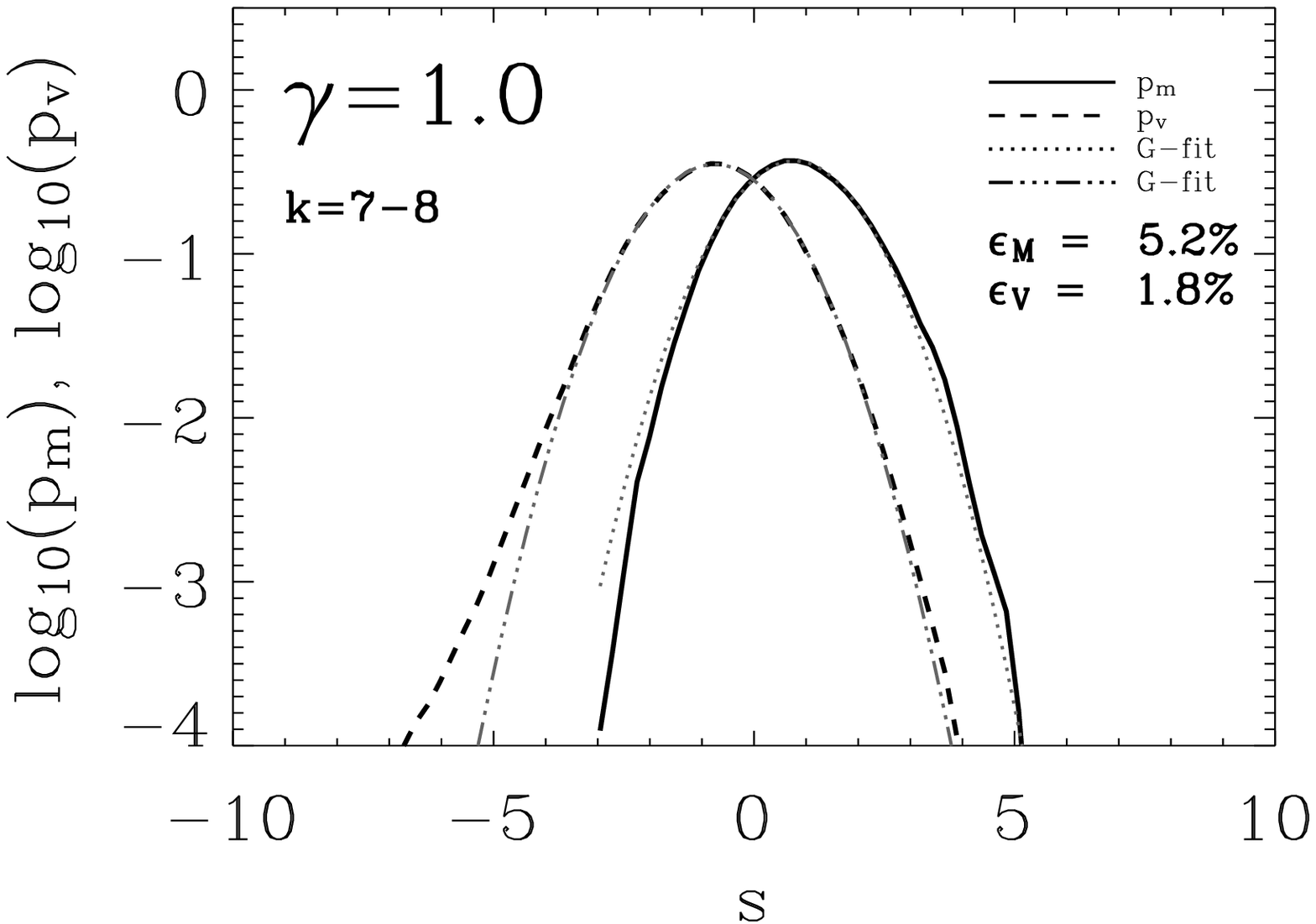}\\
\includegraphics[height=2.0in]{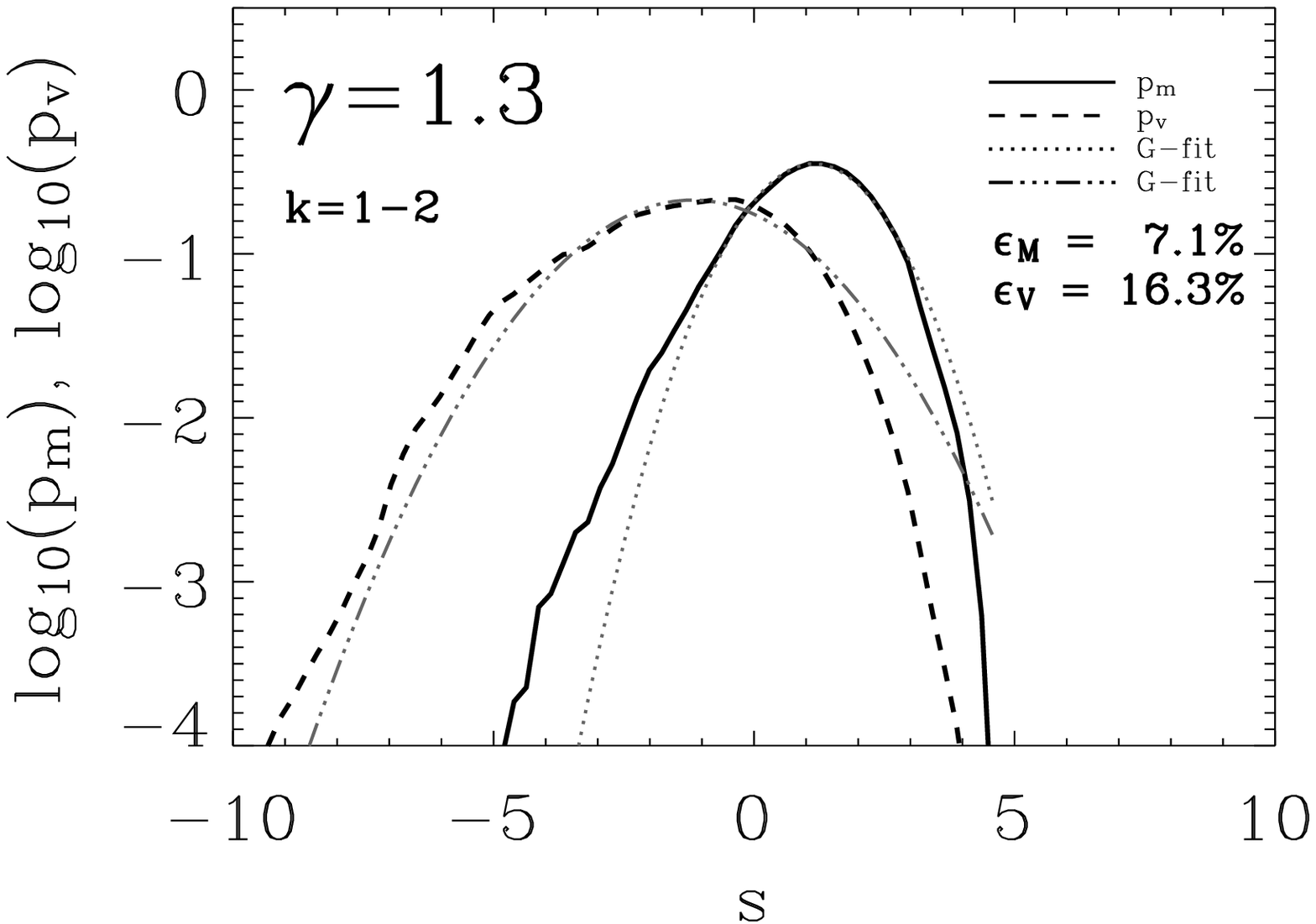}
\includegraphics[height=2.0in]{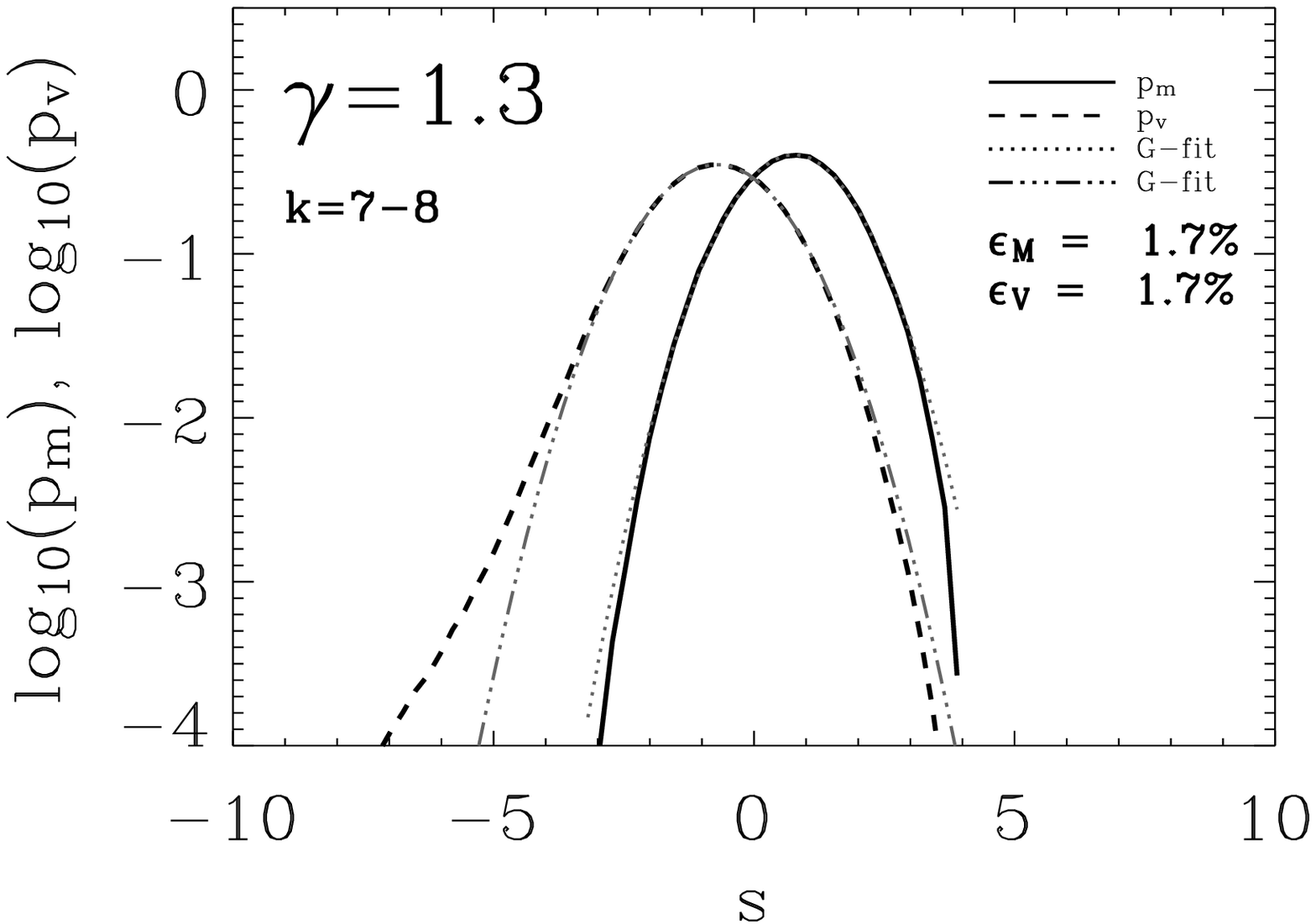}
\caption{\label{fig_pdf_int}Mass-weighted (\textit{thick solid-line})
and volume-weighted (\textit{thick dashed-line}) gas density PDFs
$p_m$ and $p_v$ of fully-developed turbulence prior to turning on
self-gravity.  Models with driving wavenumber $k=1-2$ (\textit{left
column}) and $k=7-8$ (\textit{right column}) are shown. Gaussian fits
are shown with \textit{thin dotted} and \textit{thin dot-dashed} lines
correspondingly, and the errors $\epsilon_M$ and $\epsilon_V$
summed over the parts of the curves above $10\%$ of the peak values,
are quoted. Note that $s=\ln (\rho/\rho_0)$.}
\end{center}
\end{figure}

\newrsk{In a polytropic gas cloud, any temperature fluctuation corresponds to
a density fluctuation, $\Delta\log_{10}(T) \simeq \Delta\log_{10}(\rho)(1 -
\gamma) -0.6$. We note that from the gas density PDFs in Figure
\ref{fig_pdf_int}, the maximum width of the profiles is about 12, which
corresponds to $\Delta\log_{10}(\rho) \simeq 5$. So for the extreme  case
$\gamma = 0.2$, $\Delta\log_{10}(T) \simeq 3.5$. For a cloud with an average
temperature of about 10 K  the maximum temperature of the low-density gas
therefore  could reach values up to $10^4$ K, if $\gamma$ is extremely
low. This still falls in the temperature range observed in star forming regions.}

\newrsk{The principal motivation of this paper is to investigate how the
fragmentation behavior of a turbulent cloud and the subsequent
formation of stars are affected by physical state of the cloud, as
described by the EOS. Since $\gamma$ reflects the balance between
heating and cooling, $\gamma$ is expected bo be low in regions of
strong cooling, while $\gamma$ may be large where the cooling
mechanism is insufficient \citep{ss00}. Unfortunately, the true chemical state
and composition of interstellar gas is difficult to assess, so exact values
of $\gamma$ are difficult to estimate observationally.}

The possibility that isolated, massive stars form when $\gamma > 1$ is
of interest, not only because ionizing radiation, stellar winds, and
supernovae from massive stars play an important role in the
interstellar medium, but also because massive stars are rarely
observed in isolation.  Rather, they are usually found as members of
rich stellar clusters.  Recently, however, \citet{lamers02} has
reported observations of massive stars found in isolation or
associated only with very small groups of lower-mass stars in the
bulge of M51. \citet{mas02} also reported finding apparently isolated,
massive, field stars in both the Large and Small Magellanic Clouds.
Our simulations suggest that molecular cloud regions forming isolated,
high-mass stars have $\gamma > 1$. In this case turbulent cloud
fragmentation tends to produce isolated, massive collapsing cores that
will feed onto isolated high-mass stars. \citet{ys02}
demonstrated that high-mass stars can form in collapsing gas clumps
via disk accretion very much the same way as low mass stars are
believed to do. There is thus no need for a dense cluster environment
where high-mass stars could build up by collisions of lower-mass stars
(as suggested by \citealt{bbz98}).
 
\begin{figure} 
\plottwo{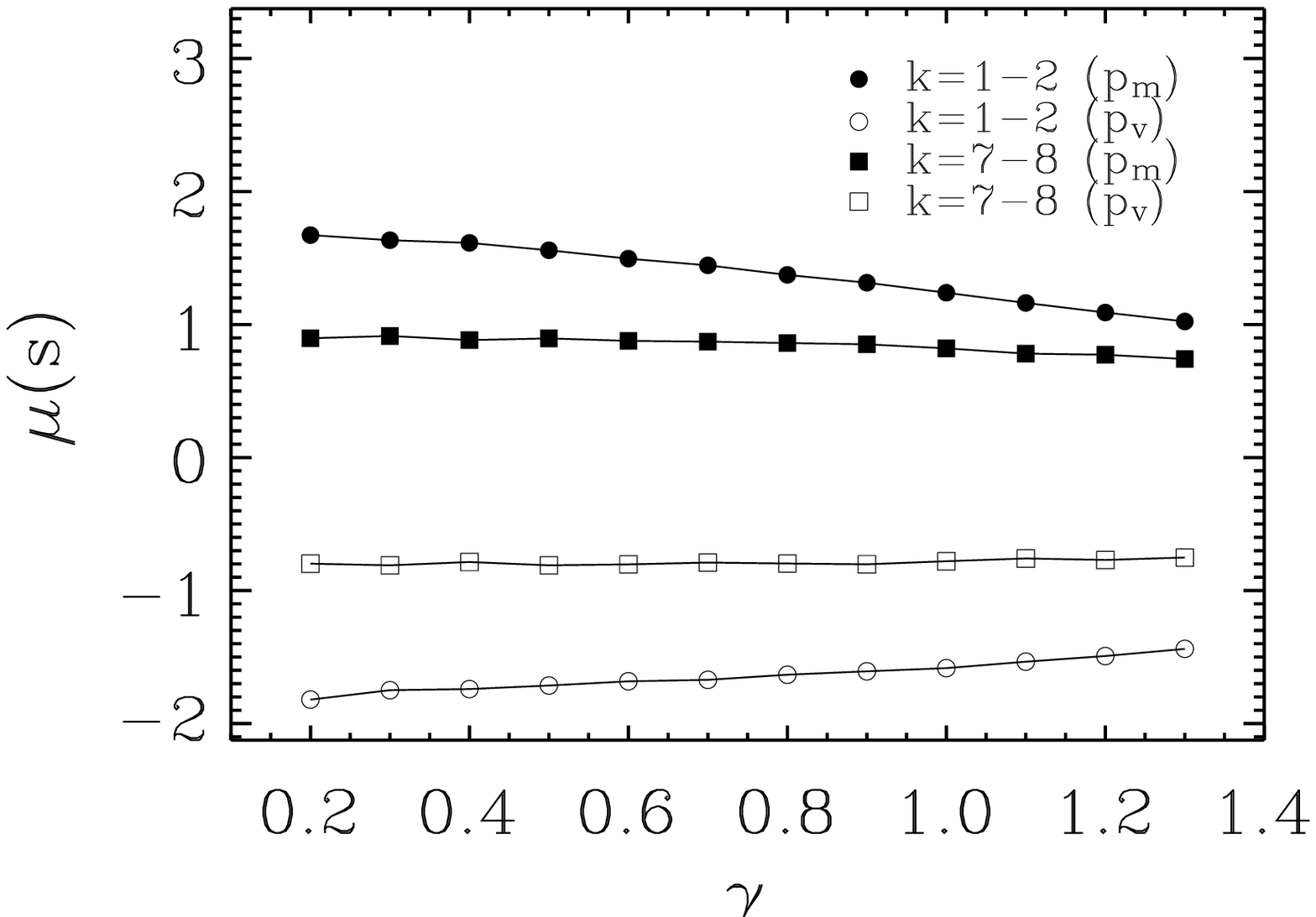}{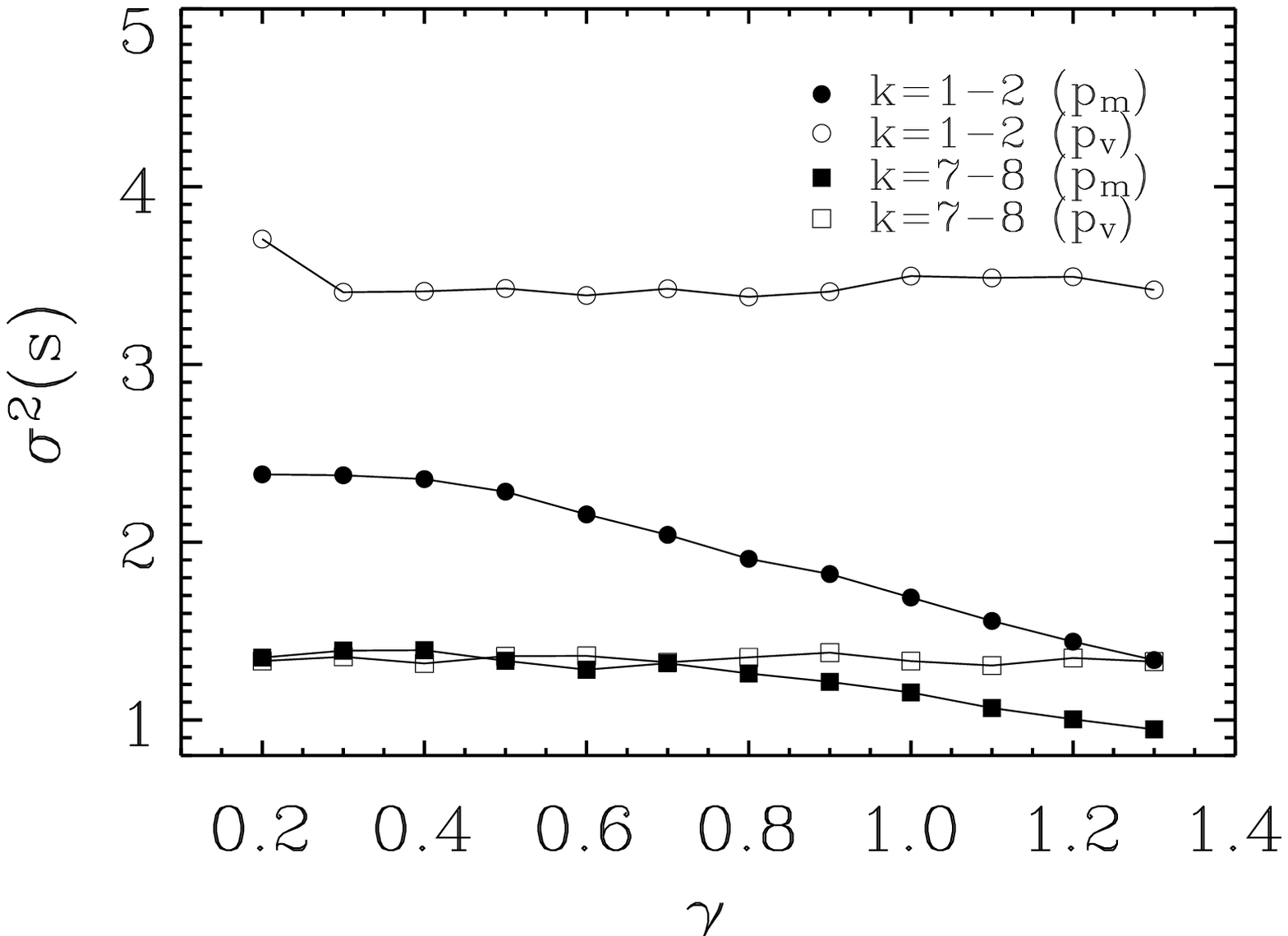}
\caption{\label{fig_mom_k1278} First moment (mean) $\mu$
(\textit{top}) and second moment (variance) $\sigma^2$
(\textit{bottom}), of both $p_m$ (\textit{filled}) and $p_v$
(\textit{open}) as functions of $\gamma$, for models driven with
wavenumbers $k=1-2$ (\textit{circle}) and $k=7-8$ (\textit{square}),
where $s=\ln (\rho/\rho_0)$.}
\end{figure}

A theory of galactic-scale star formation partially based on the
universality of log-normal density PDFs has been proposed by Elmegreen
(2002). Our results do show a log-normal form at least for the peak of
the PDFs, containing most of the gas.  However, the dispersion of the
PDF varies substantially as the EOS changes
(Fig.~\ref{fig_mom_k1278}), so that the fraction of gas available for
star formation would depend significantly on the EOS in such a theory.
This raises the question of how it could still be able to explain a
global Schmidt Law across many different galaxies 
with different metallicities, radiation fields, interstellar
pressures, and so forth.

High resolution simulations by \citet{abn00, abn02} of the
formation of the first star suggest that fragmentation during the
collapse of metal-free pregalactic halos is rather inefficient,
resulting in the formation of single, massive stars rather than
clusters of lower-mass objects.  In the absence of metals, inefficient
cooling may result in high $\gamma$, perhaps helping to explain these
results. 

We do note, however, that simulations by \citet{bcl99, bcl02} show greater
multiplicity.  They use similar chemistry, 
but examine more massive, more isolated halos, with substantially
larger limiting mass resolution.  The EOS will have less effect on
these larger scales, which are more dominated by gravity.  If the
metal-free gas indeed shows a high effective $\gamma$, our models
would suggest that their collapsing regions will show little further
fragmentation if followed down to stellar mass scales.

\acknowledgments We thank M. Fall, H. Lamers, H. Zinnecker, \new{and
S. Glover} for valuable discussions, and D. Janies and W. Wheeler for their
work on the AMNH Parallel Computing Facility, which we used for the 
computations presented here.  \new{We are grateful to the anonymous referee for
useful comments and corrections}. YL thanks the AIP for its warm
hospitality, and the Kade Foundation for support of her visits there.
M-MML acknowledges partial support by NASA ATP grant NAG5-10103, and
by NSF CAREER grant AST99-85392.  RSK acknowledges support by the Emmy
Noether Program of the Deutsche Forschungsgemeinschaft (DFG,
KL1358/1).

\clearpage

\clearpage

\clearpage

\clearpage

\clearpage

\clearpage

\begin{figure}
\begin{center}
\includegraphics[height=1.5in]{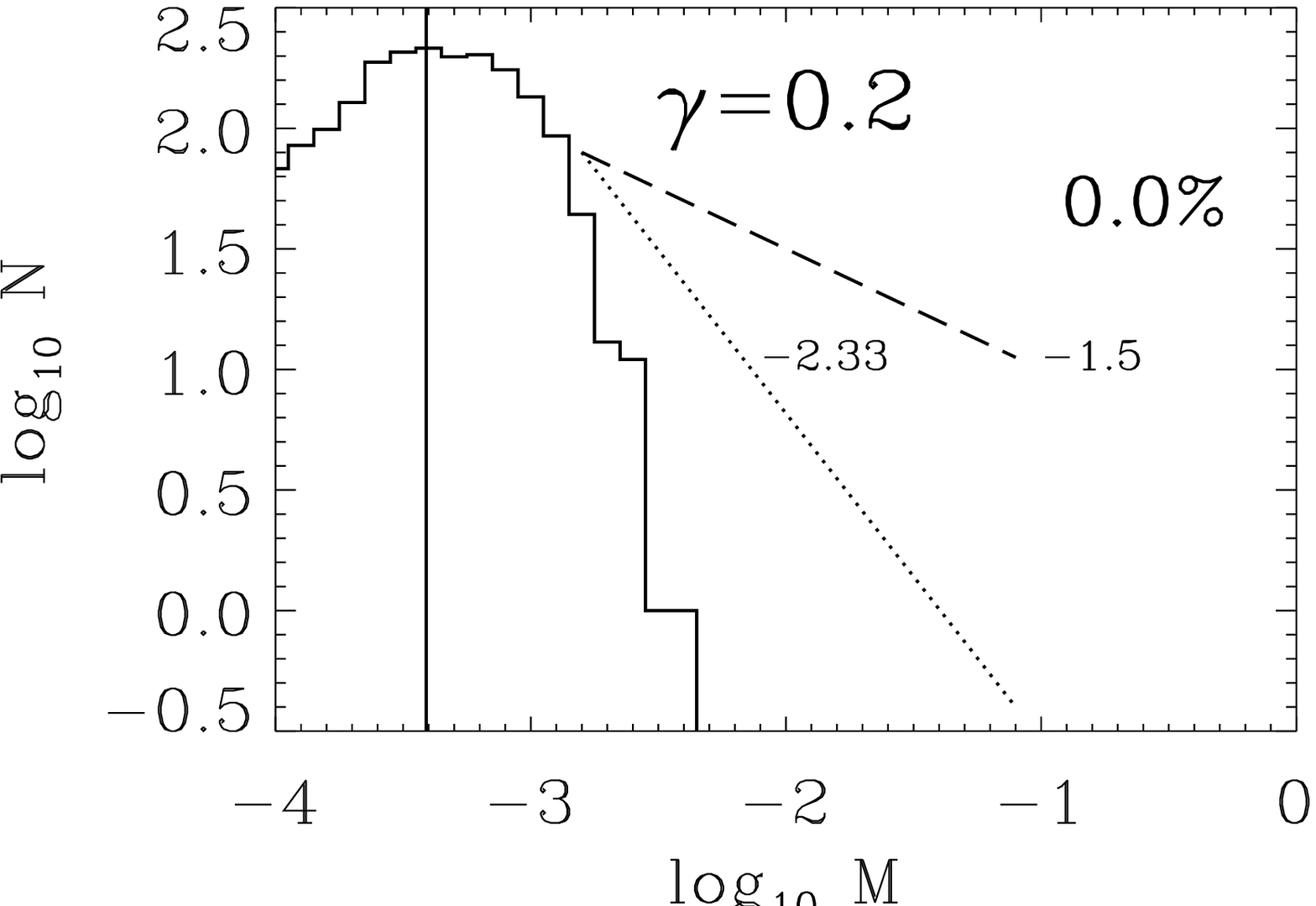}
\includegraphics[height=1.5in]{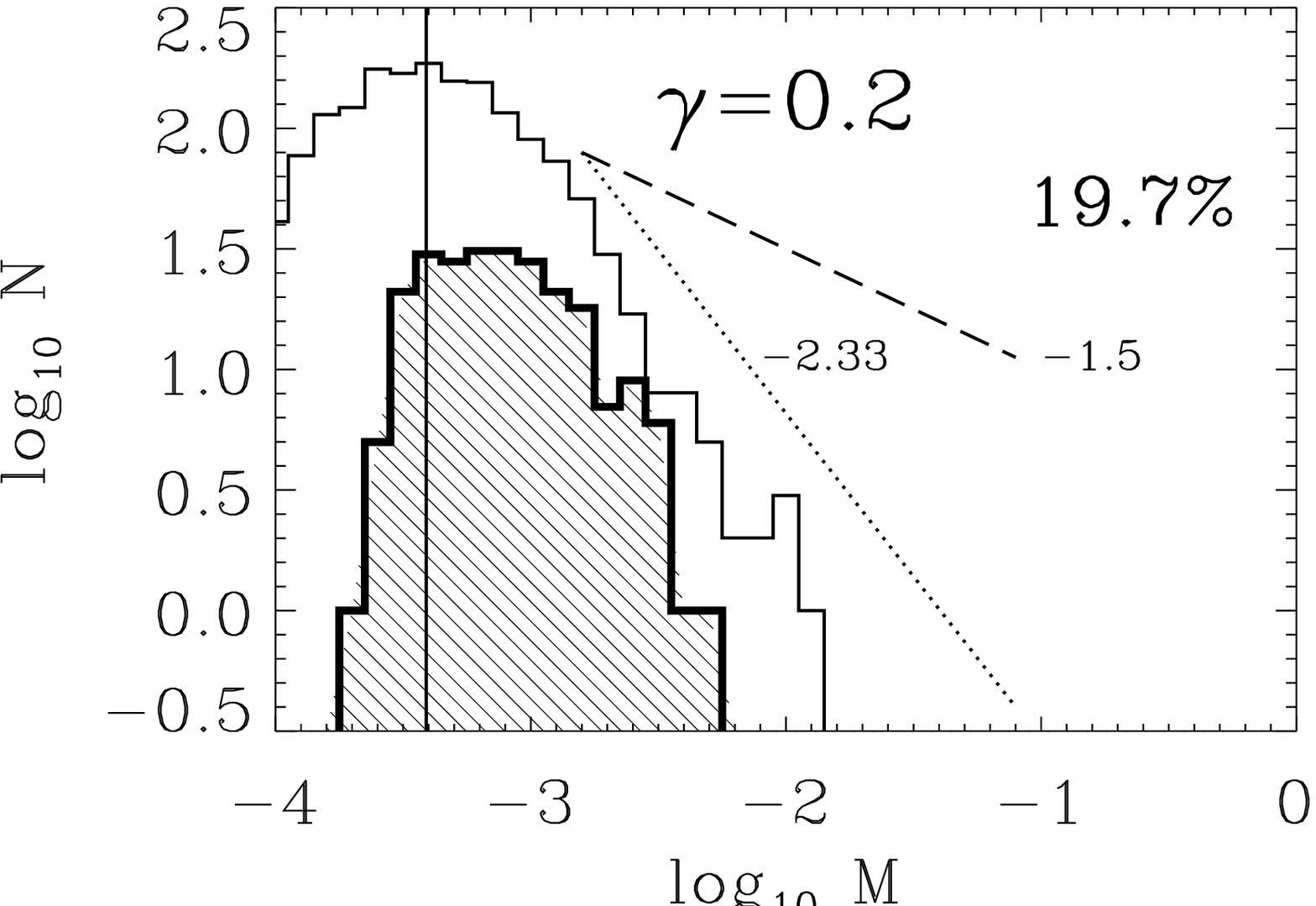}
\includegraphics[height=1.5in]{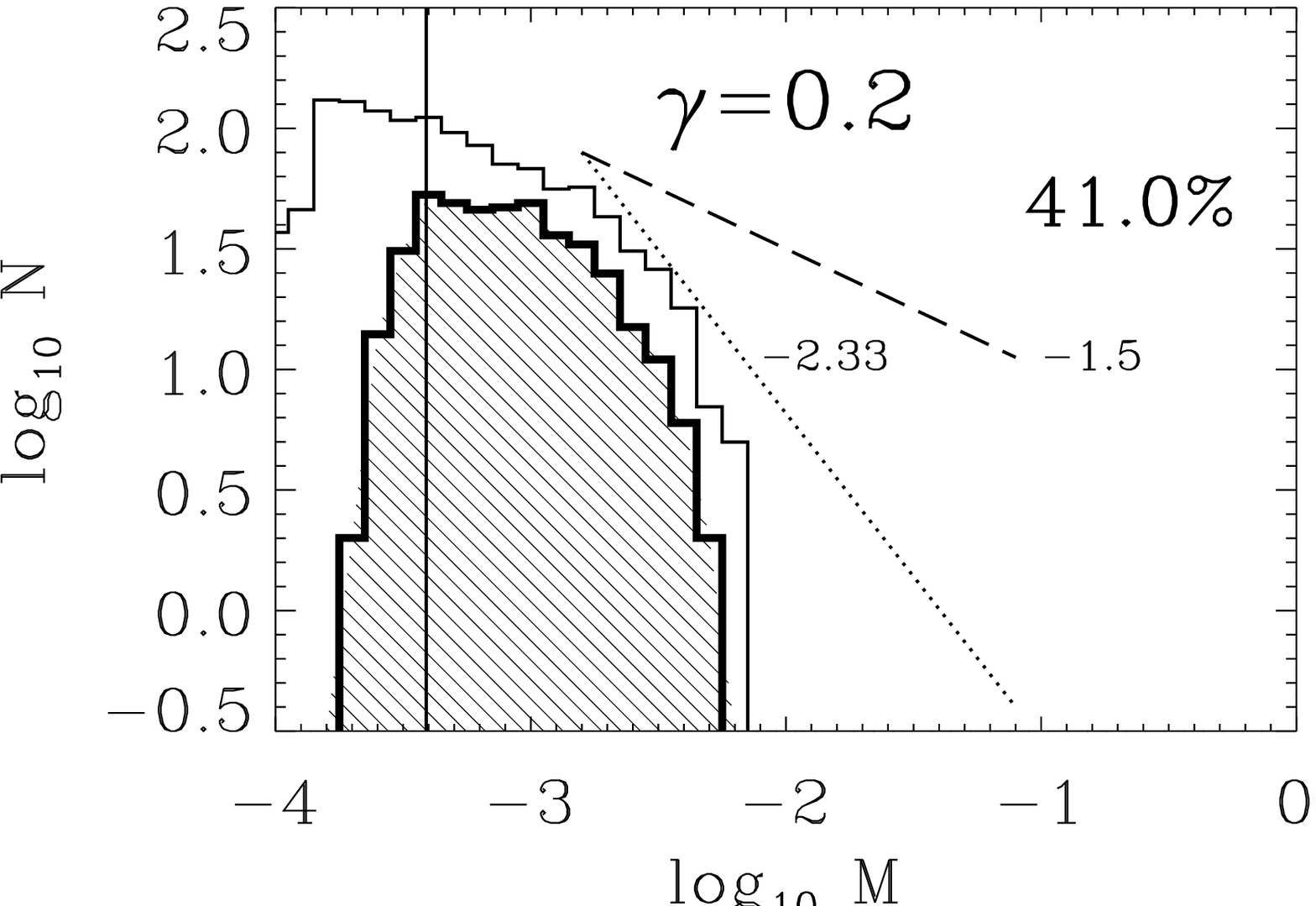}\\
\includegraphics[height=1.5in]{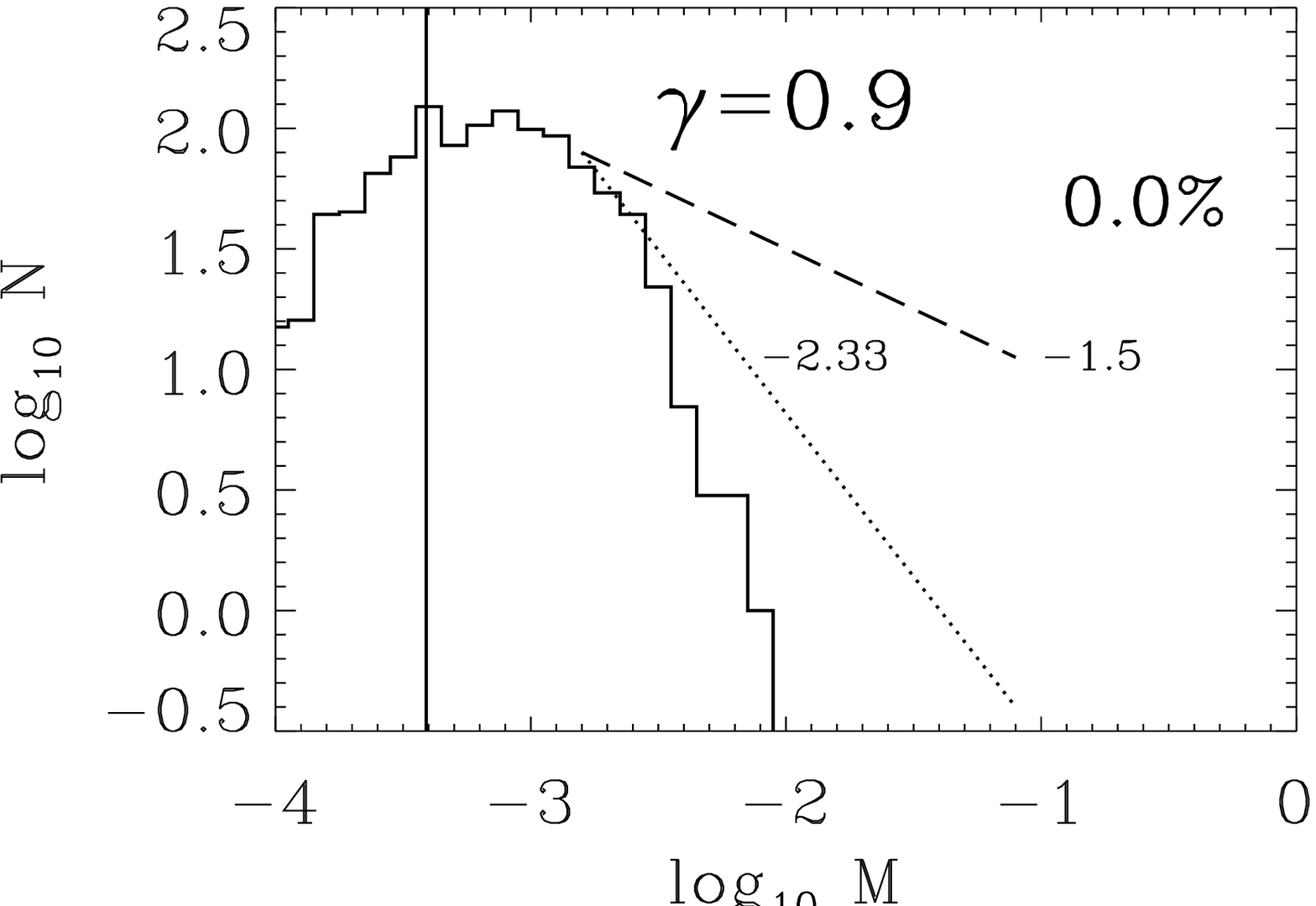}
\includegraphics[height=1.5in]{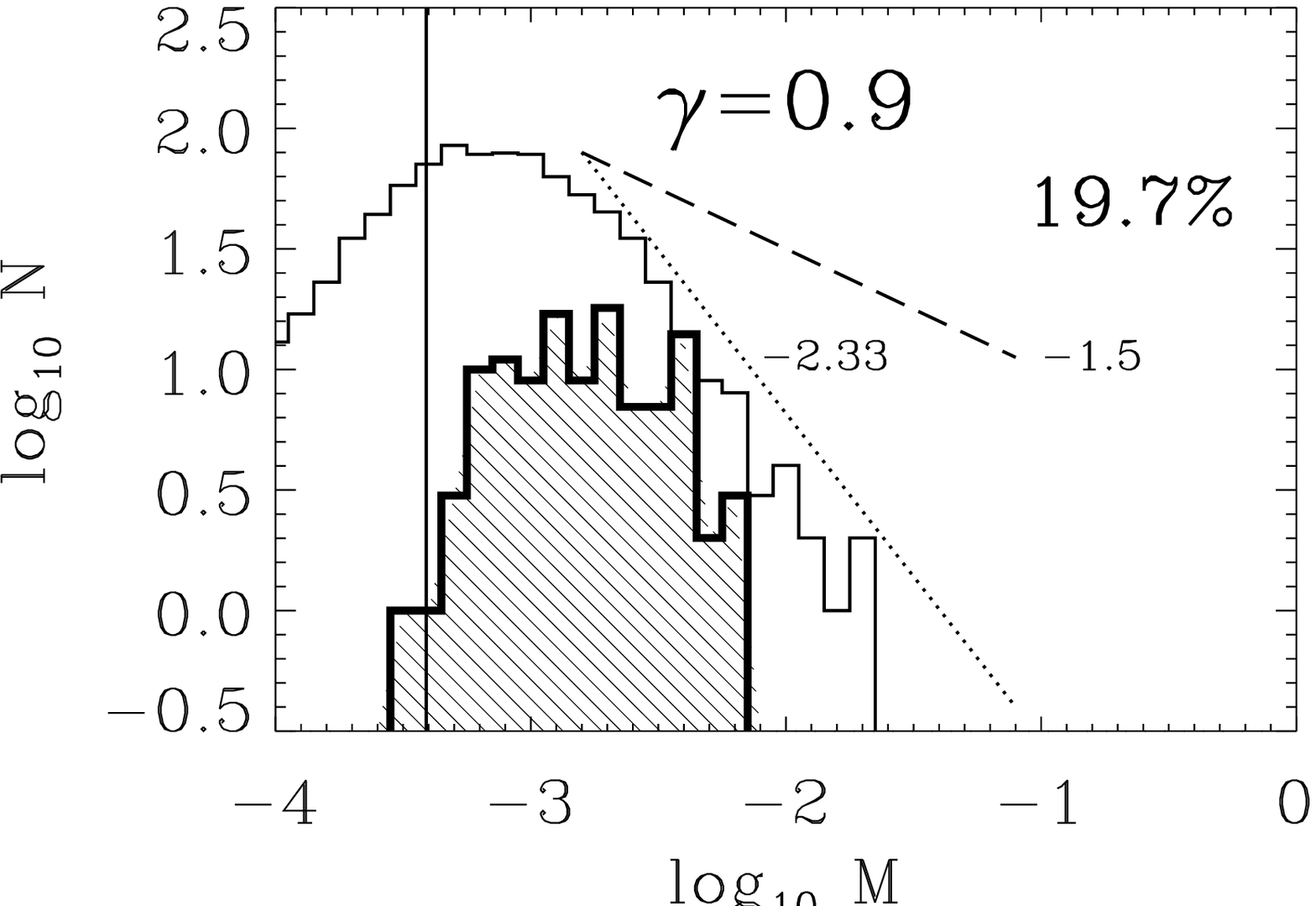}
\includegraphics[height=1.5in]{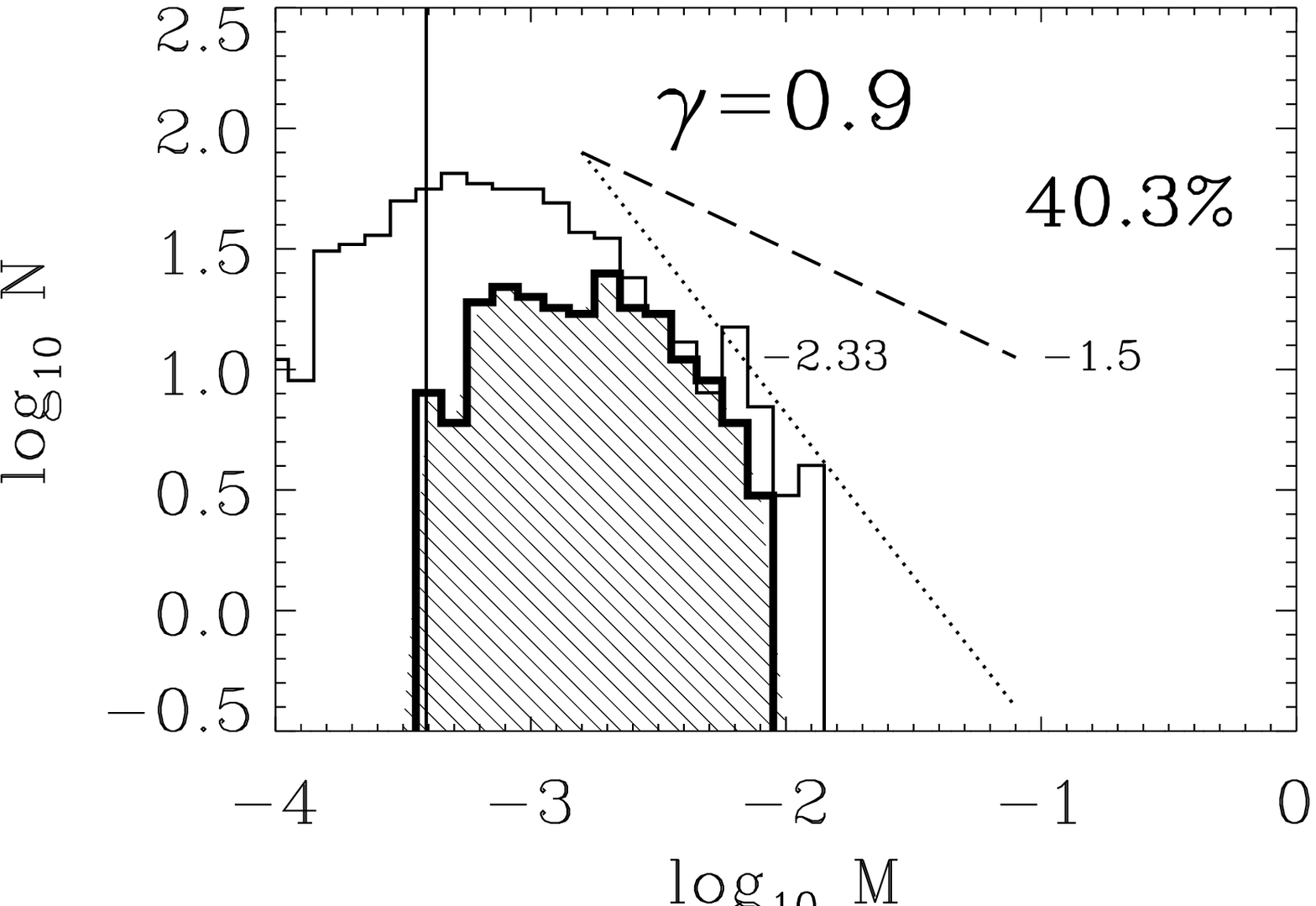}\\
\includegraphics[height=1.5in]{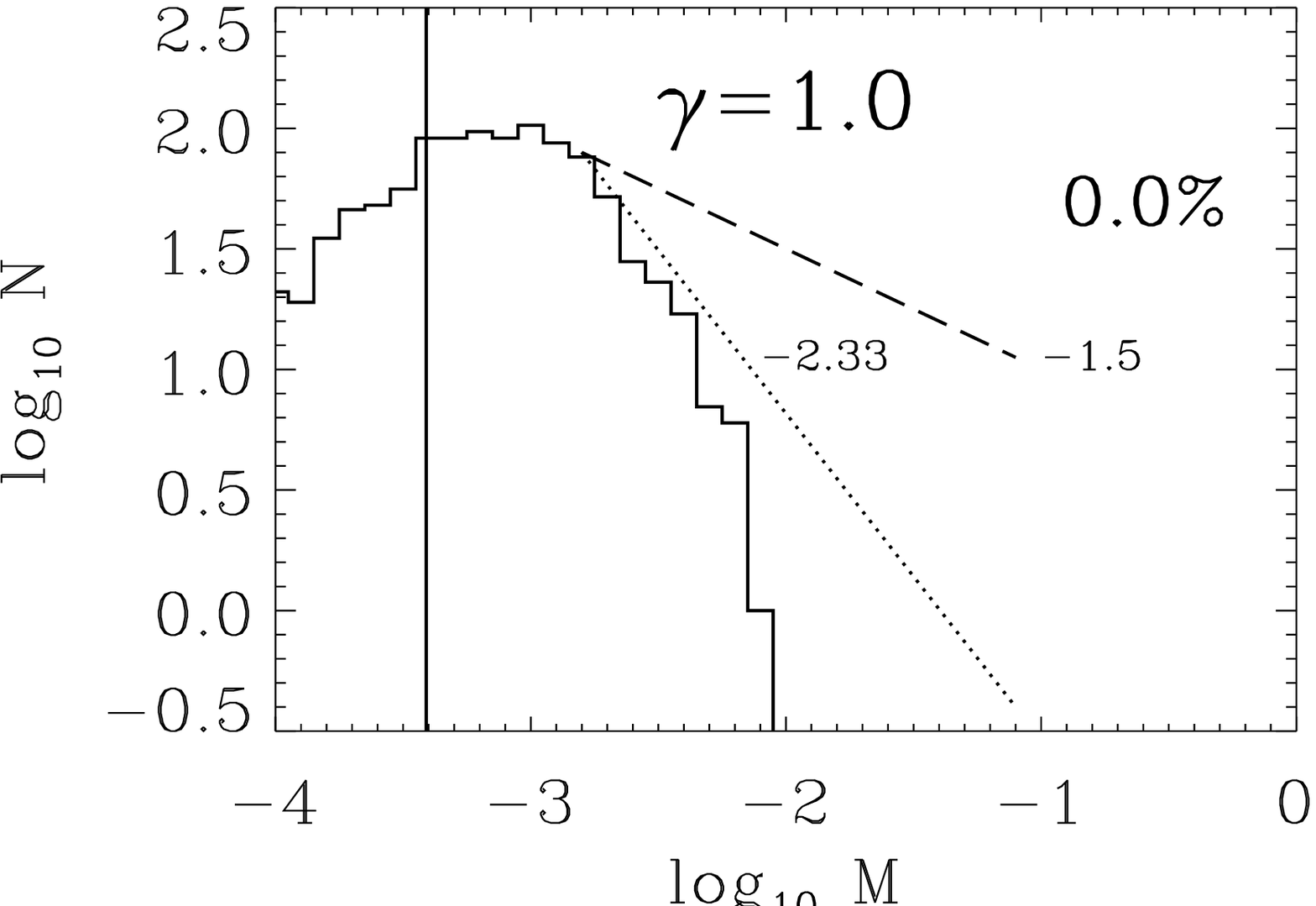}
\includegraphics[height=1.5in]{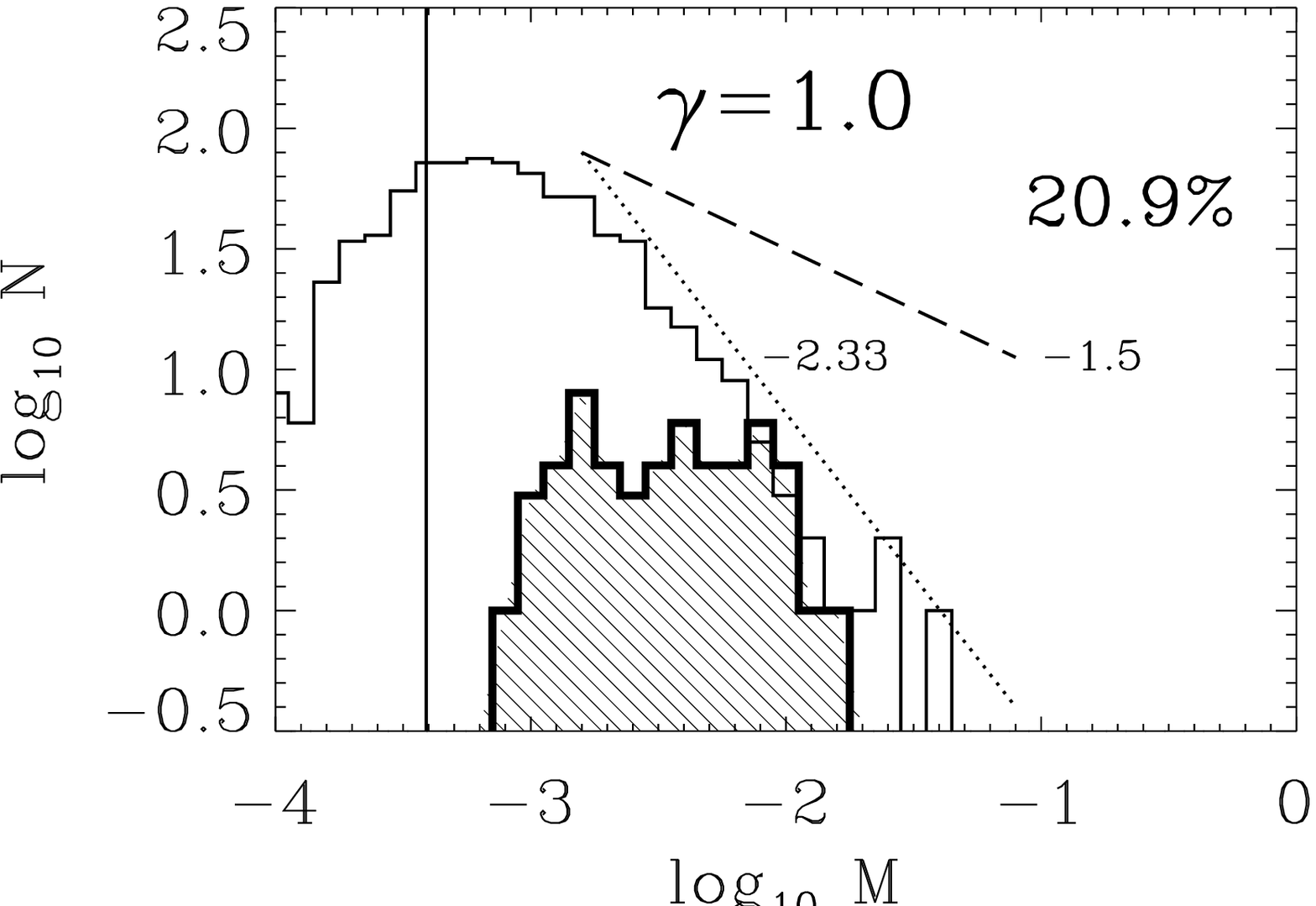}
\includegraphics[height=1.5in]{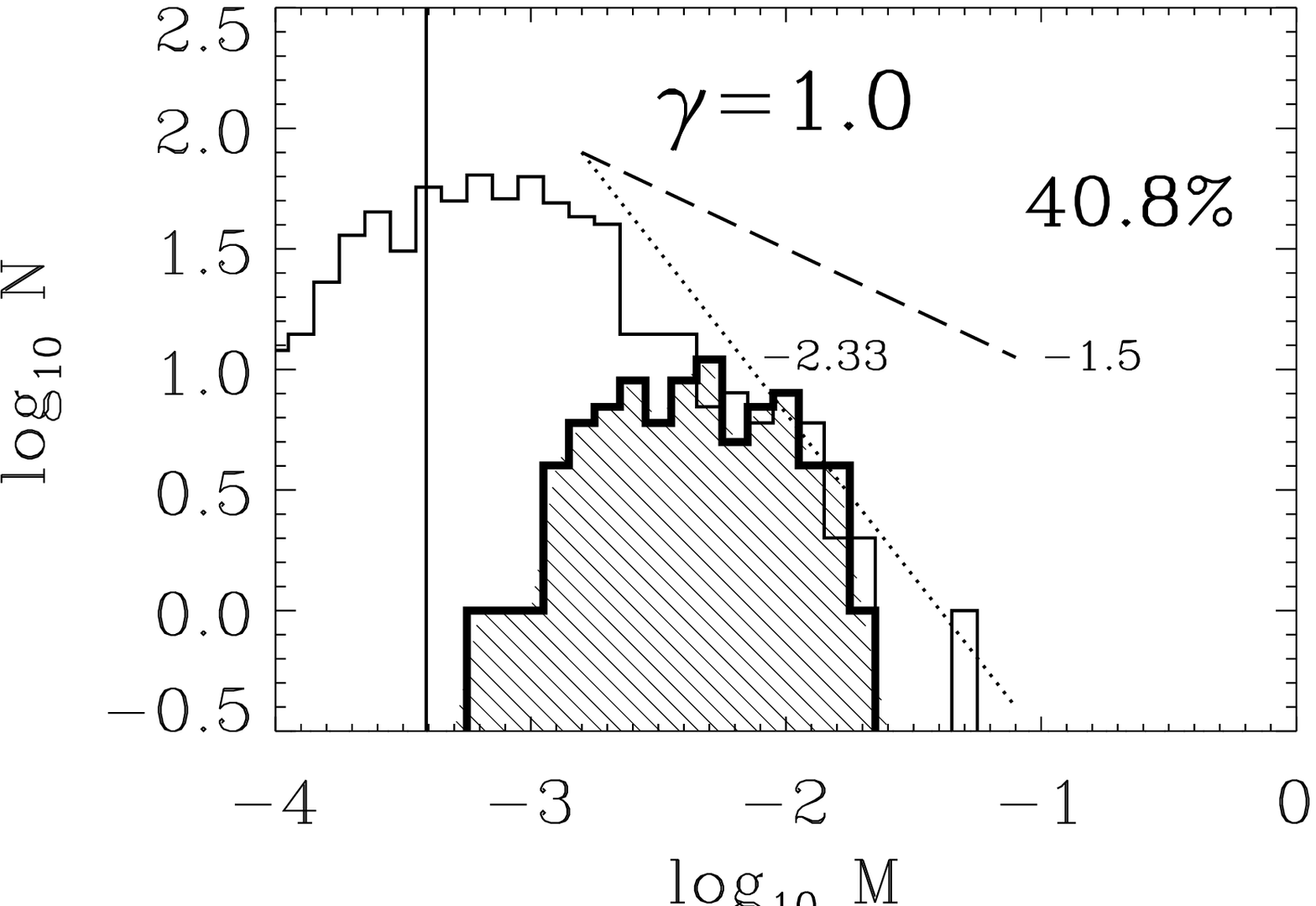}\\
\includegraphics[height=1.5in]{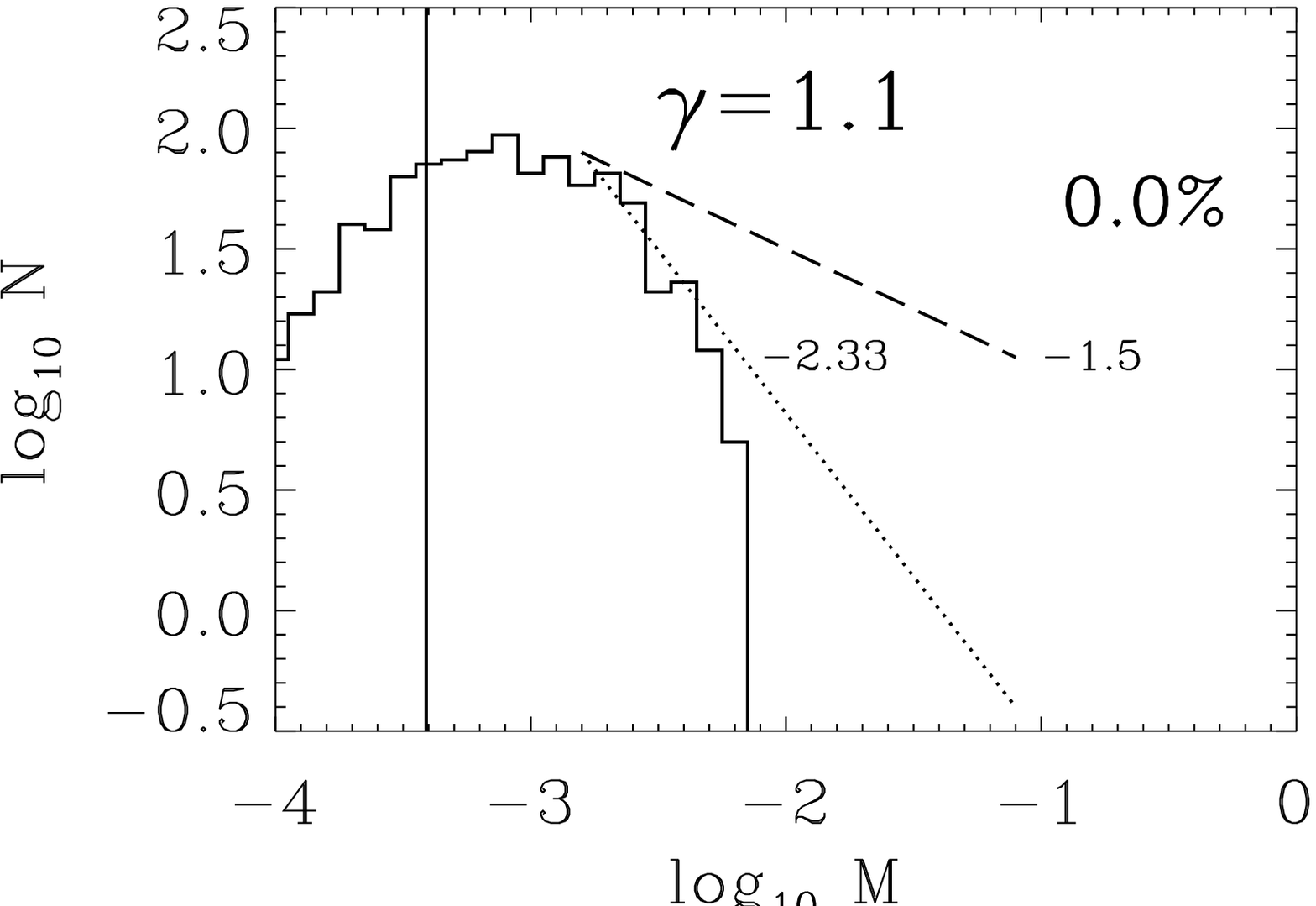}
\includegraphics[height=1.5in]{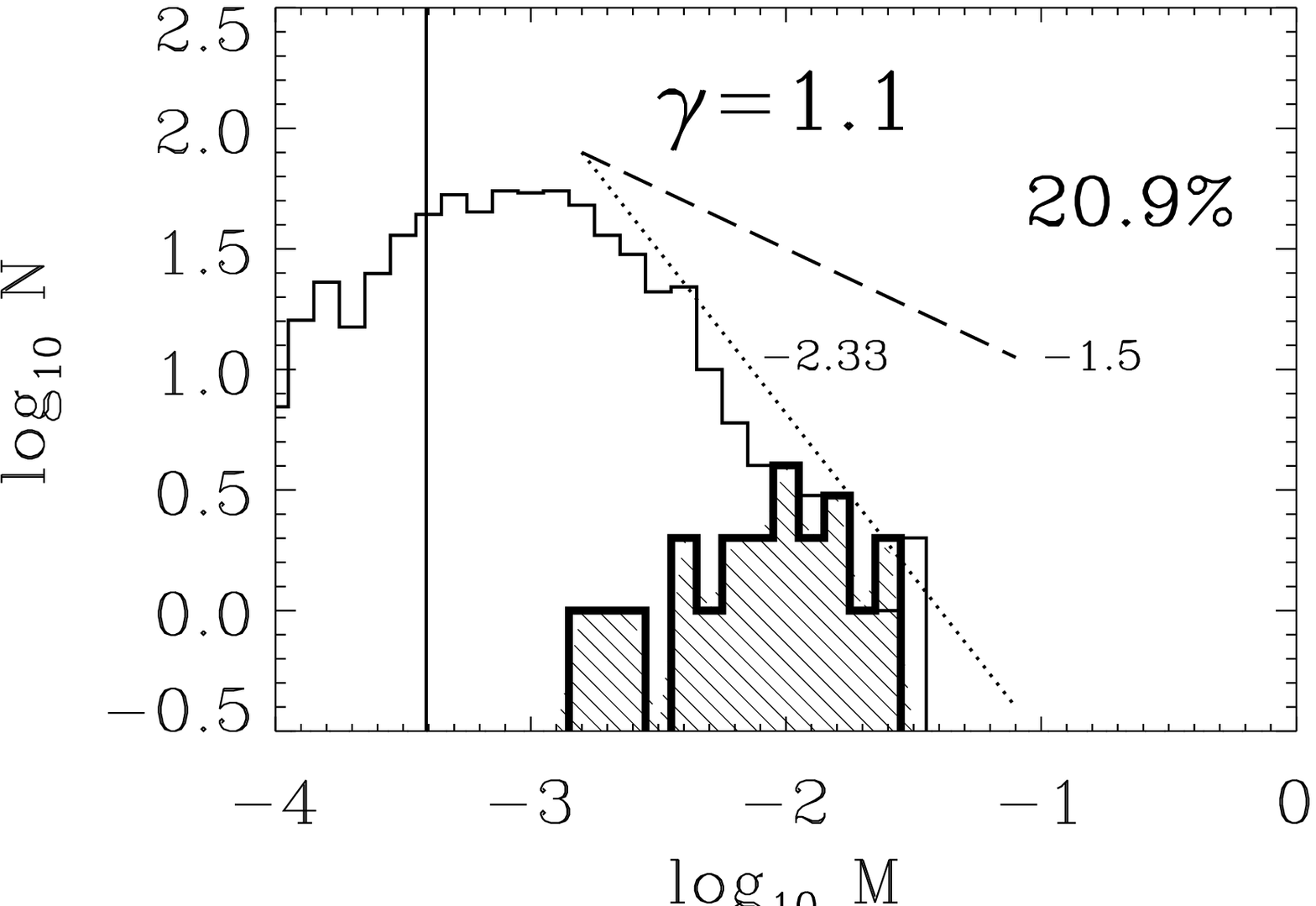}
\includegraphics[height=1.5in]{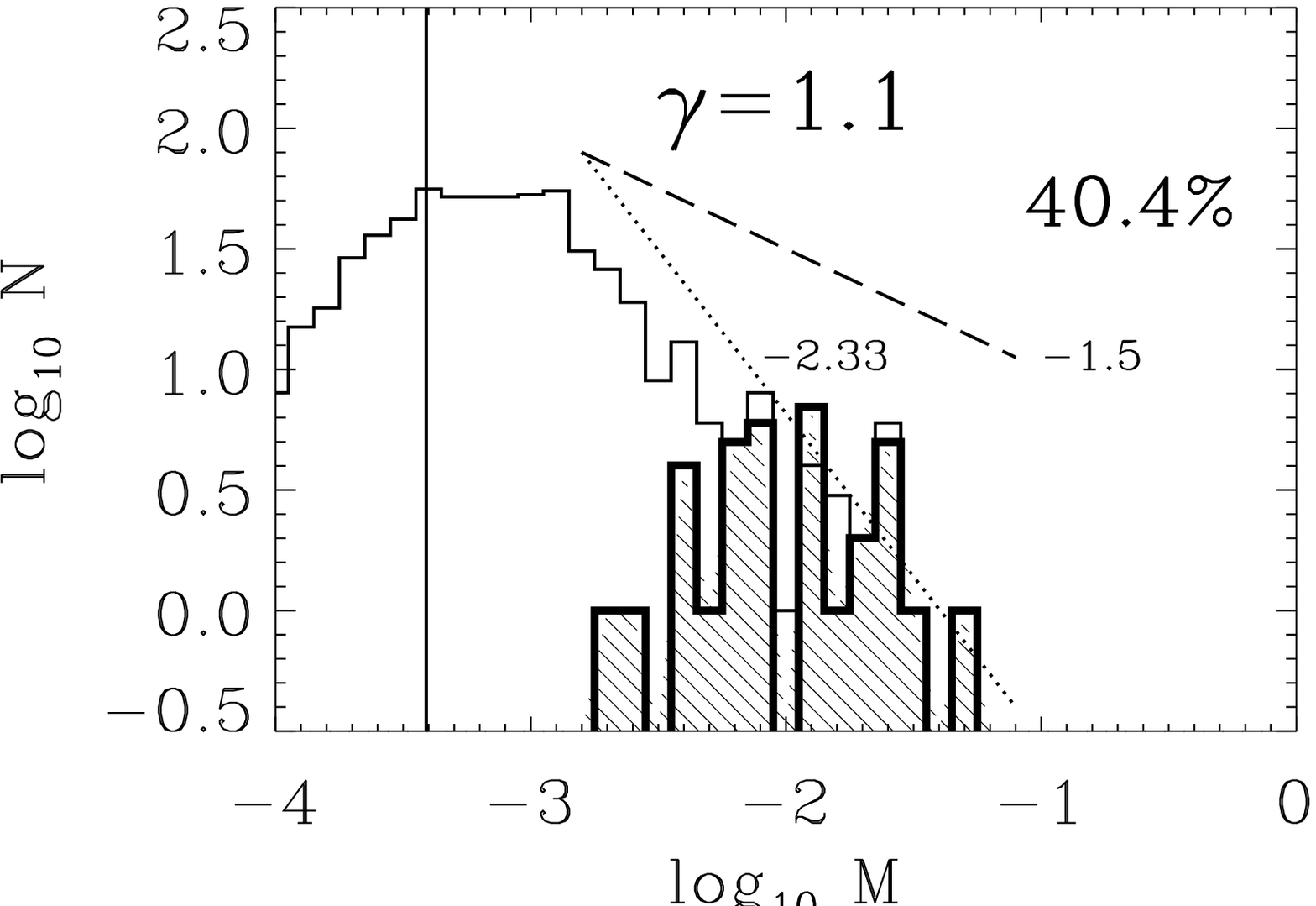}\\
\includegraphics[height=1.5in]{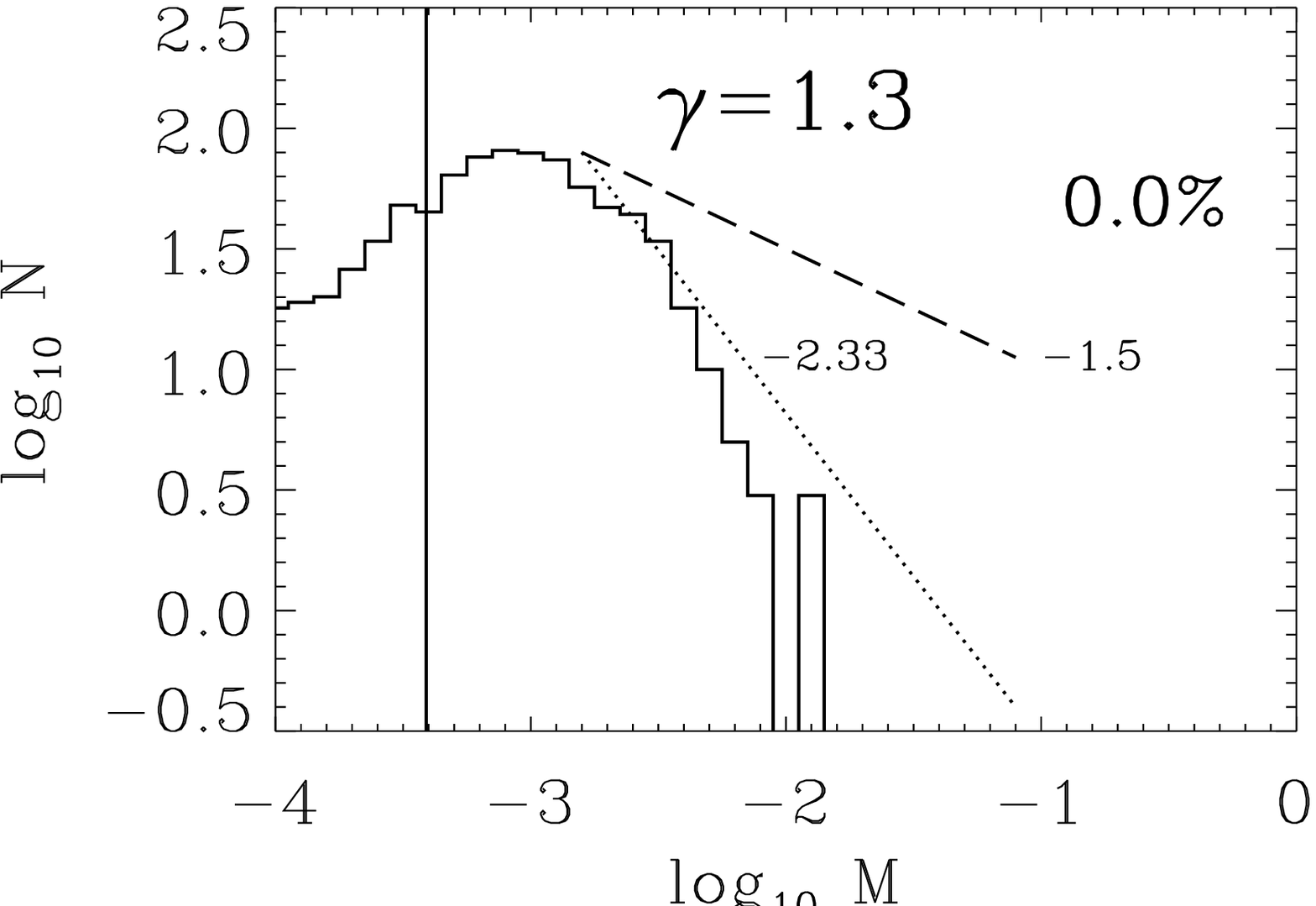}
\includegraphics[height=1.5in]{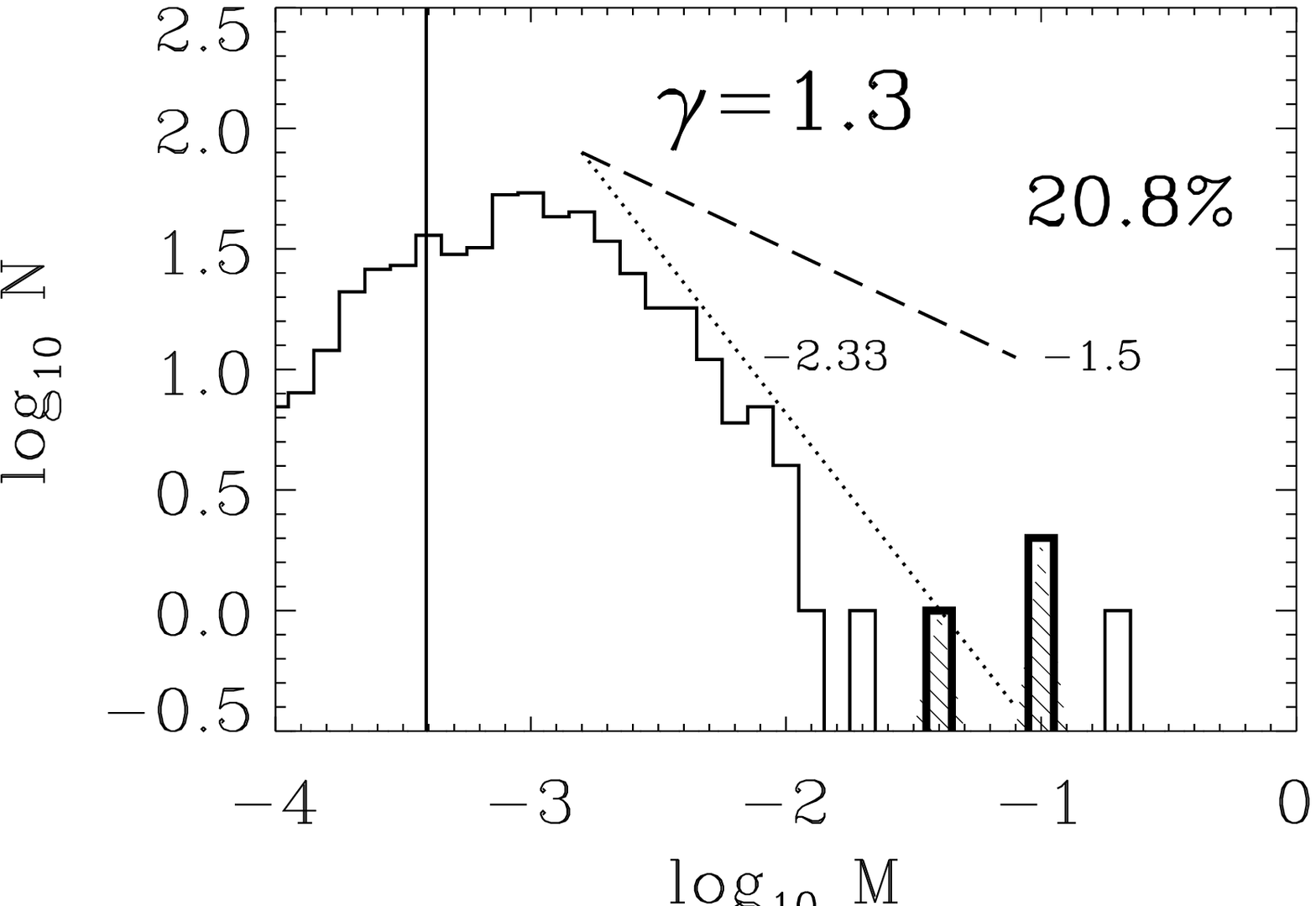}
\includegraphics[height=1.5in]{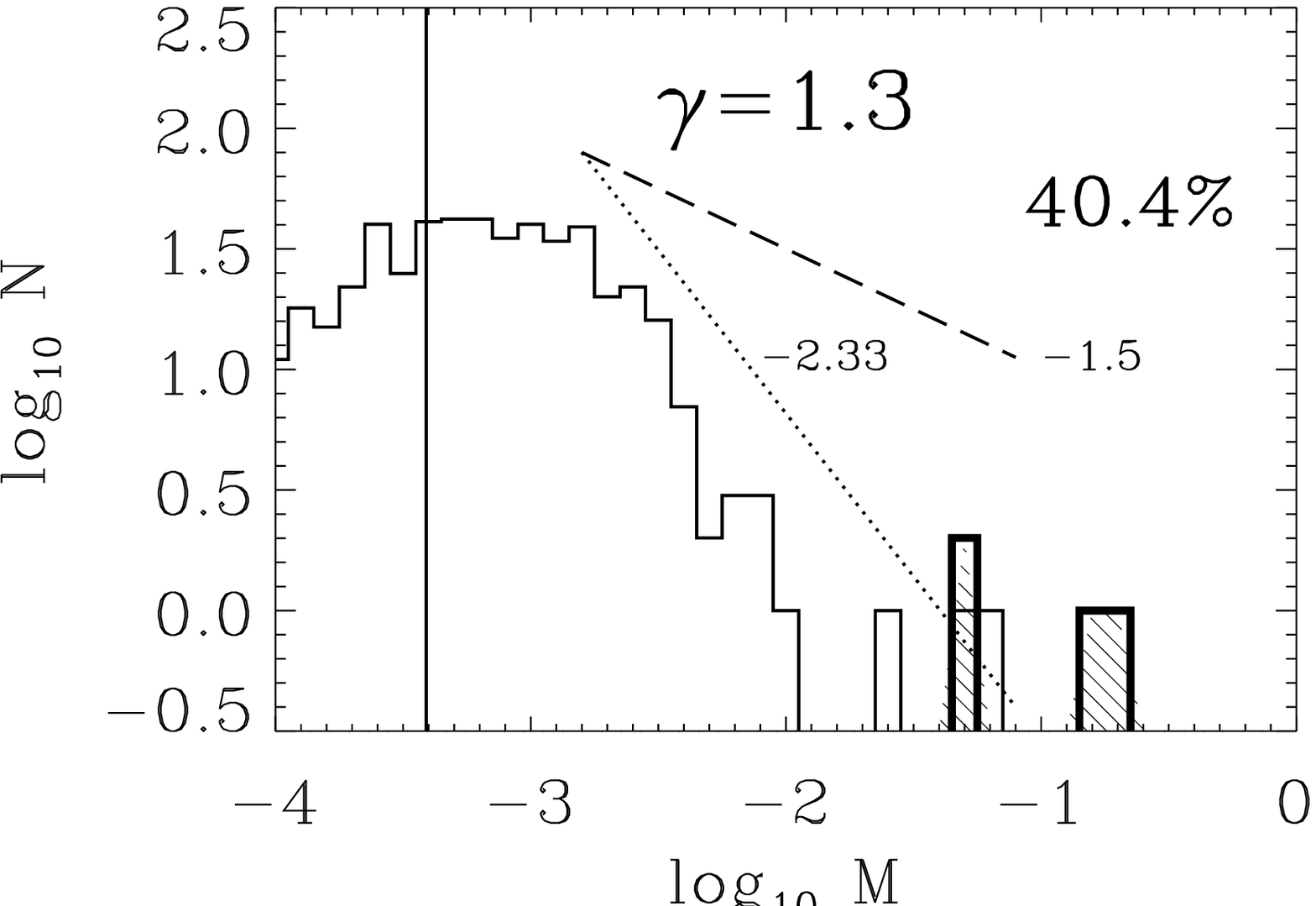}
\caption{\label{fig_mas}Mass spectra of gas clumps (\textit{open
histogram}), and of collapsed cores (\textit{filled histogram}) for
models with different $\gamma$ and driving with $k=1$--2. The fraction
of mass accreted into collapsed cores (sink particles) is given for
the right two columns. The vertical line shows the SPH resolution
limit. Also shown are two power-law spectra with $\nu = -1.5$
(dashed-line) and $\nu = -2.33$ (dotted line).}
\end{center}
\end{figure}

\clearpage

\clearpage

\clearpage

\clearpage

\clearpage

\begin{thebibliography}{JUNK}

\bibitem[Abel, Bryan, \& Norman(2000)]{abn00} Abel, T, Bryan, G., \& Norman,
  M. L. 2000, ApJ, 540, 39 
\bibitem[Abel, Bryan, \& Norman(2002)]{abn02} Abel, T, Bryan, G., \& Norman,
  M. L. 2002, Science, 295, 93 
\bibitem[Bate, Bonnell, \& Price(1995)]{bbp95} Bate, M. R., Bonnell, I. A., \&
  Price, N. M. 1995, MNRAS, 277, 362
\bibitem[Bate \& Burkert(1997)]{bab97} Bate, M. R., \& Burkert, A. 1997,
  MNRAS, 288, 1060 
\bibitem[Benz(1990)]{benz90} Benz, W. 1990, in \textit{The Numerical Modeling
  of Nonlinear Stellar Pulsations}, ed. J. R. Buchler (Dordrecht: Kluwer), 269
\bibitem[Bonnell \& Bastien(1991)]{bob91} Bonnell, I. \& Bastien, P. 1991,
  ApJ, 374, 610  
\bibitem[Bonnell \& Bastien(1993)]{bob93} Bonnell, I. \& Bastien, P. 1993,
  ApJ, 406, 614  
\bibitem[Bonnell, Bate, \& Zinnecker(1998)]{bbz98} Bonnell, I. A., Bate,
  M. R., \& Zinnecker, H. 1998, MNRAS, 298, 93   
\bibitem[Bromm, Coppi, \& Larson(1999)]{bcl99} Bromm, V., Coppi, P. S., \&
  Larson, R. B. 1999, ApJ, 527, L5 
\bibitem[Bromm, Coppi, \& Larson(2002)]{bcl02} Bromm, V., Coppi, P. S., \&
  Larson, R. B. 2002, ApJ, 564, 23 
\bibitem[Elmegreen(1993)]{elm93} Elmegreen, B. G. 1993, ApJ, 419, L29 
\bibitem[Elmegreen(2002)]{elm02} Elmegreen, B. G. 2002, ApJ, 577, 206 
\bibitem[Galli et al.(2001)]{galli01} Galli, D., Shu, F. H., Laughlin, G., \&
  Lizano, S. 2001, ApJ, 551, 367  
\bibitem[Hartmann(1998)]{hart98} Hartmann, L. 1998, \textit{Accretion
  Processes in Star Formation} (Cambridge: Cambridge Univ. Press), 33
\bibitem[Heitsch, Mac Low, \& Klessen(2001)]{hmk01} Heitsch, F., Mac Low,
  M.-M. \& Klessen, R. S. 2001, ApJ, 547, 280 (Paper III)
\bibitem[Jeans(1902)]{jeans02} Jeans, J. H. 1902, Phil. Trans.A., 199, 1   
\bibitem[Klessen(1997)]{rsk97} Klessen, R. S. 1997, MNRAS, 292, 11 
\bibitem[Klessen(2000)]{rsk00} Klessen, R. S. 2000, ApJ, 535, 869
\bibitem[Klessen(2001)]{rsk01} Klessen, R. S. 2001, ApJ, 556, 837 (Paper IV)
\bibitem[Klessen \& Burkert(2000)]{kb00} Klessen, R. S., \& Burkert,
  A. 2000, ApJS, 28, 287 (Paper I) 
\bibitem[Klessen, Burkert, \& Bate(1998)]{kbb98} Klessen, R. S., Burkert, A.,
  \& Bate, M. R. 1998, ApJ, 501, L205
\bibitem[Klessen, Heitsch, \& Mac Low(2000)]{khm00} Klessen, R. S., Heitsch,
  F., \& Mac Low, M.-M. 2000, ApJ, 535, 887 (Paper II)
\bibitem[Kroupa(2002)]{kroupa02} Kroupa, P. 2002, Science, 295, 82
\bibitem[Lamers et al.(2002)]{lamers02} Lamers, H. J. G. L. M. et al. 2002,
  ApJ, 566, 818  
\bibitem[Low \& Lynden-Bell(1976)]{low76} Low, C., \& Lynden-Bell, D. 1976,
  MNRAS, 176, 367 
\bibitem[Mac Low et al.(1998)]{mm98} Mac Low, M.-M., Klessen, R. S., Burkert,
  A., \& Smith, M. D. 1998, Phys.\ Rev.\ Lett., 80, 2754  
\bibitem[Mac Low(1999)]{mm99} Mac Low, M.-M. 1999, ApJ, 524, 169 
\bibitem[ Mac Low \& Klessen(2003)]{mk03} Mac Low, M.-M., \& Klessen,
  R. S. 2003, Rev.\ Mod.\ Phys., submitted (astro-ph/0301093)
\bibitem[Massey(2002)]{mas02} Massey, P. 2002, ApJS, 141, 81 
\bibitem[Ostriker, Stone, \& Gammie(2001)]{osg01} Ostriker, E. C., Stone,
  J. M., \& Gammie, C. F. 2001, ApJ 546, 980 
\bibitem[Padoan \& Nordlund(1999)]{pn99} Padoan, P., \& Nordlund, A. 1999,
  ApJ, 526, 279   
\bibitem[Passot \& V\'azquez-Semadeni(1998)]{pv98} Passot, T., \&
  V\'azquez-Semadeni, E. 1998, Phys.\ Rev.\ E, 58, 4501  
\bibitem[Pudritz(2002)]{pud02} Pudritz, R., 2002, Science, 295, 68 
\bibitem[Rouleau \& Bastien(1990)]{rb90} Rouleau, F. \& Bastien, P., 1990,
  ApJ, 355, 172  
\bibitem[Salpeter(1955)]{salp55} Salpeter, E. E. 1955, ApJ, 121, 161 
\bibitem[Scalo(1998)]{scalo98} Scalo, J. 1998, in ASP Conf. Ser. 142,
  \textit{The Stellar Initial Mass Function} (38th Herstmonceux Conference),
  ed. G. Gilmore \& D. Howell (San Francisco: ASP), 201  
\bibitem[Smith, Mac Low, \& Heitsch(2000)]{smh00}  Smith, M.~D., Mac Low,
  M.-M., \& Heitsch, F.\ 2000, \aap, 362, 333 
\bibitem[Spaans \& Silk(2000)]{ss00} Spaans, M., \& Silk, J. 2000, ApJ, 538,
  115  
\bibitem[V\'azquez-Semadeni \& Garcia(2001)]{vg01} V\'azquez-Semadeni, E. \&
  Garcia, N. 2001, ApJ, 557, 727 
\bibitem[V\'azquez-Semadeni,  Ballesteros-Paredes, \& Klessen (2003)]{vbk03}
  V\'azquez-Semadeni, E., Ballesteros-Paredes, J., \& Klessen, R.\ S. 2003,
  ApJ, in press (astro-ph/0301546) 
\bibitem[Ward-Thompson(2002)]{ward02} Ward-Thompson, D. 2002, Science, 295, 76 
\bibitem[Williams, De Geus, \& Blitz(1994)]{wdb94} Williams, J. P., De Geus,
  E. J., \& Blitz, L. 1994, ApJ, 428, 693 
\bibitem[Williams,  Blitz, \& McKee(2000)]{wbm00} Williams, J. P., Blitz, L.,
  \& McKee, C. F. 2000, in \textit{Protostars and Planets IV},
  ed. V. Mannings, A. P. Boss, \& S. S. Russell (Tucson: Univ. Arizona Press),
  97  
\bibitem[Yorke \& Sonnhalter(2002)]{ys02} Yorke, H.~W.~\& Sonnhalter, C.\ 2002,
  ApJ, 569, 846  

\end{thebibliography}
\end{document}